\documentclass{LMCS}

\usepackage{amssymb,graphicx,enumerate,wrapfig}
\usepackage{hyperref}
\usepackage[all]{xy}
\usepackage{colortbl}
\usepackage{ifpdf}

\usepackage{comment,changebar}

\newenvironment{proofof}[1]{
\begin{trivlist}\item[\hskip\labelsep{\em Proof~of~{#1}.\ }]}
{\hspace*{\fill} \qed\end{trivlist}}

\newcommand{\precocong}{precocongruence}

\newcommand{\precocongs}{precocongruences}
\newcommand{\Precocongs}{Precocongruences}

\newcommand{\eqcl}[1]{#1^{e}} 
\newcommand{\inrel}[2]{{#1}_{#2}} 

\newcommand{\predlif}{\lambda} 
\newcommand{\transp}[1]{\hat{#1}} 

\newcommand{\Mon}{\functor{Mon}}
\newcommand{\MonIF}{\Mon_{\omega}}
\newcommand{\TTwoIF}{{\TTwo}_{\!\!\omega}}

\newcommand{\aug}[1]{#1^\text{aug}}
\newcommand{\krp}[1]{#1^\text{krp}}

\newcommand{\vareps}{\varepsilon} 
\newcommand{\Fct}[1]{\functor{#1}}    

\DeclareMathOperator{\gra}{Gr}  

\newcommand{\mop}[1]{[#1]}  

\newcommand{\ufhat}[1]{\hat{#1}} 
\newcommand{\SetOp}{\cat{Set}^\mathrm{op}} 

\renewcommand{\k}{k}
\newcommand{\Nbk}{{\functor{N}^\k}}
\newcommand{\card}[1]{\lvert{#1}\rvert}  

\newcommand{\str}[1]{\mathcal{#1}}   
\newcommand{\nstr}[1]{\mathcal{#1}}  
\newcommand{\fstr}[1]{\mathfrak{#1}}  
\newcommand{\cla}[1]{\mathbf{#1}}   
\newcommand{\functor}[1]{\mathsf{#1}}

\newcommand{\cat}[1]{\mathsf{#1}}
\newcommand{\bb}[1]{\mathbb{#1}}
\newcommand{\ax}[1]{\mathsf{#1}}    
\newcommand{\logic}[1]{\mathbf{#1}}    

\newcommand{\Set}{\mathsf{Set}}   
\newcommand{\Pow}{\mathcal{P}}
\newcommand{\cmp}[1]{#1^{c}} 
\newcommand{\Two}{\functor{2}}
\newcommand{\TTwo}{\Two^{\Two}}

\newcommand{\up}[1]{{\uparrow\!{#1}}}		

\newcommand{\Coalg}[1]{\mathsf{Coalg}(\functor{#1})}

\newcommand{\ra}{\rightarrow}		
\newcommand{\lra}{\leftrightarrow}		
\newcommand{\Lra}{\Leftrightarrow}		

\newcommand{\Ra}{\Rightarrow}		

\newcommand{\surjR}{\twoheadrightarrow}

\newcommand{\dom}{{\mathsf{dom}}}  
\newcommand{\rng}{{\mathsf{rng}}}  
\newcommand{\inc}{\iota}     

\newcommand{\tup}[1]{\langle #1 \rangle} 
\newcommand{\ext}[1]{[\![ {#1}]\!]}   

\newcommand{\langML}{\mathcal{L}} 
\newcommand{\langFO}{\mathcal{L}_1} 

\newcommand{\modeq}{\equiv}   
\newcommand{\modcls}[1]{[#1]_\modeq}       
\newcommand{\MOC}{\mathrm{MOC}}
\newcommand{\modeledby}{=\joinrel\mathrel|}

\newcommand{\st}{\mathit{st}}

\newcommand{\nbm}[1]{{#1_\circ}}  
\newcommand{\fom}[1]{{#1^\circ}}  

\newcommand{\pb}{\mathrm{pb}}
\newcommand{\id}{\mathrm{id}}
\newcommand{\Fun}{\Fct{F}}
\newcommand{\sse}{\subseteq}

\newcommand{\coloneqq}{\mathrel{:=}}
\newcommand{\bis}{\mathrel{\mathchoice
{\raisebox{.3ex}{$\,
  \underline{\makebox[.7em]{$\leftrightarrow$}}\,$}}
{\raisebox{.3ex}{$\,
  \underline{\makebox[.7em]{$\leftrightarrow$}}\,$}}
{\raisebox{.2ex}{$\,
  \underline{\makebox[.5em]{\scriptsize$\leftrightarrow$}}\,$}}
{\raisebox{.2ex}{$\,
  \underline{\makebox[.5em]{\scriptsize$\leftrightarrow$}}\,$}}}}
\newcommand{\beh}{\bis_b}
\newcommand{\rbeh}{\bis_p}
\renewcommand{\TTwo}{\contra^\contra}
\newcommand{\contra}{\mathsf{2}}

\newcommand{\range}{\mathsf{rng}}

\newcommand{\wlift}[1]{\mathit{Lif}({#1})}

\newcommand{\Uf}{\mathrm{Uf}}
\newcommand{\U}{\mathbf{u}}

\newcommand{\black}{\boxtimes}

\newcommand{\princ}{\mathsf{u}}

\newcommand{\At}{\mathsf{At}}

\newcommand{\Prop}{\mathsf{At}}
\newcommand{\Cons}{\mathsf{Cons}}

\newcommand{\fof}[1]{\mathsf{#1}} 
\newcommand{\sort}[1]{\mathsf{#1}}

\def\doi{5 (2:2) 2009}
\lmcsheading%
{\doi}
{1--38}
{}
{}
{Dec.~19, 2007}
{Apr.~\phantom{0}9, 2009}
{}   

\begin{document}

\title[Neighbourhood Structures: Bisimilarity and Basic Model
  Theory]{Neighbourhood Structures:\\Bisimilarity and Basic Model
  Theory\rsuper*}

\author[H.~H.~Hansen]{Helle Hvid Hansen\rsuper a}	
\address{{\lsuper a}Eindhoven University of Technology, FM group, P.O. Box 513,
5600~MB Eindhoven, Netherlands.}	
\email{h.h.hansen@tue.nl}  
\thanks{{\lsuper a}Supported by NWO grant 612.000.316.}	

\author[C.~Kupke]{Clemens Kupke\rsuper b}	
\address{{\lsuper b}Imperial College London, Department~of~Computing,
  180 Queen's Gate,
  London~SW7~2AZ, UK.}	
\email{ckupke@doc.ic.ac.uk}  
\thanks{{\lsuper b}Supported by NWO under FOCUS/BRICKS grant 642.000.502.}	

\author[E.~Pacuit]{Eric Pacuit\rsuper c}	
\address{{\lsuper c}Stanford University, Department of Philosophy, Stanford, CA 94305-2155, USA.}	
\email{pacuit@stanford.edu}  
\thanks{{\lsuper c}Supported by NSF grant OISE 0502312.}	

\titlecomment{{\lsuper*}This is an extended and revised version of \cite{HanKupPac07:CALCO-nbis}.}

\begin{abstract}
Neighbourhood structures are the standard semantic tool
used to reason about non-normal modal logics.
The logic 
of all neighbourhood models is called classical modal logic.
In coalgebraic terms, a neighbourhood frame is a 
coalgebra for the contravariant powerset functor composed with itself, 
denoted by $\TTwo$.
We use this coalgebraic modelling to derive notions of 
equivalence between neighbourhood structures.
$\TTwo$-bisimilarity and behavioural equivalence are 
well known coalgebraic concepts,
and they are distinct, since $\TTwo$ does not preserve weak pullbacks.
We introduce a third, intermediate notion whose witnessing relations
we call {\precocongs} (based on pushouts).
We give back-and-forth style characterisations for $\TTwo$-bisimulations
and {\precocongs}, we 
show that on a single coalgebra, 
{\precocongs} capture behavioural equivalence, and 
that between neighbourhood structures,
{\precocongs} are a better approximation of behavioural equivalence
than $\TTwo$-bisimulations.
We also introduce a notion of modal saturation for neighbourhood
models, and investigate its relationship with definability and
image-finiteness. 
We prove a Hennessy-Milner theorem 
for modally saturated and for image-finite
neighbourhood models.
Our main results are an analogue of 
Van Benthem's characterisation theorem 
and a model-theoretic proof of
Craig interpolation for classical modal logic.
\end{abstract}

\maketitle



\section{Introduction}

Neighbourhood semantics \cite{Chellas} 
forms a generalisation of Kripke semantics,
and it has become the standard tool for reasoning about 
non-normal modal logics in which (Kripke valid) principles such as
$\Box p \land \Box q \,\ra\, \Box(p \land q)$ and 
$\Box p \,\ra\, \Box(p \lor q)$ 
are considered not to hold.
In a neighbourhood model, with each state one associates a collection of 
subsets of the universe (called its neighbourhoods), and a modal formula 
$\Box\varphi$ is true at a state $s$ if 
the truth set of $\varphi$ is a neighbourhood of $s$. 
The modal logic of all neighbourhood models is
called classical modal logic.

Neighbourhood semantics was invented in 1970 by Scott and Montague
(independently in \cite{Scott:advice} and \cite{Montague:univ-grammar});
and Segerberg~\cite{Segerberg71:classic-ML} presents some basic results
about neighbourhood models and the classical modal logics that 
correspond to them. These and other salient results were incorporated
by Chellas in his textbook~\cite{Chellas}.
During the past 15-20 years, non-normal modal logics have emerged
in the areas of computer science and social choice theory,
where system (or agent) properties are 
formalised in terms of various notions of ability in strategic games
(e.g.~\cite{AlurHenzKupf02:ATL,Pau02:ML-coal-pow}).
These logics have in common that they are monotonic,
meaning they contain the above-mentioned formula 
$\Box p \,\ra\, \Box(p \lor q)$.
The corresponding property of neighbourhood models is that
neighbourhood collections are closed under supersets.
Non-monotonic modal logics occur in deontic logic 
(see e.g. \cite{Goble:Murder}) where monotonicity
can lead to paradoxical obligations, and in the modelling of 
knowledge and related epistemic notions 
(cf.~\cite{Vardi86:epistemic,PadGovSu07:KRAQ}).
Furthermore, the topological semantics of modal logic
can be seen as neighbourhood semantics 
(see \cite{CateGabSus:topoML} and references).

Neighbourhood frames are easily seen to be coalgebras for 
the contravariant powerset functor composed with itself, 
denoted $\TTwo$.
From a coalgebra point of view, neighbourhood structures are interesting 
since they constitute a general framework for 
studying coalgebraic modal logics in the style of 
Pattinson~\cite{Patt03:coalg-ML},
where modalities are defined in terms of predicate liftings.
It can easily be shown that any (unary) modality defined in this way,
can be viewed as a neighbourhood modality. 
Furthermore, in much work on coalgebra (cf. \cite{Rut00:TCS-univ-coal}) 
it is often assumed that the functor preserves weak pullbacks,
however, it is not always clear whether this requirement is really needed.
In \cite{GummSchr05:types-coal}, weaker functor requirements for congruences 
are studied, 
and $\TTwo$ provides an example of a functor which
does not preserve weak pullbacks in general, 
but only the special ones consisting of kernel pairs.

From the modal logic point of view,
coalgebra is interesting since it offers an abstract theory
which can be instantiated
to neighbourhood models, and help us generalise the well-known
Kripke notions such as bisimilarity and image-finiteness  
to neighbourhood models. 
For monotonic neighbourhood structures, these questions 
have already been addressed 
(cf.~\cite{Pau99:mon-bis,Han03:math-thesis,HanKup04:CMCS-UpP}),
but as mentioned in \cite{Pau99:mon-bis},
if one starts from elementary intuitions,
it is not immediately clear how to generalise monotonic bisimulation
to arbitrary neighbourhood structures.
The theory of coalgebra
provides us not with one, but with several notions of 
state equivalence in $\Fun$-coalgebras for an arbitrary
functor $\Fun$.
$\Fun$-bisimilarity and behavioural equivalence
are well known concepts, 
and it is generally known that
the two notions coincide 
if and only if the functor $\Fun$ preserves weak pullbacks 
\cite{Rut00:TCS-univ-coal}.
This is, for example, the case over Kripke frames
which are coalgebras for the covariant powerset functor $\Pow$, and
it explains some of the fundamental properties of 
Kripke bisimulation:
(i) Kripke bisimulations are characterised by 
back-and-forth conditions, 
which makes it possible to efficiently compute Kripke bisimilarity
over finite models as a greatest fixed point.
(ii)
The Hennessy-Milner theorem for normal modal logic
states that over the class of finite Kripke models,
two states are Kripke bisimilar if and only if they satisfy the same
modal formulas.
(iii)
Van Benthem's characterisation 
theorem~\cite{Benthem:phd,Benthem:Correspondence}
tells us that Kripke bisimilarity characterises the modal fragment of
first-order logic. 
These properties of Kripke bisimulations
form the starting points of our investigation into
equivalence notions in
neighbourhood structures and classical modal logic.

As neighbourhood structures are coalgebras for a functor that 
does not preserve weak pullbacks, 
it is to be expected that only behavioural equivalence will 
give rise to a Hennessy-Milner theorem for classical modal logic.
However, it turns out to be very difficult to give a 
back-and-forth style characterisation of behavioural equivalence.
This motivates our introduction of a third equivalence notion
whose witnessing relations we call \precocongs, since they
can be seen as a two-coalgebra analogue of
the precongruences from~\cite{AczMen:final}.

The main contributions of this paper are:
(1) the introduction of {\precocongs} and basic results
which relate them to bisimulations and behavioural equivalence.
In particular, we show that on a single coalgebra,
the largest {\precocongs} is behavioural equivalence
(Theorem~\ref{t:on}), 
and that over neighbourhood models, 
{\precocongs} are a better approximation of
behavioural equivalence than $\TTwo$-bisimilarity;
(2) the definition of a notion of modal saturation
    for neighbourhood models, which leads to a
    behavioural-equivalence-somewhere-else result
    (Theorem~\ref{thm:somewhere})
    by showing that ultrafilter extensions are a 
    Hennessy-Milner class;
(3) a Van Benthem style characterisation of the
classical modal fragment of first-order logic 
(Theorem~\ref{thm:char}); and 
(4) a model-theoretic proof of Craig interpolation
for classical modal logic (Theorem~\ref{thm:interpolation}).

In section~\ref{sec:nbis-prelims} 
we define basic notions and notation.
In section~\ref{sec:equiv-notions}, 
we define {\precocongs} and
investigate their relationship with  
bisimulations and behavioural equivalence.
We also instantiate all three notions to
the concrete case of neighbourhood frames, 
provide back-and-forth style characterisations for
$\TTwo$-bisimulations and \precocongs, and
prove the results mentioned in (1).
In section~\ref{sec:HM-classes},
we introduce our notion of modal saturation for
neighbourhood models, and use it to prove 
a Hennessy-Milner theorem for
the class of finite neighbourhood models.
We then use general coalgebraic constructions
to define image-finite neighbourhood models 
and ultrafilter extensions of neighbourhood models, 
and show that these are also Hennessy-Milner classes.
Finally, in section~\ref{sec:model-theory} we prove
our main results as described in (3) and (4) above.
In particular,
we demonstrate that $\TTwo$-bisimulations  are
a useful tool for proving Craig interpolation of classical modal logic.

Since neighbourhood structures are of general interest
outside the world of coalgebra,
we have tried to keep this paper
accessible to readers who are not familiar with 
coalgebraic modal logic.
This means that some of our results could be obtained 
by instantiating more general results in coalgebra.
When this is the case, we give a brief explanation
in the form of a remark of how
the general coalgebraic framework instantiates to 
neighbourhood structures. 
However, these remarks are not necessary for understanding 
the main results of the paper.
On the other hand, we also hope that these remarks will inspire
readers to study the more general results.

\section{Preliminaries and notation}
\label{sec:nbis-prelims}

In this section, we settle on notation, 
define the necessary set-theoretic and coalgebraic notions, and
introduce neighbourhood semantics for modal logic.
For further reading on coalgebra
we refer to \cite{Rut00:TCS-univ-coal,Ven06:Handbook-ML-AC}.
We assume the reader is familiar with 
the Kripke semantics and
the basic model theory of normal modal logic.
Some knowledge of more advanced topics
such as modal saturation and ultrafilter extensions
will be useful.
All the necessary background information 
can be found in \cite{BdRV:ML-book}.
Extensive discussions on neighbourhood semantics can be found in 
\cite{Segerberg71:classic-ML,Chellas,Han03:math-thesis}.

\subsection{Functions and relations}
\label{ssec:functions-relations}

Let $X$ and $Y$ be sets. 
We denote by $\Pow(X)$ the powerset of $X$,
and by $X+Y$ the disjoint union of $X$ and $Y$.
If $Y \subseteq X$, then we write $\inc_Y$
for the inclusion map  $\inc_Y\colon Y \hookrightarrow X$;
$\cmp{Y}$ for the complement $X \setminus Y$ of $Y$ in $X$;
$Y \subseteq_\omega X$ if $Y$ is a finite subset of $X$; and
$\up{Y} = \{ Y' \subseteq X \mid Y \subseteq Y'\}$
for the {\em upwards closure of $\{Y\}$ in $\Pow(X)$}.

For a function $f\colon X \to Y$ and subsets $U \subseteq X$ and $V \subseteq
Y$ we define the {\em direct $f$-image of $U$} and the 
{\em $f$-preimage of $V$} by
putting $f[U] \coloneqq \{ f(x) \mid x \in U\}$  
and $f^{-1}[V] \coloneqq \{ x \in X \mid f(x) \in V\}$,
respectively. Furthermore we call $\dom(f)\coloneqq X$ the domain 
of $f$ and we call $\rng(f) \coloneqq f[X]$ the {\em range 
of $f$}. More generally, we also define the notions image, 
preimage, domain and range
for a relation $R \subseteq X \times Y$. 
For $U \subseteq X$ and $V \subseteq Y$,
we denote the {\em $R$-image of $U$} 
by $R[U] = \{ y \in Y \mid \exists x \in U : xRy \}$,
and the {\em $R$-preimage of $V$} by
$R^{-1}[V] = \{x \in X \mid \exists y \in V : xRy \}$.
The {\em domain of $R$} is $\dom(R) = R^{-1}[Y]$, and the 
{\em range of $R$} is $\rng(R) = R[X]$.
We will often work with a relation in terms of its projection maps.
Let $R \subseteq X_1 \times X_2$ be a relation.
The maps $\pi_1: R \to X_1$ and $\pi_2: R \to X_2$
denote the {\em projections} defined for all $\tup{x_1,x_2} \in R$ 
by $\pi_i(\tup{x_1,x_2}) = x_i$, for $i=1,2$.
$R$ is called a {\em bitotal} relation if $\pi_1$ and $\pi_2$ are surjective.
Note that for $U_i \subseteq X_i$, $i=1,2$, we have
$R[U_1] = \pi_2[\pi_1^{-1}[U_1]]$ and
$R^{-1}[U_2] = \pi_1[\pi_2^{-1}[U_2]]$.

If $R \subseteq X \times X$, 
then we denote by $\eqcl{R}$ the 
smallest equivalence relation on $X$ which contains $R$, 
and if $R$ is an equivalence relation on $X$ then $X/R$
is the set of $R$-equivalence classes.
A relation $R\subseteq X_1 \times X_2$,
can be viewed as a relation 
$\inrel{R}{X_1+X_2}$ on $X_1 + X_2$
by composing the projections with the canonical inclusion maps
$\inc_1: X_1 \to X_1 + X_2$
and $\inc_2: X_2 \to X_1 + X_2$.
More precisely, 
$\inrel{R}{X_1+X_2} = 
\{ \tup{\inc_1(x_1), \inc_2(x_2)} \mid \tup{x_1,x_2} \in R\}$.

Throughout this paper the notion of coherence will be used extensively.

\begin{defi}\label{defi:R-coh}
Let $X_1$ and $X_2$ be sets,
$R \subseteq X_1 \times X_2$ a relation, 
$U_1 \subseteq X_1$ and $U_2 \subseteq X_2$.
The pair {\em $\tup{U_1,U_2}$ is $R$-coherent} 
if: $R[U_1] \subseteq U_2$
and $R^{-1}[U_2] \subseteq U_1$.
For a set $X$, a relation 
$R \subseteq X \times X$ and $U \subseteq X$, 
we say that {\em $U$ is $R$-coherent},
if $\tup{U,U}$ is $R$-coherent.
\end{defi}

If $R \sse X_1 \times X_2$, then trivially, 
$\tup{\emptyset,\emptyset}$ and $\tup{X_1,X_2}$ are $R$-coherent.
Note that if
$R$ is an equivalence relation,
then an $R$-coherent subset $U$ is often called {\em $R$-closed}.
We list a number of useful properties of $R$-coherence
in the following two lemmas.
Their easy, but instructive, proofs are left to the reader.

\begin{lem}\label{lem:R-coh-1}\label{l:relate-coherency}
Let $R \subseteq X_1 \times X_2$ be a relation with projections 
$\pi_i : R \to X_i$, $i=1,2$.
For all
$U_1 \subseteq X_1$ and $U_2 \subseteq X_2$,
the following are equivalent:
\begin{enumerate}[\em(1)]
\item  $\tup{U_1,U_2}$ is $R$-coherent.
\item for all $\tup{x_1,x_2} \in R$: $x_1 \in U_1 \Lra x_2 \in U_2$.
\item $\pi_1^{-1}[U_1] = \pi_2^{-1}[U_2]$.
\item $U_1 + U_2$ is $\inrel{R}{X_1+X_2}$-coherent.
\end{enumerate}
\end{lem}

\begin{lem}\label{lem:R-coh-2}
Let $R \subseteq X \times X$ be a relation
and $U \subseteq X$.
The following are equivalent:
\begin{enumerate}[\em(1)]
\item  $U$ is $R$-coherent.
\item  $U$ is $\eqcl{R}$-coherent, i.e. $\eqcl{R}$-closed.
\item  $U$ is a union of $\eqcl{R}$-equivalence classes.
\item  $\cmp{U}$ is $\eqcl{R}$-coherent.
\end{enumerate}
\end{lem}

\subsection{Classical modal logic and neighbourhood semantics}

Let $\At=\{p_j\mid j\in\omega\}$ be a countable set of atomic sentences.
The {\em basic modal language over $\At$}, denoted $\langML(\At)$,  
is defined by the grammar: 
\[ \varphi ::= \bot \mid p_j\ |\ \neg\varphi\ |\ \varphi\wedge\varphi\ |\ \Box\varphi,\] 
where $j\in\omega$.  
We define $\top$, $\ra$ and $\lra$ in the usual way.
We will assume $\At$ to be fixed, and to ease notation, 
we write $\langML$ instead of $\langML(\At)$.

\begin{defi}\label{d:nbhd} 
A {\em neighbourhood frame} is a pair $\tup{S, \nu}$ 
where $S$ is a set of states and 
$\nu\colon S \to \Pow(\Pow(S))$ 
is a neighbourhood function which assigns to 
each state $s \in S$ its collection of neighbourhoods $\nu(s)$.   
A {\em neighbourhood model} based on a neighbourhood frame 
$\tup{S,\nu}$ 
is a triple $\tup{ S, \nu, V}$ 
where $V\colon\At\rightarrow \Pow(S)$ 
is a valuation function.  
\end{defi}

Given a neighbourhood model $\nstr{M}$, 
a state $s$ in $\nstr{M}$
and an $\langML$-formula $\varphi$, 
we write $\nstr{M},s \models \varphi$ 
to denote that $\varphi$ is true at $s$ in $\nstr{M}$,
and $\nstr{M},s \not\models \varphi$, 
if $\varphi$ is not true at $s$ in $\nstr{M}$.
Truth of the atomic propositions is defined via the valuation:
$\nstr{M}, s \models p_j$ iff $s \in V(p_j)$, and 
inductively over the boolean connectives as usual.
Truth of modal formulas is given by,
\begin{equation}\label{eq:def-box-truth}
\nstr{M},s\models\Box\phi \quad\text{ iff }\quad 
\ext{\phi}^\nstr{M}\in \nu(s),
\end{equation} 
where $\ext{\phi}^\nstr{M} = \{t \in S \ |\ \nstr{M},t \models\phi\}$ 
denotes the {\em truth set of $\phi$ in $\nstr{M}$}.     
Let also $\nstr{N}$ be a neighbourhood model.
Two states, $s$ in $\nstr{M}$ and $t$ in $\nstr{N}$, 
are {\em modally equivalent} 
(notation: $\nstr{M},s \modeq \nstr{N},t$ or simply $s \modeq t$),
if they satisfy the same modal $\langML$-formulas, 
i.e., $s \modeq t\;$ {if and only if} 
for all $\varphi \in \langML$: $\;\nstr{M},s \models \varphi\;$ iff
$\;\nstr{N},t \models \varphi$.
A subset $X \subseteq S$ is {\em modally coherent}\label{m:mod-coh}, 
if
for all $s, t \in S$ such that $s \modeq t$: $s \in X$ iff $t \in X$
i.e., $X$ is $\modeq$-coherent.

Let $\Phi \cup \{\varphi\} \subseteq \langML$.
We write $\Phi \models \varphi$ if 
$\varphi$ is a {\em local semantic consequence} of $\Phi$ over the class of 
all neighbourhood models, i.e., 
for any neighbourhood model $\nstr{M}$ and state $s$ in $\nstr{M}$,
if $\nstr{M}, s \models \Phi$ then $\nstr{M}, s \models \varphi$.
In particular,
if $\Phi \not\models\bot$ 
then $\Phi$ is called {\em consistent}, 
which means that $\Phi$ is satisfiable
in some neighbourhood model, and 
$\models \varphi$
means that $\varphi$ is valid in all neighbourhood models.
We define {\em classical modal logic} $\logic{E}$ to be the
theory of neighbourhood models, that is,
for all $\langML$-formulas $\varphi$: $\varphi \in \logic{E}$ iff $\models \varphi$.
We will not be concerned with proof theory or axiomatics.
For these matters, the reader is referred to \cite{Chellas}.

The structure preserving maps between neighbourhood structures 
will be referred to as bounded morphisms.
These have previously been studied in the context of algebraic duality
\cite{Dosen:nbhd-dua},
and monotonic neighbourhood structures 
(which we define in Remark~\ref{rem:mon-nbhd-coalg} below).

\begin{defi}\label{d:boundedmorph} 
If $\nstr{M}_1=\tup{ S_1, \nu_1, V_1}$ and
 $\nstr{M}_2=\tup{ S_2, \nu_2, V_2}$
are neighbourhood models, and $f \colon S_1 \to S_2$ is a function, 
then $f$ is a {\em (frame) bounded morphism from 
$\tup{S_1,\nu_1}$ to $\tup{S_2,\nu_2}$} 
(notation: $f \colon \tup{S_1,\nu_1} \to \tup{S_2,\nu_2}$)
if for all $s \in S_1$ and all $X\subseteq S_2$:
\begin{equation}\label{eq:b-m}
f^{-1}[X]\in \nu_1(s) \;\text{ iff }\; X\in \nu_2(f(s)).
\end{equation}
If also $s\in V_1(p_j)$ iff $f(s) \in V_2(p_j)$,
for all $p_j \in \At$, and all $s \in S_1$,
then $f$ is a {\em bounded morphism from $\nstr{M}_1$ to $\nstr{M}_2$}
(notation: $f \colon \nstr{M}_1 \to \nstr{M}_2$). 
\end{defi}
Bounded morphisms preserve truth of modal formulas.

\begin{lem}\label{l:boundedmorph}  
Let $\nstr{M}_1=\tup{S_1, \nu_1,V_1}$ and 
$\nstr{M}_2=\tup{S_2,\nu_2,V_2}$ be two neighbourhood models 
and $f \colon \nstr{M}_1\rightarrow \nstr{M}_2$ a bounded morphism.  
For each modal formula $\varphi\in\langML$ and state $s\in S_1$, 
$\nstr{M}_1,s\models\varphi$ iff $\nstr{M}_2,f(s)\models\varphi$.
\end{lem}
\proof
By a straightforward induction on the formula structure.
Details left to the reader.
\qed

Neighbourhood frames and bounded (frame) morphisms
form a category which we denote by $\cat{NbhdFr}$.
Similarly, neighbourhood models and bounded morphisms 
form a category $\cat{Nbhd}$.
This can easily be verified directly, but it also follows
from the straightforward coalgebraic modelling of 
neighbourhood strcutures which we describe now.

\subsection{Coalgebraic modelling}

We will work in the category $\Set$ of sets and functions.
Let $\Fun\colon\Set \to \Set$ be a functor.
An {\em $\Fun$-coalgebra} is a pair $\tup{X,\xi}$ 
where $X$ is a set, and $\xi\colon X \to \Fun(X)$ is a function,
sometimes called the {\em coalgebra map}.
Given two $\Fun$-coalgebras, $\tup{X_1,\xi_1}$ and $\tup{X_2,\xi_2}$,
a function $f\colon X_1 \to X_2$ is an {\em $\Fun$-coalgebra morphism}
if $\Fun(f) \circ \xi_1 = \xi_2 \circ f$, that is, 
the following diagram commutes:
\[ \xymatrix{
X_1 \ar[d]_-{\xi_1} \ar[r]^-{f} & X_2 \ar[d]^-{\xi_2}\\
\Fun(X_1) \ar[r]^-{\Fun(f) }& \Fun(X_2)
}\]
The category of
$\Fun$-coalgebras and $\Fun$-coalgebra morphisms 
is denoted by $\Coalg{\Fun}$. 
All notions pertaining to $\Fun$-coalgebras are parametric 
in the functor $\Fun$,
but if $\Fun$ is clear from the context or immaterial, 
we will often leave it out
and simply speak of coalgebras, coalgebra morphisms, and so on.
Several examples of systems which can be modelled as coalgebras 
can be found in~\cite{Rut00:TCS-univ-coal,Schr08:TCS-expr}.

   The contravariant powerset functor $\Two \colon \Set \to \Set$ 
   maps a set $X$ to $\Pow (X)$,
   and a function
   $f\colon X \to Y$ to the inverse image function 
   $f^{-1}[\_\,] \colon \Pow(Y) \ra \Pow(X)$. 
   The functor $\TTwo$ is defined as 
   the composition 
   of $\Two$ with itself.
That is, for any set $X$ and any function $f\colon X \to Y$,
\[\begin{array}{lcl}
\TTwo(X) &=& \Pow(\Pow(X)), \\
\TTwo(f)(U) &=& \{ D \sse Y \mid f^{-1}[D] \in U\} \text{ for all } U \in \TTwo(X).
\end{array}\]
It should be clear that $\cat{NbhdFr}$ and $\Coalg{\TTwo}$
have the same objects.
Similarly,
given a neighbourhood model $\tup{S,\nu,V}$,
we can view the valuation $V \colon\At \to \Pow(S)$
in its transposed form $\transp{V}\colon S \to \Pow(\At)$ 
where $p_j \in \transp{V}(s)$ iff $s \in V(p_j)$.
It is now easy to see that $\tup{S,\nu,V}$
uniquely corresponds to a coalgebra
$\tup{\nu,\transp{V}}\colon S \to \TTwo(S) \times \Pow(\At)$
for the functor 
$\TTwo(-) \times\Pow(\At)$.
Moreover, it is straightforward to show that
   a function $f \colon S_1 \ra S_2$ is a bounded morphism between
   the neighbourhood frames $\nstr{S}_1=\tup{S_1,\nu_1}$ and
   $\nstr{S}_2 =\tup{S_2,\nu_2}$ iff 
   $f$ is a coalgebra morphism from $\nstr{S}_1$ to $\nstr{S}_2$.
Similarly, $\TTwo(-) \times\Pow(\At)$-coalgebra morphisms
are simply the same as bounded morphisms between
neighbourhood models.
Hence $\cat{NbhdFr} = \Coalg{\TTwo}$ and 
$\cat{Nbhd} = \Coalg{\TTwo(-) \times\Pow(\At)}$.
From now on, 
we will switch freely between the coalgebraic setting
and the neighbourhood setting.

In the course of this paper, 
we will relate some of our results and definitions
to existing ones for monotonic modal logic 
and normal modal logic.
We briefly remind the reader of their definitions and
their relationship with neighbourhood structures and coalgebras.

\begin{rem}\label{rem:mon-nbhd-coalg}
A neighbourhood frame/model is {\em monotonic},
if for all $s \in S$, the collection of neighbourhoods 
$\nu(s)$ is {\em upwards closed}, i.e.,
if $U \subseteq V$ and $U \in\nu(s)$ then $V \in \nu(s)$.
Monotonic modal logic is the theory of 
monotonic neighbourhood models 
(cf.~\cite{Chellas,Han03:math-thesis}).
It was shown in~\cite{HanKup04:CMCS-UpP} that
monotonic neighboourhood frames are coalgebras
for the subfunctor 
$\Mon$ of $\TTwo$ which is defined by 
$\Mon(X) = \{ U \in \Pow(\Pow(X)) \mid U \text{ is upwards closed} \}$
on a set $X$.
\end{rem}

\begin{rem}\label{rem:krip-nbhd-coalg}
It is well known that Kripke frames and their bounded morphisms
can be seen as the category of coalgebras and coalgebra morphisms 
for the covariant powerset functor 
$\Pow\colon \Set \to \Set$ which maps a set $X$ to the powerset $\Pow(X)$,
and a function $f\colon X \to Y$  to the direct image function 
$f[\_\,]\colon \Pow(X) \to \Pow(Y)$. 

Kripke frames/models are in 1-1 correspondence with
so-called augmented neighbourhood frames/models
(cf.~\cite{Chellas}).
A neighbourhood frame $\tup{S,\nu}$ is {\em augmented}, 
if it is monotonic and
for all $s \in S$, $\bigcap\nu(s) \in \nu(s)$. 
In other words, in an augmented neighbourhood frame,
each neighbourhood collection is the upwards closure
of a unique, smallest neighbourhood.
Given a Kripke model $\nstr{K}= \tup{S,R,V}$, we obtain an
augmented neighbourhood model $\aug{\nstr{K}} = \tup{S,\nu,V}$,
by taking $\nu(s) = \up{R[s]}$ for all $s \in S$.
Conversely, 
given an augmented neighbourhood model $\nstr{M} = \tup{S,\nu,V}$,
we define the Kripke model  $\krp{\nstr{M}} = \tup{S,R,V}$
by taking $R[s] = \bigcap\nu(s)$ for all $s \in S$.
It shold be easy to see that these transformations 
are inverses of each other.
It is also straightforward to show that
for any two Kripke models
$\nstr{K}_1$ and $\nstr{K}_2 $,
a function 
is a Kripke bounded morphism from $\nstr{K}_1$ to $\nstr{K}_2$
iff
$f$ is a (neighbourhood) bounded morphism
from $\aug{\nstr{K}_1}$ to $\aug{\nstr{K}_2}$. 
Hence the category of Kripke frames is isomorphic
to the category of augmented neighbourhood frames.
Moreover, 
a Kripke model $\nstr{K}$ and its corresponding augmented
model $\aug{\nstr{K}}$ are pointwise equivalent,
i.e.,
for all states $s$ in $\nstr{K}$
and any $\langML$-formula $\varphi$:
$\nstr{K},s \models \varphi$ iff $\aug{\nstr{K}},s \models \varphi$.
This can be proved by an easy induction on $\varphi$ (cf.~\cite{Chellas}).
Normal modal logic is the logic of all Kripke models,
or equivalently, of all augmented neighbourhood models.
\end{rem}

\subsection{Basic constructions}

Finally, we will need a number of technical constructions.
Disjoint unions of neighbourhood structures
lift disjoint unions of sets to
neighbourhood structures such that the inclusion maps are bounded morphisms.
Disjoint unions are instances of the category theoretical notion of 
{\em coproducts}, and hence they satisfy a {\em universal property}
(which we will use in several proofs).
We give the concrete definition of disjoint unions 
neighbourhood models and their universal property,
The definition for neighbourhood frames is obtained by leaving out 
the part about the valuations. 

\begin{defi}\label{d:disjointunion} 
Let  $\nstr{M}_1=\tup{S_1,\nu_1,V_1}$ 
and  $\nstr{M}_2=\tup{S_2,\nu_2,V_2}$ be two neighbourhood models.
The {\em disjoint union of $\nstr{M}_1$ and $\nstr{M}_2$}  
is the neighbourhood model $\nstr{M}_1+\nstr{M}_2=\tup{ S_1+S_2,\nu,V}$ 
where
for all $p_j\in\At$, $V(p_j)=V_1(p_j)+ V_2(p_j)$; 
and  for $i=1,2$,
for all $X\subseteq S_1 +S_2$, and $s\in S_i$:
$X\in\nu(s)$ iff $X\cap S_i\in\nu_i(s)$.
$\nstr{M}_1+\nstr{M}_2$ has the following universal property:
If $\nstr{N}$ is a neighbourhood model
and $f_j\colon \nstr{M}_j \to \nstr{N}$, $j=1,2$, are bounded morphisms,
then there is a unique bounded morphism 
$f\colon \nstr{M}_1+\nstr{M}_2 \to \nstr{N}$ 
such that for $j=1,2$, $f_j=f \circ \inc_j$,
where $\inc_j\colon \nstr{M}_j \to \nstr{M}_1+\nstr{M}_2$ 
is the inclusion map.
\end{defi}

In the sequel we will also use coequalisers, pushouts and pullbacks.
The general definition of these notions can be found in any standard book
on category theory (for example~\cite{AdaHerStr90:ACC}).
We are interested in particular instances of these notions in $\Set$, and
we therefore only give the concrete definitions using the well 
known constructions.
We also give the universal property of coequalisers and pushouts,
which we will also use.

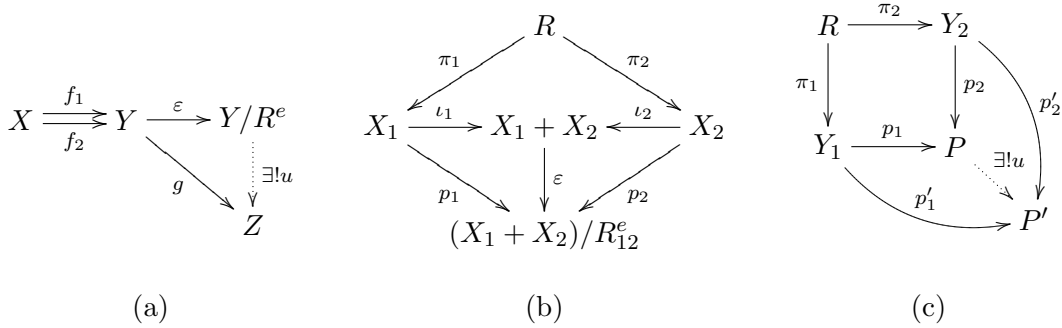
\begin{figure}[h!]
\caption{Coequalisers and Pushouts.}
\label{fig:pushout}
\[\begin{array}{ccccc}
\xymatrix{ \\
X \ar@<.5ex>[r]^-{f_1} \ar@<-.3ex>[r]_-{f_2} &
Y \ar[r]^-{\vareps} \ar[dr]_-{g} & Y/\eqcl{R} \ar@{.>}[d]^-{\exists ! u}\\
 & & Z
}
& \qquad &
\xymatrix@C=1pc{ 
& R \ar[dl]_{\pi_1} 
   \ar[dr]^{\pi_2} \\
X_1 \ar[r]^-{\inc_1} \ar[dr]_{p_1} &
X_1+X_2 \ar[d]^-{\vareps} &  
X_2 \ar[l]_-{\inc_2} \ar[dl]^{p_2} \\
& (X_1 + X_2)/\eqcl{R_{12}}
}& \qquad &
\xymatrix@R=1pc@C=1pc{
R \ar[dd]_-{\pi_1} \ar[rr]^-{\pi_2} && 
Y_2 \ar[dd]^-{p_2} \ar@/^1.2pc/[dddr]^-{p_2'}\\\\
Y_1 \ar[rr]^-{p_1} \ar@/_1.2pc/[rrrd]^-{p_1'} &&
P \ar@{.>}[dr]^(.4){\exists ! u}\\
&&& P'
} 
\\\\
 \text{(a)} && \text{(b)} && \text{(c)}
\end{array}
\] 
\end{figure}

\begin{defi}
(coequaliser) 
Let $f_1,f_2\colon X \to Y$ be a pair of functions.
The {\em coequaliser of $f_1$ and $f_2$ in $\Set$} is
the natural quotient map $\vareps\colon Y \to Y/\eqcl{R}$
where $R = \{\tup{f_1(x),f_2(x) \mid X}\}$.
For any function $g \colon Y \to Z$ such that $g\circ f_1 = g\circ f_2$
there is a unique function $u \colon X/\eqcl{R} \to Z$
such that $u\circ\vareps = g$.
The coequaliser of a relation $R \sse X \times X$ is 
the coequaliser of its projections $\pi_1,\pi_2\colon R \to X$.
 
(pushout) Let $R \sse X_1 \times X_2$ be a
relation with projections
$\pi_1\colon R \ra X_1$ and $\pi_2\colon R \ra X_2$.
The {\em pushout of $R$ in $\Set$} is the triple
$\tup{P,p_1,p_2}$,
where
$P \coloneqq \big( X_1 + X_2 \big)/\eqcl{R_{12}}$, 
$R_{12} = \inrel{R}{X_1+X_2} = 
\{ \tup{\inc_1(x_1),\inc_2(x_2)} \mid \tup{x_1,x_2} \in R\}$,
$\vareps\colon X_1 + X_2 \to (X_1 + X_2)/\eqcl{R_{12}}$
is the coequaliser of $\inc_1\circ\pi_1$ and $\inc_2\circ\pi_2$,
and
$p_i = \vareps\circ\inc_i$, $i\in \{1,2\}$.
The construction is illustrated in Figure~\ref{fig:pushout}(b).
Moreover, if $P'$, $p_1'\colon Y_1 \to P'$ and 
$p_2' \colon Y_2 \to P'$ are such that
$p_1'\circ \pi_1 = p_2'\circ \pi_2$, then there exists a unique
function $u \colon P \to P'$  such that
$p_1'= u \circ p_1$ and $p_2'= u \circ p_2$, 
as illustrated in Figure~\ref{fig:pushout}(c).

(pullback)   
Let $f_1\colon X_1 \ra Y$ and $f_2\colon X_2 \ra Y$ be functions.
   The {\em pullback of $f_1$ and $f_2$} in $\Set$
   is the triple $\tup{\pb(f_1,f_2),\pi_1,\pi_2}$,
   where
   $\pb(f_1,f_2) \coloneqq \{\tup{x_1,x_2} \in X_1 \times X_2 
   	\mid f_1(x_1) = f_2(x_2) \}$; and
   $\pi_1\colon \pb(f_1,f_2) \ra X_1$ and $\pi_2\colon\pb(f_1,f_2) \ra X_2$
   are the projections.
\end{defi}

Coproducts and coequalisers are a special form of colimit. 
It is known that for any functor $\Fun\colon\Set\to\Set$, 
all colimits exist in $\Coalg{\Fun}$ and they are constructed
essentially as in $\Set$,
see \cite[Section~4.4]{Rut00:TCS-univ-coal}.
We have already seen how this works for coproducts.
For coequalisers, 
it means that the coequaliser of two $\Fun$-coalgebra morphisms 
$f_1, f_2 \colon \tup{X,\xi} \to \tup{Y,\gamma}$
in $\Coalg{\Fun}$ is the same map $e \colon Y \to Y/\eqcl{R}$ which
is the coequaliser of $f_1$ and $f_2$ in $\Set$,
and there is a coalgebra structure 
$\lambda\colon Y/\eqcl{R} \to \Fun(Y/\eqcl{R})$
such that  $e$ is an $\Fun$-coalgebra morphism from
$\tup{Y,\gamma}$ to $\tup{Y/\eqcl{R},\lambda}$.

\section{Equivalence notions}
\label{sec:equiv-notions}

In this section we will study various notions of ``observational
equivalence" for neighbourhood frames in detail. In the first
part we list the three coalgebraic equivalence notions 
that we are going to consider. 
In the second part
we work out in detail what 
these three equivalence notions mean on neighbourhood frames. 

\subsection{Three coalgebraic notions of equivalence}
The main observation for defining equivalences between coalgebras
is that coalgebra morphisms preserve the behaviour of coalgebra 
states.
This basic idea motivates the well-known coalgebraic definitions of 
bisimilarity and behavioural equivalence. In the following
$\Fun$ denotes an arbitrary $\Set$ functor. 

\begin{defi}\label{d:fbis}
Let $\tup{X_1,\xi_1}$ and $\tup{X_2,\xi_2}$
be $\Fun$-coalgebras. 

(1)
A relation $R \sse X_1 \times X_2$ is 
   an {\em ($\Fun$-)bisimulation} between
   $\tup{X_1,\xi_1}$ and $\tup{X_2,\xi_2}$,
   if there exists a function 
   $\rho\colon R \to \Fun (R)$
   such that the projections
   $\pi_i\colon R \to X_i$  are $\Fun$-coalgebra morphisms 
   from $\tup{R,\rho}$ to $\tup{X_i,\xi}$, $i \in \{1,2\}$.
   Two states $x_1$ and $x_2$
   are {\em ($\Fun$-)bisimilar} (notation: $x_1 \bis x_2$),
   if they are linked by some $\Fun$-bisimulation.
   The relation $\bis$ is called $\Fun$-bisimilarity.

(2)   Two states $x_1 \in X_1$ and $x_2 \in X_2$
   are {\em behaviourally equivalent} 
   (notation: \mbox{$x_1 \beh x_2$}),
   if there exists an $\Fun$-coalgebra $\tup{Y,\gamma}$ and
   $\Fun$-coalgebra morphisms $f_i\colon\tup{X_i,\xi_i} \to 
   \tup{Y,\gamma}$ for $i=1,2$
   such that 
   $f_1(x_1) = f_2(x_2)$.
The triple $\tup{\tup{Y,\gamma},f_1,f_2}$ 
is called a {\em cocongruence} between 
$\tup{X_1,\xi_1}$ and $\tup{X_2,\xi_2}$.
If $\tup{\tup{Y,\gamma},f_1,f_2}$ is a cocongruence,
then we also refer to $R=\pb(f_1,f_2)$
as a cocongruence between 
$\tup{X_1,\xi_1}$ and $\tup{X_2,\xi_2}$.
The relation $\beh$ is called {\em behavioural equivalence}.
\end{defi}

\begin{rem}
Cocongruences were introduced by Kurz in \cite{Kurz:diss}.
In {\em loc.cit.}, Kurz refers to (the kernel of) an epimorphism 
as a behavioural equivalence. 
We have chosen to follow the terminology of 
\cite{AczMen:final,GummSchr05:types-coal}
and use the word congruence for kernels. 
We reserve behavioural equivalence
to denote the equivalence notion associated with congruences and 
cocongruences.
\end{rem}

For any functor $\Fun$,
$\Fun$-bi\-si\-mi\-la\-ri\-ty implies behavioural equivalence
(this fact will also follow from 
Proposition~\ref{p:equiv-compare}).
However, the converse only holds 
if $\Fun$ preserves weak pullbacks.
Precongruences were introduced in~\cite{AczMen:final}
as an alternative to bisimulations 
for functors that do not preserve weak pullbacks.

\begin{defi}\label{def:cong}
   Let $\tup{X,\xi}$ be an $\Fun$-coalgebra and
   $R \subseteq X\times X$ a relation.
   $R$ is a {\em congruence} on $\tup{X,\xi}$
   if 
the coequaliser $\vareps\colon X \to X/R$
of $R$ is an $\Fun$-coalgebra morphism,
i.e., there exists a unique coalgebra structure 
$\lambda\colon X/R \to \Fun(X/R)$
such that $\vareps$
is a coalgebra morphism
from $\tup{X,\xi}$ to $\tup{X/R,\lambda}$.
We call $\tup{X,\xi}/R := \tup{X/R,\lambda}$ 
the {\em quotient of $\tup{X,\xi}$ with $R$}.
   $R$ is a {\em precongruence} on $\tup{X,\xi}$
   if $\eqcl{R}$ is a congruence.
\end{defi}


Since any $\Fun$-coalgebra morphism
$f\colon \tup{X,\xi} \to \tup{Y,\gamma}$
factors through $X/\ker(f)$, 
it follows that
$R$ is a congruence on $\tup{X,\xi}$
iff 
$R = \ker(f) = \pb(f,f)$ 
for some $\Fun$-coalgebra morphism 
$f\colon \tup{X,\xi} \to \tup{X',\xi'}$.   

\begin{lem}\label{lem:cong-approx}
Let $\tup{X,\xi}$ be an $\Fun$-coalgebra. 
Behavioural equivalence, the largest congruence and the largest
precongruence on $\tup{X,\xi}$ all coincide.
\end{lem}
\proof
The lemma follows from results in
\cite{AczMen:final}~and~\cite[Lemma 5.10]{GummSchr05:types-coal},
but we also provide a quick argument here.
Clearly, a congruence is also a precongruence and
a precongruence is contained in a congruence.
Hence the largest congruence is the largest precongruence.
We refer to~\cite{AczMen:final} for more details.
Similarly, a congruence is clearly a cocongruence,
and any cocongruence is contained in a congruence,
since the category of $\Fun$-coalgebras has coequalisers:
if $R = \pb(f_1,f_2)$ for $\Fun$-coalgebra morphisms
$f_1,f_2\colon X \to Y$,
then $R \sse \ker(e \circ f_1)$,
where $e$ is the coequaliser of $f_1$ and $f_2$.
See also~\cite[Lemma 5.10]{GummSchr05:types-coal}.
Hence the largest congruence is behavioural equivalence
\qed

{\Precocongs} can be seen as a generalisation of 
precongruences to relations between coalgebras
obtained by replacing coequalisers by pushouts.

\begin{defi}\label{defi:precocong}
	Let 
	$\tup{X_1,\xi_1}$ and 
	$\tup{X_2,\xi_2}$ be $\Fun$-coalgebras, and
	let $R \subseteq X_1 \times X_2$ 
	be a 
\begin{minipage}[t]{9cm}
        relation with pushout
	$\tup{P,p_1,p_2}$.
	The relation $R$ is called a {\em \precocong} between
	$\tup{X_1,\xi_1}$ 
        and $\tup{X_2,\xi_2}$,
	if there exists a 
	coalgebra map 
        $\lambda\colon P \to \Fun(P)$ such that 
	the pushout maps $p_1\colon X_1 \to P$ and 
	$p_2\colon X_2 \to P$ are $\Fun$-coalgebra 
	morphisms, 
	i.e., 
        the diagram on right commutes.
In other words, $R$ is a {\precocong} if and only if
its pushout $\tup{P,p_1,p_2}$
\end{minipage}
\begin{minipage}[t]{6.5cm}
\centerline{
\xymatrix@R=1pc {& R \ar[ld]_{\pi_1} \ar[rd]^{\pi_2}& \\
	X_1 \ar[dd]_{\xi_1}
	\ar[r]^{p_1} & P \ar@{.>}[dd]_{\exists \lambda} &
	X_2 \ar[dd]^{\xi_2} \ar[l]_{p_2}\\\\
	\Fun (X_1) \ar[r]^{\Fun (p_1)} & \Fun (P) & \ar[l]_{\Fun (p_2)}
	\Fun (X_2)}
}
\end{minipage}
is a cocongruence.
If two states $x_1$ and $x_2$ are related by some {\precocong},
we write $x_1 \rbeh x_2$.

\end{defi}

The following lemma tells us that we can think of {\precocongs}
as the relations that are precongruences on the coproduct (disjoint union),
and it provides a useful criterion for proving that a
relation is a {\precocong}.

\begin{lem}\label{lem:precocongs}
	Let 
	$\tup{X_1,\xi_1}$ and 
	$\tup{X_2,\xi_2}$ be $\Fun$-coalgebras,
	and let $R \subseteq X_1 \times X_2$ 
	be a relation with pushout
	$\tup{P,p_1,p_2}$.
	The following are equivalent:
\begin{enumerate}[\em(1)]
\item $R$ is a {\precocong} between $\tup{X_1,\xi_1}$ and 
	$\tup{X_2,\xi_2}$.
\item $\Fun(p_1)\circ\xi_1\circ\pi_1 = \Fun(p_2)\circ\xi_2\circ\pi_2$,
  i.e., $R \sse \pb(\Fun(p_1)\circ\xi_1,\Fun(p_2)\circ\xi_2)$.
\item $\inrel{R}{X_1+X_2}$ is a precongruence on 
        $\tup{X_1,\xi_1}+\tup{X_2,\xi_2}$.
\end{enumerate}
\end{lem}

\proof
($1 \Lra 2$):
  Item 2 holds
  iff the outer part of the diagram in Def.~\ref{defi:precocong} 
  commutes,
  so the implication ($1 \Ra 2$) is immediate.
  Conversely, if item 2 holds,
  then
  by the universal property of the pushout
  $\tup{P,p_1,p_2}$ there is 
  a (unique) function $\lambda \colon P \to \Fun(P)$ such that 
  $\lambda \circ p_1 = \Fun(p_1) \circ \xi_1$ 
  and  $\lambda \circ p_2 = \Fun(p_2) \circ \xi_2$.
  Hence $R$ is a {\precocong}, 

($1 \Ra 3$):
If the pushout maps are morphisms,
there exists by the universal property of the disjoint union
$\tup{X_1,\xi_1}+\tup{X_2,\xi_2}$ in $\Coalg{F}$,
a unique $\Fun$-coalgebra morphism 
$u \colon X_1 +X_2 \to P$ such that 
$p_i = u \circ \inc_i$, $i \in \{1,2\}$.
By the definition of the pushout
(cf. Figure~\ref{fig:pushout}(b)), it must be the case that
$u$ is equal to the natural quotient map 
$\vareps\colon X_1 +X_2 \to P$, 
and hence $\inrel{R}{X_1 + X_2}$ is a precongruence.

($3 \Ra 1$):
If $\inrel{R}{X_1 + X_2}$ is a precongruence
on the disjoint union, then the quotient map 
$\vareps \colon X_1 +X_2 \to (X_1 +X_2)/\eqcl{\inrel{R}{X_1 + X_2}}$ is
an $\Fun$-coalgebra morphism.
Since $p_i = \vareps\circ\inc_i$, $i \in \{1,2\}$,
and the canonical inclusions
$\inc_i \colon X_i \to X_1 + X_2$, $i \in\{1,2\}$,
are also $\Fun$-coalgebra morphisms,
it follows that the pushout maps are $\Fun$-coalgebra morphisms.
\qed


An interesting property of {\precocongs}, is that,
like bisimulations,
they can be characterised by a form of relation lifting.
\begin{defi}\label{d:rellift}
	Let $R \subseteq X_1 \times X_2$ be a relation and let 
	$\tup{P,p_1,p_2}$ be the pushout of $\tup{R,\pi_1,\pi_2}$. 
	We define the $\Fun$-lifting 
$\wlift{\Fun}(R) \subseteq \Fun(X_1) \times \Fun(X_2)$ of $R$ by 
\[
\wlift{\Fun} (R) \coloneqq \pb(\Fun(p_1),\Fun (p_2)). 
\] 
\end{defi}
Note that $\wlift{\Fun}$
is independent of the concrete
representation of the pushout.
This follows easily from
the fact that pushouts are unique up-to isomorphism. 
The definition of $\wlift{\Fun}$ 
goes back to an idea by Kurz (\cite{Kurz:personal}) 
for defining a relation lifting of functors that do not 
preserve weak pullbacks. 

\begin{lem}\label{lem:wlift-char}
Let 
$\tup{X_1,\xi_1}$ and 
$\tup{X_2,\xi_2}$ be $\Fun$-coalgebras,
and let $R \subseteq X_1 \times X_2$ be a relation. 
$R$ is {\precocong} iff 
for all $\tup{x_1,x_2} \in R$: 
       $\;\;\tup{\xi_1(x_1),\xi_2(x_2)} \in \wlift{\Fun}(R)$.
\end{lem}
\proof
Immediate from Lemma~\ref{lem:precocongs} 
and the definition of $\wlift{\Fun}$.
\qed

The characterisation of {\precocongs} in Lemma~\ref{lem:wlift-char}
makes it easy to show that
between any two coalgebras, there exists a largest,
and necessarily unique, \precocong.
First, note that for any relations $R' \sse R \sse X_1 \times X_2$
with pushouts $\tup{P',p_1',p_2'}$ and $\tup{P,p_1,p_2}$,
respectively, there exists 
by the universal property of $P'$
a unique map $u \colon P' \to P$ such that
$p_i = u \circ p_i'$, $i \in \{1,2\}$.
Consequently,
$\Fun(p_i) = \Fun(u) \circ \Fun(p_i')$, $i \in \{1,2\}$,
and 
for all $t_1 \in \Fun(X_1)$, $t_2 \in \Fun(X_2)$:
$\Fun(p_1')(t_1) = \Fun(p_2')(t_2)$
implies that
$\Fun(p_1)(t_1) = \Fun(p_2)(t_2)$.
Hence,
\begin{equation}\label{eq:wlift-sse}
R'\sse R \quad\Ra\quad \wlift{\Fun}(R') \sse \wlift{\Fun}(R).
\end{equation}

\begin{lem}\label{lem:large-precocong}
Let $\tup{X_1,\xi_1}$ and 
$\tup{X_2,\xi_2}$ be $\Fun$-coalgebras.
The union of all {\precocongs}
between $\tup{X_1,\xi_1}$ and 
$\tup{X_2,\xi_2}$ is again a \precocong.
\end{lem}
\proof
Let $R$ be the union of all {\precocongs}
between $\tup{X_1,\xi_1}$ and $\tup{X_2,\xi_2}$,
and $\tup{P,p_1,p_2}$ the pushout of $R$.
If $\tup{x_1,x_2} \in R$,
then there is a {\precocong} $R' \sse R$ such that
$\tup{x_1,x_2} \in R'$. 
Letting $\tup{P',p_1',p_2'}$ be the pushout of $R'$,
it follows 
that
$\tup{\xi_1(x_1),\xi_2(x_2)} \in \wlift{\Fun}(R')$,
and hence by \eqref{eq:wlift-sse} that
$\tup{\xi_1(x_1),\xi_2(x_2)} \in \wlift{\Fun}(R)$.
We conclude by Lemma~\ref{lem:wlift-char} 
that $R$ is a {\precocong}.
\qed

In the following proposition we give a first comparison
between {\precocongs}, bisimulations 
and cocongruences. 

\begin{prop}\label{p:equiv-compare}
Let $\tup{X_1,\xi_1}$ and 
$\tup{X_2,\xi_2}$ be $\Fun$-coalgebras,
and let $R$ be a relation between $X_1$ and $X_2$.
\begin{enumerate}[\em(1)]
\item If $R$ is a bisimulation, then $R$ is a \precocong.
\item If $R$ is a \precocong, then $R$ is contained in 
      a cocongruence. 
\end{enumerate} 
Consequently,
   for all $x_1 \in X_1$ and $x_2 \in X_2$:
\begin{center}
    $x_1 \bis x_2$ \mbox{ implies } 
    $x_1 \rbeh x_2$ \mbox{ implies } 
   $x_1 \beh x_2$.
\end{center}
\end{prop}
\proof
   Let $R \subseteq X_1 \times X_2$ be a relation with
   projections $\pi_1\colon R \to X_1$ and
   $\pi_2\colon R \to X_2$, and pushout
   $\tup{P,p_1,p_2}$.
Item 1: Assume $R$ is a bisimulation. 
  By composing the projections with the 
  canonical inclusion morphisms into the coproduct, 
  we have a pair of parallel $\Fun$-coalgebra morphisms
  $\inc_1\circ\pi_1, \inc_2\circ\pi_2 \colon R \to X_1+X_2$.
  The quotient map
  $\vareps \colon X_1 +X_2 \to (X_1 +X_2)/\eqcl{\inrel{R}{X_1+X_2}}$,
  is now the coequaliser of $\inc_1\circ\pi_1$ and $\inc_2\circ\pi_2$ 
  in $\Coalg{F}$, hence an $\Fun$-coalgebra morphism. 
  Since $p_i = \vareps\circ\inc_i$, $i=1,2$,
  $p_1$ and $p_2$ are $\Fun$-coalgebra morphisms.
   Item 2:
   If $R$ is a \precocong, then 
   the pushout maps $p_1$ and $p_2$ are 
   $\Fun$-coalgebra morphisms.
   The claim now follows from the fact that
   $R \sse \pb(p_1,p_2)$.
\qed

Proposition~\ref{p:equiv-compare} alone does not yet tell us whether
{\precocongs} are a better approximation
of behavioural equivalence than $\Fun$-bisimulations,
but in the next subsection, we will see that,
in general, the implications of Proposition~\ref{p:equiv-compare} 
are strict. 
The following lemma provides us with a criterion which ensures
that a cocongruence is a {\precocong}.

\begin{lem}\label{l:full}
   If $\tup{X_1,\xi_1}$ and $\tup{X_2,\xi_2}$
   are $\Fun$-coalgebras and $R \sse X_1 \times X_2$ is a 
   bitotal cocongruence between 
   $\tup{X_1,\xi_1}$ and $\tup{X_2,\xi_2}$,
   then $R$ is a \precocong.
\end{lem}
\proof
   Let $R$ be a cocongruence with projection
   maps $\pi_1\colon R \to X_1$ and $\pi_2 \colon R \to X_2$ and
   pushout $\tup{P,p_1,p_2}$.
   Then there exist an $\Fun$-coalgebra $\tup{Y,\gamma}$
   and
   $\Fun$-coalgebra morphisms $f_i\colon X_i \to Y$ for $i\in\{1,2\}$
   such that $R = \pb(f_1,f_2)$. 
   We are going to define a function $\lambda \colon P \to \Fun (P)$
   such that $p_i$ is an $\Fun$-coalgebra morphism from 
   $\tup{X_i,\xi_i}$ to $\tup{P,\lambda}$ for $i\in\{1,2\}$. 
\begin{minipage}[t]{10.8cm}
   By the universal property of the   
   pushout there has to be a function $j\colon P \to Y$ such that
   $j \circ p_i = f_i$ for $i\in \{1,2\}$, 
   as shown in the diagram to the right.
   We claim that this function is injective. 
   First, it follows from the definition of the pushout 
   that both $p_1$ and
   $p_2$ are surjective, because $R$ is bitotal.
   Let now $z_1,z_2 \in P$ and suppose that $j(z_1) = j(z_2)$. 
   The surjectivity of
   the $p_i$'s implies that there are $s_1 \in X_1$ and $s_2 \in X_2$
   such that $p_1(s_1) = z_1$ and $p_2(s_2) = z_2$.
   Hence $j(p_1(s_1)) = j(p_2(s_2))$ which in turn yields
   $f_1(s_1) = f_2(s_2)$. This implies that $\tup{s_1,s_2} \in R$ and 
   consequently,
\end{minipage}
\begin{minipage}[t]{4.5cm}
\[\xymatrix@R=1pc@C=1pc{
R \ar[dd]_-{\pi_1} \ar[rr]^-{\pi_2} && 
X_2 \ar[dd]^-{p_2} \ar@/^1.2pc/[dddr]^-{f_2}\\\\
X_1 \ar[rr]^-{p_1} \ar@/_1.2pc/[rrrd]^-{f_1} &&
P \ar@{^{(}.>}[dr]^(.5){\exists ! j}\\
&&& Y
}\]
\end{minipage}
 $p_1(s_1) = p_2(s_2)$, i.e., $z_1 = z_2$.
   This demonstrates that $j$ is injective and thus there is
   some surjective map $e\colon Y \to P$ with $e \circ j = \id_{P}$.
   Now define $\lambda \coloneqq \Fun (e) \circ \lambda \circ j$.
   It is straightforward to check that for $i\in\{1,2\}$, the function
   $p_i\colon \tup{X_i,\xi_i} \to \tup{P,\lambda}$ is an 
   $\Fun$-coalgebra morphism.
\qed

We will now show that on a single $\Fun$-coalgebra,
an equivalence relation is a {\precocong}
iff it is a congruence.
It then follows immediately that 
the largest congruence is a {\precocong}.

\begin{thm}\label{t:on}
   Let 
   $\tup{X,\xi}$ an $\Fun$-coalgebra.
   \begin{enumerate}[\em(1)]
   \item  If $R \sse X \times X$ is an equivalence relation then:
   $R$ is a {\precocong} on $\tup{X,\xi}$ iff 
   $R$ is a congruence on $\tup{X,\xi}$.
   \item
   For all $x_1,x_2 \in X$: $x_1 \beh x_2$ iff $x_1 \rbeh x_2$. 
   \end{enumerate}
\end{thm}
\proof
To prove item 1,
first, observe that if $R \sse X \times X$ is an equivalence 
relation, then $\tup{x,x} \in R$ for all $x \in X$,
hence $p_1(x) = p_2(x)$ for all $x \in X$, i.e., $p_1=p_2$.
It follows that the pushout of  $R$ is of the form $\tup{P,p,p}$ 
and 
$R = \ker(p)$.  
Hence if $R$ is also a \precocong, 
then $p$ is a coalgebra morphism and $R=\ker(p)$ is a congruence. 
Conversely, if $R$ is a congruence, then
$R$ is clearly a bitotal cocongruence on $\tup{X,\xi}$ 
and so by Lemma~\ref{l:full}, a \precocong.
Item 2 of the lemma follows
from item 1 and Lemma~\ref{lem:cong-approx}.
\qed

We have introduced {\precocongs} 
as a generalisation
of precongruences to relations between different coalgebras. 
However, we point out that this generalisation is conceptual
rather than set-theoretic, since on a single coalgebra, 
a precongruence is not necessarily a {\precocong}
(as we will see in Example~\ref{ex:beheq-releq-not-bis} below).
In fact, one might say that 
{\precocongs} specialise precongruences in the one-coalgebra case,
since the converse does hold.

\begin{lem}
Let $\tup{X,\xi}$ be an $\Fun$-coalgebra and $R \sse X \times X$.
If $R$ is a {\precocong} on $\tup{X,\xi}$,
then $R$ is also a precongruence on $\tup{X,\xi}$.
\end{lem}

\proof
Let $\tup{P,p_1,p_2}$ be the pushout of $R$, and
let $\vareps_R\colon X \to X/\eqcl{R}$ be the natural quotient map
(i.e., the coequaliser of $R$).
By the universal property of the pushout in $\Set$,
there is a unique map $u \colon P\to X/\eqcl{R}$ such that 
$u \circ p_1 = \vareps_R = u\circ p_2$.
It follows that 
$\Fun(u) \circ \Fun(p_1) = \Fun(\vareps_R) = \Fun(u)\circ\Fun(p_2)$, 
and hence for all $x,y \in X$:
$\Fun(p_1)(\xi(x)) = \Fun(p_2)(\xi(y))$ implies that
$\Fun(\vareps_R)(\xi(x)) = \Fun(\vareps_R)(\xi(y))$.
Consequently, using Lemma~\ref{lem:precocongs}(2) and the fact that
$R$ is a precongruence iff $R \sse \ker(\Fun(\vareps_R)\circ \xi)$
(this can easily be shown using the universal property of coequalisers, 
see also~\cite{AczMen:final}), we conclude that if $R$ is a {\precocong},
then $R$ is also a precongruence.
\qed

\subsection{Equivalences between neighbourhood frames}

In this subsection, we will investigate 
behavioural equivalence, bisimilarity and
the equivalence notion arising from {\precocongs}
over $\TTwo$-coalgebras, i.e., neighbourhood frames.
First,
we obtain set-theoretic, back-and-forth style predicates
for $\TTwo$-bisimulations and $\TTwo$-{\precocongs}.
Next, we provide examples which show that
the implications from Proposition~\ref{p:equiv-compare}
are strict.
However, we also show that on a single neighbourhood frame
all three equivalence notions coincide.
Finally, we compare the three equivalence notions with 
bisimulations over monotonic neighbourhood frames
and Kripke frames.

\begin{rem} 
For simplicity of presentation, we have chosen to only treat
equivalence notions on neighbourhood frames, but 
the results of this section can easily be extended to neighbourhood 
{\em models}, i.e., $\TTwo(-) \times\Pow(\At)$-coalgebras.
For example, 
working out the details of the definition
of $\TTwo(-) \times\Pow(\At)$-bisimulation
results in the expected characterisation:
A relation $R$ is  $\TTwo(-) \times\Pow(\At)$-bi\-si\-mu\-la\-tion and 
if and only if 
$R$ is a $\TTwo$-bi\-si\-mu\-la\-tion and 
for all $\tup{s,t} \in R$, 
$s$ and $t$ satisfy the same atomic propositions.
Similar statements hold for cocongruences
and \precocongs.
\end{rem}

Let us start out by considering $\TTwo$-bisimulations.
Recall from Def.~\ref{d:fbis}
that a relation $R \sse S_1 \times S_2$ is a $\TTwo$-bisimulation
between two $\TTwo$-coalgebras $\str{S}_1=\tup{S_1,\nu_1}$ 
and $\str{S}_2=\tup{S_2,\nu_2}$ if the projection maps $\pi_1$
and $\pi_2$
are bounded morphisms ($\TTwo$-coalgebra
morphisms) from some
$\TTwo$-coalgebra $\tup{R,\mu}$ to $\str{S}_1$ and $\str{S}_2$ respectively.
By Definition~\ref{d:boundedmorph} of a bounded morphism 
this means that for $\tup{s_1,s_2} \in R$ and $i=1,2$:
\[ U \in \nu_i(s_i) \quad \mbox{iff} \quad \pi_i^{-1}[U] \in \mu(\tup{s_1,s_2}) 
\qquad
\mbox{for} \; U \sse S_i.\] 
This leads to two ``minimal requirements'' on the neighbourhood
functions $\nu_1$ and $\nu_2$ for pairs $\tup{s_1,s_2}$
contained in a $\TTwo$-bisimulation. For all $U_i,U_i' \sse S_i$, $i=1,2$:
\begin{enumerate}[(1)]
   \item $\pi_i^{-1}[U_i] = \pi_i^{-1}[U_i']$ implies
      $U_i \in \nu_i(s_i)$ iff $U_i' \in \nu_i(s_i)$,
   \item 
     $\pi_1^{-1}[U_1] = \pi_2^{-1}[U_2]$ implies $U_1 \in \nu_1(s_1)$
   iff $U_1' \in \nu_2(s_2)$. 
\end{enumerate}

Using the notion of $R$-coherence 
we can reformulate the previous requirements and prove that
they in fact characterise $\TTwo$-bisimulations.

\begin{prop}\label{prop:TTwo-bis-char}
 Let $\str{S}_1 = \tup{S_1,\nu_1}$ and 
 $\str{S}_2 = \tup{S_2,\nu_2}$ be neighbourhood frames.
 A relation $R \subseteq S_1 \times S_2$ is a
 {\em $\TTwo$-bisimulation} between $\str{S}_1$ and
 $\str{S}_2$ iff for all $\tup{s_1,s_2} \in R$, 
 for all $U_1,U_1' \sse S_1$ and for all
 $U_2,U_2' \sse S_2$ the 
 following two conditions are satisfied: 
  \begin{enumerate}[\em(1)]
    \item 
       \begin{enumerate}[\em(a)]
       \item if $\dom(R) \cap U_1 = \dom(R) \cap U_1'$
      then $U_1 \in \nu_1(s_1)$ iff 
      $U_1' \in \nu_1 (s_1)$, and
       \item if $\range(R) \cap U_2 = \range(R) \cap U_2'$
      then $U_2 \in \nu_2(s_2)$ iff 
      $U_2' \in \nu_2 (s_2)$.
      \end{enumerate}
    \item if the pair $\tup{U_1,U_2}$ is $R$-coherent, then: 
      $U_1 \in \nu_1(s_1)$ iff $U_2 \in \nu_2(s_2)$.
  \end{enumerate} 
\end{prop}
\proof
   It is a matter of routine checking
   that  every $\TTwo$-bisimulation $R$ fulfills
   conditions 1 and 2. 
   Let now $R \subseteq S_1 \times S_2$ be a relation that
   fulfills the conditions 1 and 2 for all $\tup{s_1,s_2} \in R$. 
   We define the neighbourhood function $\mu\colon R  \ra  \TTwo (R)$
   by $\mu(\tup{s_1,s_2}) :=  \{ \pi_1^{-1}[U] \mid U \in 
		\nu_1(s_1) \} \cup \{ \pi_2^{-1}[V] \mid V \in 
		\nu_2(s_2) \}$.
   In order to show that $R$ is a $\TTwo$-bisimulation it suffices
   to prove that for $i=1,2$ 
   the projection functions $\pi_i\colon \tup{R,\mu}
   \to \str{S}_i$ are bounded morphisms. We only provide the
   details for the proof that $\pi_1$ is a bounded morphism.
   We have to demonstrate that for all $\tup{s_1,s_2} \in R$
   and all $U \subseteq S_1$ we have 
   \begin{equation}\label{equ:pi1bound}
   	  U \in \nu_1(s_1) \quad \mbox{iff} \quad \pi_1^{-1}[U]
	  \in \mu(\tup{s_1,s_2}).
   \end{equation}
   Let $\tup{s_1,s_2} \in R$ and $U \subseteq S_1$. By definition
   of $\mu(\tup{s_1,s_2})$ the direction from left to right in 
   (\ref{equ:pi1bound}) is immediate.
   In order to prove the other implication in 
   (\ref{equ:pi1bound}) suppose that $\pi_1^{-1}[U] \in 
   \mu(\tup{s_1,s_2})$ for some $U \subseteq S_1$. According to
   the definition of $\mu(\tup{s_1,s_2})$ the following cases can occur:
   \begin{enumerate}[\bf C{a}se:]
   	\item $\pi_1^{-1}[U] = \pi_1^{-1}[U']$ for some
		$U' \in \nu_1(s_1)$. Then $\dom(R) \cap U = 
		\dom(R) \cap U'$ and hence 
		$U$ must be also in $\nu_1(s_1)$
		by condition 1 of the proposition.
	\item $\pi_1^{-1}[U] = \pi_2^{-1}[V]$ for some
		$V \in \nu_2(s_1)$, i.e., 
		the pair $\tup{U,V}$ is $R$-coherent. Condition
		2 therefore yields  $U \in \nu_1(s_2)$ as required.\qed
   \end{enumerate}

Another way of formulating
condition 1a in Proposition~\ref{prop:TTwo-bis-char},
is to say that if $U_1 \in \nu_1(s_1)$ and $U_1' \notin \nu_1(s_1)$,
then there is a $u \in (U_1\setminus U_1')\cup(U_1'\setminus U_1)$
such that $u \in \dom(R)$. Similarly for condition 1b.
Informally, one can say that condition 1 requires
that the relation $R$ must witness the difference between 
subsets when one is a neighbourhood and the other is not.
We will now show that {\precocongs}
are characterised by condition 2 only,
hence condition 1 is unnecessary (unwanted even)
for the purpose of approximating behavioural equivalence.

Let $\tup{S_1,\nu_1}$ and $\tup{S_2,\nu_2}$ 
be two $\TTwo$-coalgebras and
$R \sse S_1 \times S_2$ a relation
with pushout $\tup{P,p_1,p_2}$.
We have:
\begin{eqnarray}
&& \text{$R$ is a \precocong} \nonumber\\
& \text{ iff }&
\forall \tup{s_1,s_2} \in R: 
        \TTwo(p_1)(\nu_1(s_1)) = \TTwo(p_2)(\nu_2(s_2)) \nonumber\\
& \text{ iff }&
\forall \tup{s_1,s_2} \in R\,.\,
     \forall V \sse P: p_1^{-1}[V] \in \nu_1(s_1) \;\Lra\;
                       p_2^{-1}[V] \in \nu_2(s_2)\qquad \label{eq:TTwo-wlift}
\end{eqnarray}

We now show that, in fact, \eqref{eq:TTwo-wlift} is equivalent
with condition 2 of Proposition~\ref{prop:TTwo-bis-char}.

\begin{prop}\label{prop:rel-equ-char} 
     Let $\nstr{S}_1 = \tup{S_1,\nu_1}$
     and 
     $\nstr{S}_2 = \tup{S_2,\nu_2}$ 
     be neighbourhood frames, and
      $R \subseteq S_1 \times S_2$ a relation.
     We have:
     $R$ is a {\precocong} 
     between $\nstr{S}_1$ and $\nstr{S}_2$ 
     if and only if 
     for all $\tup{s_1,s_2} \in R$ and for all 
     $U_1 \sse S_1$ and $U_2 \sse S_2$
     such that $\tup{U_1,U_2}$ is $R$-coherent:
     $U_1 \in \nu_1(s_1) \text{ iff } U_2 \in \nu_2(s_2)$.
\end{prop}

\proof
Let $\nstr{S}_1, \nstr{S}_2$ and $R$ be as stated.
Furthermore, let $\pi_i\colon R \to S_i$, $i \in \{1,2\}$, 
be the projections of $R$,
$R_{12} = \inrel{R}{S_1+S_2}$,
and
$\tup{P,p_1,p_2}$ the pushout of $R$.
We will prove that for all $U_1 \sse S_1$ and $U_2 \sse S_2$:
\begin{equation}\label{eq:p-rel}
\text{$\tup{U_1,U_2}$ is $R$-coherent} \quad \text{ iff } \quad
U_1 = p_1^{-1}[Y] \text{ and } U_2 = p_2^{-1}[Y] \text{ for some } Y \sse P.
\end{equation}
The proposition then follows from \eqref{eq:TTwo-wlift}
and \eqref{eq:p-rel}.
To prove the direction from left to right in \eqref{eq:p-rel},
assume $U_1 \sse S_2$, $U_2 \sse S_2$ and $\tup{U_1,U_2}$ is $R$-coherent.
From Lemmas~\ref{lem:R-coh-1} and \ref{lem:R-coh-2}, we get that 
$U_1 + U_2$ is $\eqcl{R_{12}}$-coherent.
Let $\vareps\colon S_1+S_2 \to P$ be the quotient map associated 
with $\eqcl{R_{12}}$.
We claim that we can take $Y = \vareps[U_1 + U_2]$,
the set of $\eqcl{R_{12}}$-equivalence classes intersecting $U_1+U_2$.
To see that $p_1^{-1}[\vareps[U_1 + U_2]] = U_1$
and  $p_2^{-1}[\vareps[U_1 + U_2]] = U_2$,
we have for all $i \in \{1,2\}$ and $s_i \in S_i$:
\[ \begin{array}{rcl}
s_i \in p_i^{-1}[\vareps[U_1 + U_2]] & \iff &
p_i(s_i) \in \vareps[U_1 + U_2] \\ 
& \iff & \exists s' \in U_1 + U_2 : \tup{s_i,s'}\in \eqcl{R_{12}}\\
\text{\scriptsize ($U_1+U_2$ $\eqcl{R_{12}}$-coh.)} & \iff  & s_i \in U_1+U_2 \\
& \iff & s_i \in U_i.
\end{array}
\]
To prove the direction from right to left in \eqref{eq:p-rel},
let $Y \subseteq P$ be arbitrary.
We have for all $\tup{s_1,s_2} \in R$: 
\[ \tup{s_1,s_2}\in \pi_1^{-1}[p_1^{-1}[Y]]
\;\text{ iff }\; p_1(s_1) \in Y
\;\text{ iff }\; p_2(s_2) \in Y
\;\text{ iff }\; \tup{s_1,s_2}\in \pi_2^{-1}[p_2^{-1}[Y]].
\]
where the middle equivalence follows from the fact that
$\tup{s_1,s_2} \in R$ implies $p_1(s_1) = p_2(s_2)$.
We have now shown that
$\pi_1^{-1}[p_1^{-1}[Y]] = \pi_2^{-1}[p_2^{-1}[Y]]$,
hence by Lemma~\ref{lem:R-coh-1}, the pair
$\tup{p_1^{-1}[Y],p_2^{-1}[Y]}$ is $R$-coherent.
\qed

Since we know that on a single coalgebra,
congruences are {\precocongs} (Theorem~\ref{t:on}),
we get the following characterisation.

\begin{cor}\label{cor:congruence}
Let $\tup{S,\nu}$ be a neighbourhood frame and 
$R \subseteq S \times S$ an equivalence relation.
We have: $R$ is a congruence on $\tup{S,\nu}$ iff
\begin{equation}\label{eq:congruence}
\text{for all $\tup{s_1,s_2} \in R$ and all $R$-coherent 
$U \subseteq S$: $U \in \nu(s_1)$ iff $U \in \nu(s_2)$.}
\end{equation}
\end{cor}

\proof
Let $R \sse S\times S$ be an equivalence relation.
We first prove a small claim:
{\em Claim:} 
A pair $\tup{U_1,U_2}$ is $R$-coherent iff
$U_1 = U_2 = U$ for some $R$-coherent subset $U \sse S$.
{\em Proof of Claim:}          
Recall that a pair $\tup{U_1,U_2}$ is $R$-coherent
iff $R[U_1] \sse U_2$ and $R^{-1}[U_2] \sse U_1$.
Since $R$ is an equivalence relation, 
$R$ is reflexive, and it follows that
if $\tup{U_1,U_2}$ is $R$-coherent,
then $U_1 \sse R[U_1] \sse U_2$ and
$U_2 \sse R^{-1}[U_2] \sse U_1$, hence $U_1 = U_2$.
Conversely, if  $U$ is some $R$-coherent subset of $S$,
then by definition, $\tup{U,U}$ is $R$-coherent.

We now have:
$R$ is a congruence iff (Thm.~\ref{t:on})
$R$ is a {\precocong} iff (Prop.~\ref{prop:rel-equ-char}) 
for all $\tup{s_1,s_2} \in R$ and
for all $U_1, U_2 \subseteq S$ such that $\tup{U_1,U_2}$ is $R$-coherent:
  $U_1 \in \nu(s_1)$ iff $U_2 \in \nu(s_2)$.
Using the above claim, this last statement is equivalent with
\eqref{eq:congruence}.
\qed

We will now demonstrate with two examples 
that $\TTwo$-bisimilarity, {\precocongs}
and behavioural equivalence differ on neighbourhood frames.
It is tempting to think of the elements
of neighbourhoods as successor states, but these examples
show that this leads to wrong intuitions.
For example,
contrary to the intuition we have from Kripke bisimulations,
behavioural equivalence in neighbourhood frames does not require that
nonempty neighbourhoods are somehow matched by nonempty
neighbourhoods.
Moreover, states that are not contained in any neighbourhood 
of some state $s$,
can influence the existence of a bisimulation or 
cocongruence at $s$.

\begin{exa}\label{ex:beheq-releq-not-bis}
   Consider the two neighbourhood frames, $\nstr{T}=\tup{T,\nu_T}$
   and $\nstr{S}=\tup{S,\nu_S}$ where
   $T = \{t_1,t_2,t_3\}$, $\nu_T(t_1) = \nu_T (t_2) = \{\{t_2\}\}$, 
   $\nu_T(t_3) := \{\emptyset\}$, and    
   $S=\{s\}$, $\nu_S(s) = \emptyset$. 
   The two states $t_1$ and $s$ are behaviourally equivalent.
   To see this, let $\nstr{U}=\tup{U,\nu_U}$
   be the neighbourhood frame 
   where $U=\{u_1,u_2\}$, $\nu_U(u_1)=\emptyset$
   and $\nu_U(u_2) =\{\emptyset\}$. Let
   $f_1\colon T \to U$  and $f_2\colon S \to U$ 
   be the functions with graphs
   $\gra(f_1) = \{\tup{t_1,u_1},\tup{t_2,u_1},\tup{t_3,u_2}\}$ and
   $\gra(f_2) = \{\tup{s,u_1}\}$, respectively,
   as illustrated in the following picture:

\medskip

\ifpdf
   \centerline{\begin{picture}(0,0)%
\includegraphics{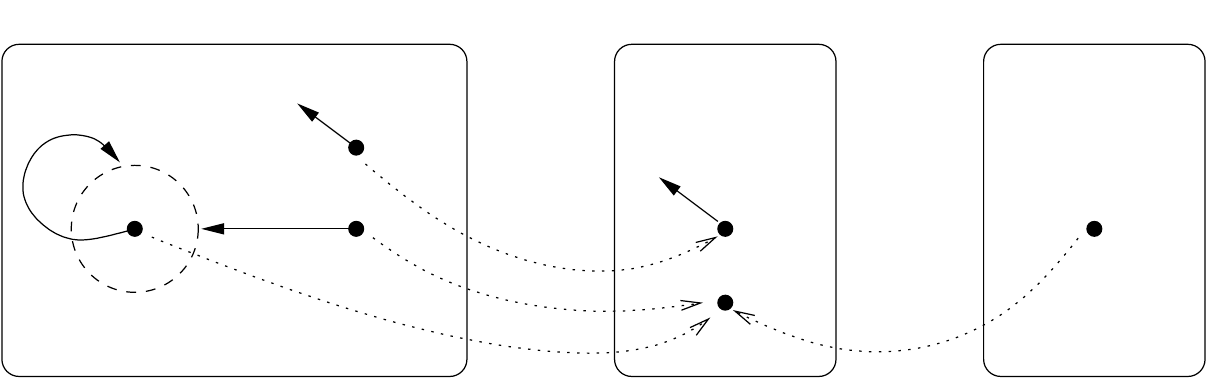}%
\end{picture}%
\setlength{\unitlength}{3108sp}%
\begingroup\makeatletter\ifx\SetFigFont\undefined%
\gdef\SetFigFont#1#2#3#4#5{%
  \reset@font\fontsize{#1}{#2pt}%
  \fontfamily{#3}\fontseries{#4}\fontshape{#5}%
  \selectfont}%
\fi\endgroup%
\begin{picture}(7359,2286)(79,-3223)
\put(811,-2176){\makebox(0,0)[lb]{\smash{{\SetFigFont{9}{10.8}{\rmdefault}{\mddefault}{\updefault}{\color[rgb]{0,0,0}$t_2$}%
}}}}
\put(1756,-1456){\makebox(0,0)[lb]{\smash{{\SetFigFont{9}{10.8}{\rmdefault}{\mddefault}{\updefault}{\color[rgb]{0,0,0}$\emptyset$}%
}}}}
\put(2386,-1861){\makebox(0,0)[lb]{\smash{{\SetFigFont{9}{10.8}{\rmdefault}{\mddefault}{\updefault}{\color[rgb]{0,0,0}$t_3$}%
}}}}
\put(4636,-2266){\makebox(0,0)[lb]{\smash{{\SetFigFont{9}{10.8}{\rmdefault}{\mddefault}{\updefault}{\color[rgb]{0,0,0}$u_2$}%
}}}}
\put(4006,-1906){\makebox(0,0)[lb]{\smash{{\SetFigFont{9}{10.8}{\rmdefault}{\mddefault}{\updefault}{\color[rgb]{0,0,0}$\emptyset$}%
}}}}
\put(1306,-1096){\makebox(0,0)[lb]{\smash{{\SetFigFont{9}{10.8}{\rmdefault}{\mddefault}{\updefault}{\color[rgb]{0,0,0}$\nstr{T}$}%
}}}}
\put(4456,-1096){\makebox(0,0)[lb]{\smash{{\SetFigFont{9}{10.8}{\rmdefault}{\mddefault}{\updefault}{\color[rgb]{0,0,0}$\nstr{U}$}%
}}}}
\put(6661,-1096){\makebox(0,0)[lb]{\smash{{\SetFigFont{9}{10.8}{\rmdefault}{\mddefault}{\updefault}{\color[rgb]{0,0,0}$\nstr{S}$}%
}}}}
\put(6706,-2581){\makebox(0,0)[lb]{\smash{{\SetFigFont{9}{10.8}{\rmdefault}{\mddefault}{\updefault}{\color[rgb]{0,0,0}$s$}%
}}}}
\put(4456,-3031){\makebox(0,0)[lb]{\smash{{\SetFigFont{9}{10.8}{\rmdefault}{\mddefault}{\updefault}{\color[rgb]{0,0,0}$u_1$}%
}}}}
\put(2161,-2581){\makebox(0,0)[lb]{\smash{{\SetFigFont{9}{10.8}{\rmdefault}{\mddefault}{\updefault}{\color[rgb]{0,0,0}$t_1$}%
}}}}
\put(3241,-2401){\makebox(0,0)[lb]{\smash{{\SetFigFont{9}{10.8}{\rmdefault}{\mddefault}{\updefault}{\color[rgb]{0,0,0}$f_1$}%
}}}}
\put(5446,-2941){\makebox(0,0)[lb]{\smash{{\SetFigFont{9}{10.8}{\rmdefault}{\mddefault}{\updefault}{\color[rgb]{0,0,0}$f_2$}%
}}}}
\end{picture}%
}
\else
   \centerline{\begin{picture}(0,0)%
\includegraphics{graphics/ex1_new.pstex}%
\end{picture}%
\setlength{\unitlength}{3108sp}%
\begingroup\makeatletter\ifx\SetFigFont\undefined%
\gdef\SetFigFont#1#2#3#4#5{%
  \reset@font\fontsize{#1}{#2pt}%
  \fontfamily{#3}\fontseries{#4}\fontshape{#5}%
  \selectfont}%
\fi\endgroup%
\begin{picture}(7359,2286)(79,-3223)
\put(811,-2176){\makebox(0,0)[lb]{\smash{{\SetFigFont{9}{10.8}{\rmdefault}{\mddefault}{\updefault}{\color[rgb]{0,0,0}$t_2$}%
}}}}
\put(1756,-1456){\makebox(0,0)[lb]{\smash{{\SetFigFont{9}{10.8}{\rmdefault}{\mddefault}{\updefault}{\color[rgb]{0,0,0}$\emptyset$}%
}}}}
\put(2386,-1861){\makebox(0,0)[lb]{\smash{{\SetFigFont{9}{10.8}{\rmdefault}{\mddefault}{\updefault}{\color[rgb]{0,0,0}$t_3$}%
}}}}
\put(4636,-2266){\makebox(0,0)[lb]{\smash{{\SetFigFont{9}{10.8}{\rmdefault}{\mddefault}{\updefault}{\color[rgb]{0,0,0}$u_2$}%
}}}}
\put(4006,-1906){\makebox(0,0)[lb]{\smash{{\SetFigFont{9}{10.8}{\rmdefault}{\mddefault}{\updefault}{\color[rgb]{0,0,0}$\emptyset$}%
}}}}
\put(1306,-1096){\makebox(0,0)[lb]{\smash{{\SetFigFont{9}{10.8}{\rmdefault}{\mddefault}{\updefault}{\color[rgb]{0,0,0}$\nstr{T}$}%
}}}}
\put(4456,-1096){\makebox(0,0)[lb]{\smash{{\SetFigFont{9}{10.8}{\rmdefault}{\mddefault}{\updefault}{\color[rgb]{0,0,0}$\nstr{U}$}%
}}}}
\put(6661,-1096){\makebox(0,0)[lb]{\smash{{\SetFigFont{9}{10.8}{\rmdefault}{\mddefault}{\updefault}{\color[rgb]{0,0,0}$\nstr{S}$}%
}}}}
\put(6706,-2581){\makebox(0,0)[lb]{\smash{{\SetFigFont{9}{10.8}{\rmdefault}{\mddefault}{\updefault}{\color[rgb]{0,0,0}$s$}%
}}}}
\put(4456,-3031){\makebox(0,0)[lb]{\smash{{\SetFigFont{9}{10.8}{\rmdefault}{\mddefault}{\updefault}{\color[rgb]{0,0,0}$u_1$}%
}}}}
\put(2161,-2581){\makebox(0,0)[lb]{\smash{{\SetFigFont{9}{10.8}{\rmdefault}{\mddefault}{\updefault}{\color[rgb]{0,0,0}$t_1$}%
}}}}
\put(3241,-2401){\makebox(0,0)[lb]{\smash{{\SetFigFont{9}{10.8}{\rmdefault}{\mddefault}{\updefault}{\color[rgb]{0,0,0}$f_1$}%
}}}}
\put(5446,-2941){\makebox(0,0)[lb]{\smash{{\SetFigFont{9}{10.8}{\rmdefault}{\mddefault}{\updefault}{\color[rgb]{0,0,0}$f_2$}%
}}}}
\end{picture}%
}
\fi
\medskip
   It can easily be verified that $f_1$ and $f_2$ are bounded morphisms.
   For example, 
   the bounded morphism condition \eqref{eq:b-m} holds for $f_1$ at 
   $t_1$ and $t_2$, since their only neighbourhood $\{t_2\}$ is 
   not the inverse $f_1$-image of any subset of $U$.
   Since $f_1(t_1)=f_2(s)$, $t_1$ and $s$
   are behaviourally equivalent.
   In fact, $R := \pb(f_1,f_2) = \{\tup{t_1,s},\tup{t_2,s}\}$ 
   is a \precocong.
   This can be verified using the characterisation given in
   Proposition~\ref{prop:rel-equ-char}.
   Note that there is no subset $U \sse S$ such that
   $\tup{\{t_2\},U}$ is $R$-coherent.

   However, $t_1$ and $s$ are not $\TTwo$-bisimilar.
   For suppose $R$ is a $\TTwo$-bisimulation 
   between $\nstr{T}$ and $\nstr{S}$, then
   $\tup{t_3,s} \notin R$, since
   $\tup{\emptyset,\emptyset}$ is $R$-coherent, 
   $\emptyset \in \nu_T(t_3)$
   and $\emptyset \notin \nu_S(s)$.
   Hence $t_3 \notin\dom(R)$, and it follows that
   $\dom(R) \cap \{t_2\}=\dom(R) \cap \{t_2,t_3\}$.
   Now, since $\{t_2\} \in \nu_T(t_1)$ and $\{t_2,t_3\} \not\in \nu_T(t_1)$,
   we can conclude from 
   condition 1a of Proposition~\ref{prop:TTwo-bis-char} 
   that $t_1$ cannot be $R$-related to any state in $\nstr{S}$,
   in particular not to $s$.
   Since $R$ was an arbitrary $\TTwo$-bisimulation,
   $t_1$ and $s$ are not $\TTwo$-bisimilar.

   Consider, now the relation $R' = \{\tup{t_1,t_2}\}$
   on the neighbourhood frame $\nstr{T}$.
   The reader can check that $R'$ is a 
   precongruence, but not a {\precocong},
   on $\nstr{T}$.
\end{exa}

The above example shows that 
between neighbourhood frames,
{\precocongs} are 
a better approximation of behavioural equivalence
than $\TTwo$-bisimilarity.
However, the next example shows that
also {\precocongs} cannot capture behavioural equivalence, in general.

\begin{exa}{\label{ex:beheq-not-releq}}
We consider now a small variation on the picture
given in Example~\ref{ex:beheq-releq-not-bis}.
The neighbourhood frames $\nstr{S}$, $\nstr{U}$ and the function $f_2$
are the same as before, but on $T$ we now take as neighbourhood function
$\nu_T'(t_1) = \{\{t_2\}\}$, $\nu_T'(t_2) = \nu_T'(t_3) = \{\emptyset\}$, and
let $\nstr{T}' = \tup{T,{\nu}_T'}$.
Instead of the function $f_1$, we take
the function $f_1'\colon T \to U$
with graph $\gra(f_1') = \{\tup{t_1,u_1},\tup{t_2,u_2},\tup{t_3,u_2}\}$.
Again, it is straightforward to check that $f_1'$ is a bounded morphism,
and hence $t_1$ and $s$ are behaviourally equivalent.

\medskip
\ifpdf
  \centerline{\begin{picture}(0,0)%
\includegraphics{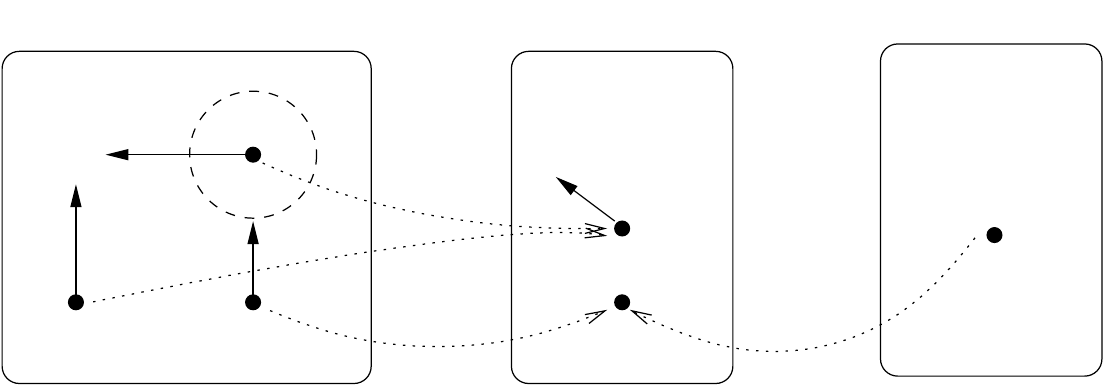}%
\end{picture}%
\setlength{\unitlength}{3108sp}%
\begingroup\makeatletter\ifx\SetFigFont\undefined%
\gdef\SetFigFont#1#2#3#4#5{%
  \reset@font\fontsize{#1}{#2pt}%
  \fontfamily{#3}\fontseries{#4}\fontshape{#5}%
  \selectfont}%
\fi\endgroup%
\begin{picture}(6729,2331)(709,-3268)
\put(4006,-1906){\makebox(0,0)[lb]{\smash{{\SetFigFont{9}{10.8}{\rmdefault}{\mddefault}{\updefault}{\color[rgb]{0,0,0}$\emptyset$}%
}}}}
\put(4456,-1096){\makebox(0,0)[lb]{\smash{{\SetFigFont{9}{10.8}{\rmdefault}{\mddefault}{\updefault}{\color[rgb]{0,0,0}$\nstr{U}$}%
}}}}
\put(6661,-1096){\makebox(0,0)[lb]{\smash{{\SetFigFont{9}{10.8}{\rmdefault}{\mddefault}{\updefault}{\color[rgb]{0,0,0}$\nstr{S}$}%
}}}}
\put(2206,-1771){\makebox(0,0)[lb]{\smash{{\SetFigFont{9}{10.8}{\rmdefault}{\mddefault}{\updefault}{\color[rgb]{0,0,0}$t_2$}%
}}}}
\put(1036,-2986){\makebox(0,0)[lb]{\smash{{\SetFigFont{9}{10.8}{\rmdefault}{\mddefault}{\updefault}{\color[rgb]{0,0,0}$t_3$}%
}}}}
\put(2161,-2986){\makebox(0,0)[lb]{\smash{{\SetFigFont{9}{10.8}{\rmdefault}{\mddefault}{\updefault}{\color[rgb]{0,0,0}$t_1$}%
}}}}
\put(1126,-1951){\makebox(0,0)[lb]{\smash{{\SetFigFont{9}{10.8}{\rmdefault}{\mddefault}{\updefault}{\color[rgb]{0,0,0}$\emptyset$}%
}}}}
\put(4546,-2221){\makebox(0,0)[lb]{\smash{{\SetFigFont{9}{10.8}{\rmdefault}{\mddefault}{\updefault}{\color[rgb]{0,0,0}$u_2$}%
}}}}
\put(4411,-3031){\makebox(0,0)[lb]{\smash{{\SetFigFont{9}{10.8}{\rmdefault}{\mddefault}{\updefault}{\color[rgb]{0,0,0}$u_1$}%
}}}}
\put(6706,-2626){\makebox(0,0)[lb]{\smash{{\SetFigFont{9}{10.8}{\rmdefault}{\mddefault}{\updefault}{\color[rgb]{0,0,0}$s$}%
}}}}
\put(3196,-2086){\makebox(0,0)[lb]{\smash{{\SetFigFont{9}{10.8}{\rmdefault}{\mddefault}{\updefault}{\color[rgb]{0,0,0}$f_1'$}%
}}}}
\put(5446,-2941){\makebox(0,0)[lb]{\smash{{\SetFigFont{9}{10.8}{\rmdefault}{\mddefault}{\updefault}{\color[rgb]{0,0,0}$f_2$}%
}}}}
\put(1756,-1141){\makebox(0,0)[lb]{\smash{{\SetFigFont{9}{10.8}{\rmdefault}{\mddefault}{\updefault}{\color[rgb]{0,0,0}$\nstr{T}'$}%
}}}}
\end{picture}%
}
\else
  \centerline{\begin{picture}(0,0)%
\includegraphics{graphics/ex2_new.pstex}%
\end{picture}%
\setlength{\unitlength}{3108sp}%
\begingroup\makeatletter\ifx\SetFigFont\undefined%
\gdef\SetFigFont#1#2#3#4#5{%
  \reset@font\fontsize{#1}{#2pt}%
  \fontfamily{#3}\fontseries{#4}\fontshape{#5}%
  \selectfont}%
\fi\endgroup%
\begin{picture}(6729,2331)(709,-3268)
\put(4006,-1906){\makebox(0,0)[lb]{\smash{{\SetFigFont{9}{10.8}{\rmdefault}{\mddefault}{\updefault}{\color[rgb]{0,0,0}$\emptyset$}%
}}}}
\put(4456,-1096){\makebox(0,0)[lb]{\smash{{\SetFigFont{9}{10.8}{\rmdefault}{\mddefault}{\updefault}{\color[rgb]{0,0,0}$\nstr{U}$}%
}}}}
\put(6661,-1096){\makebox(0,0)[lb]{\smash{{\SetFigFont{9}{10.8}{\rmdefault}{\mddefault}{\updefault}{\color[rgb]{0,0,0}$\nstr{S}$}%
}}}}
\put(2206,-1771){\makebox(0,0)[lb]{\smash{{\SetFigFont{9}{10.8}{\rmdefault}{\mddefault}{\updefault}{\color[rgb]{0,0,0}$t_2$}%
}}}}
\put(1036,-2986){\makebox(0,0)[lb]{\smash{{\SetFigFont{9}{10.8}{\rmdefault}{\mddefault}{\updefault}{\color[rgb]{0,0,0}$t_3$}%
}}}}
\put(2161,-2986){\makebox(0,0)[lb]{\smash{{\SetFigFont{9}{10.8}{\rmdefault}{\mddefault}{\updefault}{\color[rgb]{0,0,0}$t_1$}%
}}}}
\put(1126,-1951){\makebox(0,0)[lb]{\smash{{\SetFigFont{9}{10.8}{\rmdefault}{\mddefault}{\updefault}{\color[rgb]{0,0,0}$\emptyset$}%
}}}}
\put(4546,-2221){\makebox(0,0)[lb]{\smash{{\SetFigFont{9}{10.8}{\rmdefault}{\mddefault}{\updefault}{\color[rgb]{0,0,0}$u_2$}%
}}}}
\put(4411,-3031){\makebox(0,0)[lb]{\smash{{\SetFigFont{9}{10.8}{\rmdefault}{\mddefault}{\updefault}{\color[rgb]{0,0,0}$u_1$}%
}}}}
\put(6706,-2626){\makebox(0,0)[lb]{\smash{{\SetFigFont{9}{10.8}{\rmdefault}{\mddefault}{\updefault}{\color[rgb]{0,0,0}$s$}%
}}}}
\put(3196,-2086){\makebox(0,0)[lb]{\smash{{\SetFigFont{9}{10.8}{\rmdefault}{\mddefault}{\updefault}{\color[rgb]{0,0,0}$f_1'$}%
}}}}
\put(5446,-2941){\makebox(0,0)[lb]{\smash{{\SetFigFont{9}{10.8}{\rmdefault}{\mddefault}{\updefault}{\color[rgb]{0,0,0}$f_2$}%
}}}}
\put(1756,-1141){\makebox(0,0)[lb]{\smash{{\SetFigFont{9}{10.8}{\rmdefault}{\mddefault}{\updefault}{\color[rgb]{0,0,0}$\nstr{T}'$}%
}}}}
\end{picture}%
}
\fi
\medskip
  However, there is no {\precocong}
  containing the pair $\tup{t_1,s}$.
  Suppose $R' \subseteq T \times S$
  is an arbitrary {\precocong} between $\nstr{T}'$ and $\nstr{S}$.
  Since $\tup{\emptyset,\emptyset}$ is $R'$-coherent,
  $\emptyset \in \nu_T'(t_2)$ and $\emptyset \not\in \nu_S(s)$,
  it follows from Proposition~\ref{prop:rel-equ-char}
  that $\tup{t_2,s} \not\in R'$.
  This implies that
  $\tup{\{t_2\},\emptyset}$ is $R'$-coherent, but 
  $\{t_2\} \in\nu_T'(t_1)$ and $\emptyset \notin\nu_S(s)$,
  so $\tup{t_1,s} \not\in R'$.
\end{exa}

To sum it up: Example~\ref{ex:beheq-releq-not-bis} 
showed that {\precocongs} are a clear improvement when 
compared to $\TTwo$-bisimulations. 
Example~\ref{ex:beheq-not-releq}, however,
demonstrates that {\precocongs} are still 
incomplete as a proof principle
for behavioural equivalence over neighbourhood frames.

From Theorem~\ref{t:on} of the previous subsection,
we know that on a single neighbourhood frame, 
{\precocongs} do capture behavioural equivalence.
Using the results of this subsection it follows easily
that, in fact, also
$\TTwo$-bisimilarity captures behavioural equivalence
on a single structure.

\begin{prop}\label{p:allonone}
   If $\nstr{S}=\tup{S,\nu}$ is a neighbourhood frame,
   and $R \sse S \times S$ is an equivalence relation,
   then:
\begin{center}
            $R$ is a $\TTwo$-bisimulation \mbox{ iff } 
            $R$ is a {\precocong} \mbox{ iff } 
            $R$ is a congruence.
\end{center}
Consequently, for all $s_1,s_2 \in S$:
$\quad s_1 \bis s_2$ \mbox{ iff } $s_1 \rbeh s_2$ \mbox{ iff } $s_1 \beh s_2$.
\end{prop}
\proof
   If $R \subseteq S\times S$ is an equivalence
   relation, then in particular $\dom(R) = \rng(R) = S$, 
   and hence condition 1 of 
   Proposition~\ref{prop:TTwo-bis-char} is trivially satisfied.
   It follows from the characterisations in
   Propositions~\ref{prop:TTwo-bis-char} and \ref{prop:rel-equ-char}
   that
   $R$ is a $\TTwo$-bisimulation
   iff $R$ is a \precocong.
   The second equivalence 
   is an instance of the more general
   result in Theorem~\ref{t:on}.
   The final claim is an immediate consequence of the main claim
   and Lemma~\ref{lem:cong-approx}.
\qed

\begin{rem}\label{rem:weak-kernel-pairs}
   Alternatively, Proposition~\ref{p:allonone}
   follows from the result
   in \cite{GummSchr05:types-coal} that congruences are 
   $\Fct{F}$-bisimulations
   in case the functor $\Fct{F}$ weakly preserves kernel pairs - a property
   that the functor $\TTwo$ has as the following argument shows:    
   Let $f\colon S \to T$ be  a function  
   and consider its kernel $\ker(f)\coloneqq
   \{\tup{s,s'} \in S \times S \mid f(s)=f(s')\}$ with projections 
   $\pi_i\colon \ker(f) \to S$ for $i=1,2$. We have to show
   that for every pair of sets $N_1,N_2 \in
   \ker(\TTwo(f))$
   there exists a set $N \in \TTwo(\ker(f))$ such that
   $\TTwo(\pi_i)(N)=N_i$ for $i=1,2$.
   Let $N_1,N_2$ be elements
   of $\TTwo(S)$ such that $\TTwo (f)(N_1)=\TTwo(f)(N_2)$.
   We put
   $ N \coloneqq \{ \pi_1^{-1}(U_1) \mid U_1 \in N_1 \} \cup 
                 \{ \pi_2^{-1}(U_2) \mid U_2 \in N_2\}$.
   It is now easy to check that 
   $\TTwo(\pi_i)(N)=N_i$ for $i=1,2$ as required.
\end{rem}


\section{Hennessy-Milner classes}
\label{sec:HM-classes}

The Hennessy-Milner theorem for normal modal logic
states that over the class of finite Kripke models,
two states are Kripke bisimilar if and only if they satisfy the same
modal formulas.
It is well known (see e.g.~\cite{BdRV:ML-book}), 
that this Hennessy-Milner theorem can be generalised
to hold over any class of modally saturated Kripke models,
in particular, over the class of image-finite Kripke models.


In this section, we define
modal saturation and image-finiteness for 
neighbourhood models and show that each of these properties
leads to a Hennessy-Milner style theorem.
In the last subsection we describe ultrafilter extensions
of neighbourhood models, and show that they are modally saturated.

First, we make precise what we mean by a Hennessy-Milner class of
neighbourhood models.
Since we have three equivalence notions for neighbourhood models,
we have, in principle, three types of Hennessy-Milner classes.
However, 
Examples~\ref{ex:beheq-releq-not-bis} and \ref{ex:beheq-not-releq} of 
section~\ref{sec:equiv-notions} showed that even over the class of 
finite neighbourhood models, two states can be behaviourally
equivalent, and hence modally equivalent,
without being linked by a {\precocong} or a bisimulation.
This means that 
{\precocongs} and bisimulations
do not fit well with the expressivity of the modal language.
We therefore define Hennessy-Milner classes with respect to
behavioural equivalence.

\begin{defi}
A class $\cla{K}$ of neighbourhood models is a
{\em Hennessy-Milner class},
if for any $\nstr{M}_1$ and $\nstr{M}_2$ in $\cla{K}$
containing states $s_1$ and $s_2$, respectively,
we have:
$\nstr{M}_1,s_1 \modeq \nstr{M}_2,s_2$ \mbox{ iff } 
$\nstr{M}_1,s_1 \beh \nstr{M}_2,s_2$.
\end{defi}

The following lemma provides an easy, but useful, criterion
for proving that a class of models is a Hennessy-Milner class.

\begin{lem}\label{lem:m-satur-du-hm}
Let $\cla{K}$ be a class of neighbourhood models.
If for any $\nstr{M}_1, \nstr{M}_2 \in \cla{K}$,
the modal equivalence relation $\equiv$ is a congruence on
$\nstr{M}_1+\nstr{M}_2$,
then $\cla{K}$ is a Hennessy-Milner class.
\end{lem}
\proof
Let $\nstr{M}_1$ and $\nstr{M}_2$ be neighbourhood models in $\cla{K}$,
and let $\inc_i \colon \nstr{M}_i \to \nstr{M}_1 + \nstr{M}_2$
\begin{minipage}[t]{8cm}
denote the canonical inclusion morphisms.
Assume that we have states $s_1$ and $s_2$ such that
$\nstr{M}_1, s_1 \modeq \nstr{M}_2, s_2$.
Since truth is invariant under bounded morphisms,
we have $\inc_1(s_1) \modeq \inc_2(s_2)$ in
$\nstr{M}_1 + \nstr{M}_2$.
By assumption, $\modeq$ is a congruence on 
$\nstr{M}_1 + \nstr{M}_2$,
hence 
$\varepsilon \colon 
 \nstr{M}_1 + \nstr{M}_2 \to (\nstr{M}_1 + \nstr{M}_2)/\!\!\modeq$
\end{minipage}
\begin{minipage}[t]{7.5cm}
\vspace{1pt}
\centerline{\xymatrix{
{\nstr{M}_1} \ar@{->}[r]^-{\inc_1} 
 & {\nstr{M}_1 + \nstr{M}_2} \ar[d]^-{\varepsilon} 
 & {\nstr{M}_2} \ar@{->}[l]_-{\inc_2}\\
 & {(\nstr{M}_1 + \nstr{M}_2)}/\!\!\modeq
}}
\end{minipage}
is a bounded morphism (as illustrated by the diagram), and 
$\tup{s_1,s_2} \in \pb(\varepsilon\circ\inc_1,\varepsilon\circ\inc_2)$,
hence $s_1 \beh s_2$.
\qed

\subsection{Modally saturated models}
\label{ssec:HM-m-sat}

In Lemma~\ref{lem:m-satur-du-hm} we saw that 
in order to prove a Hennessy-Milner theorem,
we are interested in neighbourhood models on which modal equivalence
is a congruence. 
Let $\nstr{M}=\tup{S,\nu,V}$ be a neighbourhood model. 
By applying the characterisations of congruences on neighbourhood frames 
in Corollary~\ref{cor:congruence} and adding the condition for the 
atomic propositions, we find that $\modeq$ is a congruence
on $\nstr{M}$ iff
for all $s, t \in S$ such that $s \modeq t$:
\begin{equation}\label{eq:modeq-congruence}
\begin{array}{lll}
\text{(c1)} & \text{for all } p \in \At: & s \in V(p) \iff t \in V(p), 
\text{ and }\\
\text{(c2)} & \text{for all modally coherent } X \subseteq S: &
X \in \nu(s) \iff X \in \nu(t).
\end{array}
\end{equation} 

Clearly, condition (c1) holds in all neighbourhood models,
since modally equivalent states must make the same atomic propositions true.
One way of making condition (c2) hold, is to ensure that all 
modally coherent neighbourhoods 
are definable.

\begin{lem}\label{lem:modcoh-definable}
Let $\nstr{M} = \tup{S,\nu,V}$ be a neighbourhood model.
If for all $s \in S$ and all modally coherent $X \in \nu(s)$,
there exists a modal $\langML$-formula $\varphi$ such that 
$X = \ext{\varphi}^\nstr{M}$,
then modal equivalence is a congruence on $\nstr{M}$.
\end{lem}
\proof
Let $X$ be a modally coherent neighbourhood of some state, and
assume $X = \ext{\varphi}^\nstr{M}$.
We have for any $s,t \in S$ such that $s \modeq t$:
$X \in \nu(s)$ iff $\nstr{M},s \models \Box\varphi$
iff $\nstr{M},t \models \Box\varphi$
iff $X \in \nu(t)$.
\qed

For finite models, a standard argument shows that any modally coherent 
neighbourhood $X$ is definable by a formula of the form
$\delta = \bigvee_{i \leq n} \bigwedge_{j \leq k} \delta_{i,j}$ 
where $n,k < \omega$.
For infinite models, 
the same argument would yield a formula 
with an infinite disjunction and conjunction,
which is not a well-formed formula of our finitary language.
Modal saturation is a compactness property 
which allows us to replace infinite conjunctions and
disjunctions with finite ones
\footnote{This perspective on modal saturation was
 pointed out to us by H.P. Gumm (personal correspondence).}.
Thus we can essentially use the same argument as in finite models
to show that modally coherent neighbourhoods 
are definable (and we do so in Lemma~\ref{lem:m-satur-definability} below).
We will use the following notation.
Let $\Psi$ be a set of modal $\langML$-formulas and 
$\nstr{M} = \tup{S,\nu,V}$ a neighbourhood model.
We define
$\lnot\Psi = \{\lnot\psi \mid \psi \in \Psi\}$, 
$\ext{\bigwedge\Psi}^\nstr{M} = 
\bigcap_{\psi \in \Psi} \ext{\psi}^\nstr{M}$, and
$\ext{\bigvee\Psi}^\nstr{M} = 
\bigcup_{\psi \in \Psi} \ext{\psi}^\nstr{M}$.
A set $\Psi$ of $\langML$-formulas is {\em satisfiable in a 
subset $X \subseteq S$} of $\nstr{M}$,
if $\ext{\bigwedge\Psi}^{\nstr{M}} \cap X \neq \emptyset$.
A set $\Psi$ of $\langML$-formulas is 
{\em  finitely satisfiable in $X \subseteq S$},
if any finite subset $\Psi_0 \subseteq_\omega \Psi$ is satisfiable in $X$.

\begin{defi}\label{defi:m-saturation}
Let $\nstr{M} = \tup{S,\nu,V}$ be a neighbourhood model.
A subset $X \subseteq S$ is called {\em modally compact}
if for all sets $\Psi$ of modal $\langML$-formulas,
$\Psi$ is satisfiable in $X$ 
whenever $\Psi$ is finitely satisfiable in $X$.
The neighbourhood model $\nstr{M}$ is {\em modally saturated},
if for all $s \in S$ and all modally coherent neighbourhoods $X \in \nu(s)$,
both $X$ and the complement $\cmp{X}$ are modally compact.
\end{defi}

To see why modal compactness is really a compactness property,
note that for a subset $X$ in a neighbourhood model $\nstr{M}$,
$X \subseteq \ext{\bigvee\Psi}^{\nstr{M}}$ iff
$\{ \lnot\psi \mid \psi \in \Psi\}$ is not satisfiable in $X$.
Hence $X$ is modally compact, 
if and only if, 
for all $\Psi \subseteq \langML$ such that 
$X \subseteq \ext{\bigvee\Psi}^\nstr{M}$ 
there is a $\Psi_0 \subseteq_\omega \Psi$ such that
$X \subseteq \ext{\bigvee\Psi_0}^\nstr{M}$.
Clearly, any finite set is modally compact.
Note also that, 
in Definition~\ref{defi:m-saturation},
due to the fact that $\ext{\bigwedge\Psi}^\nstr{M} \subseteq X$
if and only if $\cmp{X} \subseteq \ext{\bigvee\lnot\Psi}^\nstr{M}$,
we have that
$\cmp{X}$ is modally compact, 
if and only if,
for all $\Psi \subseteq \langML$
such that $\ext{\bigwedge\Psi}^\nstr{M} \subseteq X$,
there is a $\Psi_0 \subseteq_\omega \Psi$ such that
$\ext{\bigwedge\Psi_0}^\nstr{M} \subseteq X$.

\begin{lem}\label{lem:m-satur-definability}
Let  $\nstr{M} = \tup{S,\nu,V}$ be a modally saturated neighbourhood
model.
For all $X \sse S$:
$X$ is modally coherent iff 
$X$ is definable by a modal $\langML$-formula.
\end{lem}

\proof
If $X = \ext{\varphi}^\nstr{M}$ for some $\varphi \in \langML$,
then clearly $X$ is modally coherent.
For the converse implication, assume $X$ is modally coherent,
i.e.,
$X$ is a union of modal equivalence classes
$X = \bigcup_{c \in C}\modcls{x_c}$.
For $c \in C$ and $y \not\modeq x_c$ there is a 
modal $\langML$-formula $\delta_{c,y}$ such that $x_c \models \delta_{c,y}$
and $y \models \lnot\delta_{c,y}$,
so by taking
$\Delta_c = \{\delta_{c,y} \mid y \not\modeq x_c\}$,
we have $\modcls{x_c} = \ext{\bigwedge\Delta_c}^{\nstr{M}}  \subseteq X$ 
for each $c \in C$.
By modal compactness of $\cmp{X}$,
for each $c \in C$ there is a finite subset 
$\Delta^0_c \subseteq_\omega \Delta_c$ such that
$\modcls{x_c} \subseteq \ext{\bigwedge\Delta^0_c}^{\nstr{M}} \subseteq X$.
Defining $\delta_c = \bigwedge\Delta^0_c$ for each $c \in C$,
we therefore have
$X = \bigcup_{c \in C} \ext{\delta_c}^{\nstr{M}}$.
Now by modal compactness of $X$, we get a finite subset
$\Delta_0 \subseteq_\omega \{\delta_c \mid c\in C\}$
such that 
$X = \ext{\bigvee\Delta_0}^{\nstr{M}}$. 
That is, $X$ is definable by the formula $\delta = \bigvee\Delta_0$.
\qed

\begin{prop}\label{prop:m-satur-mod-congr}
If $\nstr{M}$ is a modally saturated neighbourhood model,
then modal equivalence is a congruence on $\nstr{M}$.
It follows that
modally equivalent states in $\nstr{M}$
are behaviourally equivalent.
\end{prop}
\proof
Immediate consequence of 
Lemmas \ref{lem:modcoh-definable} and
\ref{lem:m-satur-definability}.
\qed

\begin{cor}\label{cor:finite-HM}
The class of finite neighbourhood models is a Hennessy-Milner class.
\end{cor}
\proof
Since the disjoint union of two finite neighbourhood models
is again finite, it suffices 
by Lemma~\ref{lem:m-satur-du-hm} and
Proposition~\ref{prop:m-satur-mod-congr}
to show that finite neighbourhood models
are modally saturated.
But this is immediate, since any set of states in a 
finite neighbourhood model $\nstr{M}$, is necessarily finite, 
and hence modally compact, so $\nstr{M}$ is modally saturated.
\qed

The question remains whether the class of all 
modally saturated neighbourhood models is a Hennessy-Milner class.
We conjecture that 
if $\nstr{M}$ and $\nstr{N}$ are modally saturated
then modal equivalence is a congruence on $\nstr{M} + \nstr{N}$.
If this is the case, then the Hennessy-Milner theorem follows from 
Lemma~\ref{lem:m-satur-du-hm}.

\begin{rem}\label{rem:mon-m-saturation}
In \cite{Pau99:mon-bis} the following definition of modal saturation
for monotonic neighbourhood models was introduced, 
and it was shown that over the class of
modally saturated monotonic neighbourhood models
modal equivalence implies monotonic bisimilarity.
A monotonic neighbourhood model $\tup{S,\nu,V}$ 
is {\em monotonic modally saturated}, 
if
for all $s \in S$ and all sets $\Psi$ of modal $\langML$-formulas
the following hold:\\

\noindent\begin{tabular}{l p{10.5cm}}
(m1-mon) & For all $X \in \nu(s)$, if $\Psi$ is finitely
  satisfiable in $X$, then $\Psi$ is satisfiable in $X$.\\
(m2-mon) & If for all $\Psi_0 \subseteq_\omega \Psi$,
  there is an $X \in \nu(s)$ such that 
  $X \subseteq (\bigwedge\Psi_0)$,
  then there is an $X \in \nu(s)$ such that 
  $X \subseteq (\bigwedge\Psi)$.\\
\end{tabular}\\

In a monotonic neighbourhood model $\nstr{M}$,
(m1-mon) clearly implies that all modally coherent neighbourhoods
are modally compact.
The converse also holds, since
for any neighbourhood $X$ of some state $s$,
the closure $X'$ of $X$ with respect to modal equivalence,
i.e., $X' = \bigcup_{x \in X} \modcls{x}$,
is also a neighbourhood of $s$ by monotonicity,
and for any $\Psi \subseteq \langML$,
$\Psi$ is satisfiable in $X$ if and only if 
$\Psi$ is satisfiable in $X'$.
However, it is not clear whether monotonic modal saturation and 
(neighbourhood) modal saturation coincide in all monotonic models.
We suspect that neither implies the other due to the following.
The condition (m2-mon) says that all neighbourhood collections are 
closed under arbitrary intersections of definable neighbourhoods, 
a property which we expect can be shown to fail in some 
modally saturated neighbourhood model.
On the other hand, it is not clear why the complements of 
modally coherent neighbourhoods should be modally compact 
in a monotonic modally saturated model.
Unfortunately, at the moment we have no examples that confirm
these intuitions. 
\end{rem}

\begin{rem}\label{rem:krip-m-saturation}
A Kripke model $\nstr{K} = \tup{S,R,V}$ is {\em Kripke modally saturated},
if for all $s \in S$ and all sets $\Psi$ of modal $\langML$-formulas:

\medskip\noindent\begin{tabular}{l p{12cm}}
(m1-krip) & If $\Psi$ is finitely
  satisfiable in $R[s]$, then $\Psi$ is satisfiable in $R[s]$,
\end{tabular}

\medskip\noindent and over the class of modally saturated 
Kripke models, modal equivalence implies Kripke bisimilarity
(see e.g. \cite{BdRV:ML-book}).
From the above definitions,
it is clear that for any augmented neighbourhood model $\nstr{M}$,
if $\nstr{M}$ is monotonic modally saturated 
or (neighbourhood) modally saturated,  
then $\krp{\nstr{M}}$ is Kripke modally saturated.
However, 
if $\krp{\nstr{M}}$ is Kripke modally saturated, 
then modally coherent neighbourhoods may fail to 
be modally compact in $\nstr{M}$.
This is shown by Example~\ref{ex:imfi} (page \pageref{ex:imfi})
in the next subsection.
Hence Kripke modal saturation does not imply monotonic modal saturation
nor (neighbourhood) modal saturation.
Note that (m2-mon) holds over any augmented neighbourhood model.
\end{rem}

As we have seen in 
Remarks~\ref{rem:mon-m-saturation}~and~\ref{rem:krip-m-saturation},
the notions of neighbourhood, monotonic and Kripke modal saturation 
do not restrict in a natural way.
Moreover, in the next subsection (Example~\ref{ex:imfi}),
we will see that
image-finite neighbourhood models are not necessarily 
modally saturated.
These observations could be interpreted as arguments 
for saying that our definition of 
modal saturation for neighbourhood models is not the right one.
On the other hand,
Definition~\ref{defi:m-saturation} arises in a natural manner,
it implies Kripke modal saturation over Kripke models,
in subsection~\ref{ssec:ultrafilter-ext} we show that
ultrafilter extensions of neighbourhood models are modally saturated,
and in subsection~\ref{ssec:char-thm} we will see that
when viewing neighbourhood models as first-order models,
then $\omega$-saturation implies modal saturation
(Lemma~\ref{lem:countable-satur-m-satur}).
We believe these are good arguments for 
Definition~\ref{defi:m-saturation} being the right notion after all.
However, further investigations are needed to support this claim.
It would be useful to have a better understanding of 
what an abstract notion of modal saturation for $\Fun$-coalgebras
should be.

\subsection{Image-finite neighbourhood models}
\label{ssec:image-finite}

In normal modal logic, we know that image-finite Kripke models
are modally saturated, and hence form a Hennessy-Milner class
with respect to Kripke bisimilarity.
In this section, we describe image-finite neighbourhood models
and prove that they form a Hennessy-Milner 
class, despite the fact that, in general, 
they are not modally saturated.

\begin{rem}
We obtain our notion of an image-finite neighbourhood model by 
instantiating a widely used categorical definition.
Similarly, we could obtain the Hennessy-Milner result 
of this section
by using a far more general theorem from coalgebraic 
modal logic. Our motivation for giving an ``elementary'' proof is 
that we want to equip the working modal logician with
some intuition concerning image-finite neighbourhood models.
We outline how the result could be 
obtained as a corollary from coalgebraic work
in Remark~\ref{rem:referee_happy} below. 
\end{rem}

In contrast with the Kripke case, 
image-finite neighbourhood models are not
necessarily modally saturated.
Instead, we will show that they satisfy the 
condition of the following lemma.

\begin{lem}\label{lem:modcoh-imfi}
Let $\nstr{M} = \tup{S,\nu,V}$ be a neighbourhood model.
If for any states $s_1, s_2 \in S$ and 
any modally coherent subset $X \subseteq S$
there is a formula $\varphi \in \langML$ such that for any $i \in \{1,2\}$,
$X \in \nu(s_1)$ if and only if $\ext{\varphi}^\nstr{M} \in \nu(s_2)$,
then modal equivalence is a congruence on $\nstr{M}$.
\end{lem}

\proof
Immediate by the characterisation given by conditions (c1) and (c2)
on page \pageref{eq:modeq-congruence}.
\qed

A Kripke model is image-finite if every state
has only finitely many successors (cf.~\cite{BdRV:ML-book}).
For neighbourhood models, the notion of image-finiteness
is less obvious,
but as with bisimilarity, universal coalgebra provides us with an 
abstract notion of image-finiteness for coalgebras
which we instantiate for the $\TTwo$-functor.
The general construction behind this definition
is that of taking the finitary part of a functor.
Recall that we denote the inclusion map of $Y \subseteq X$ 
by $\inc_Y \colon Y \hookrightarrow X$.
Given any functor $\Fun\colon \Set \to \Set$,
define the functor $\Fun_\omega$ by letting
\[ \Fun_\omega(X) = \bigcup\{ \Fun(\inc_Y)[\Fun Y] \mid 
       \inc_Y \colon Y \hookrightarrow X, Y \subseteq_\omega X \}
\]
for a set $X$, and for a function $f\colon X \to Y$,
$\Fun_\omega(f)$ is the restriction of $\Fun(f)$ to $\Fun_\omega(X)$.
It is known that $\Fun_\omega$
is the unique finitary (or $\omega$-accessible) subfunctor
of $\Fun$ which agrees with $\Fun$ on all finite sets
(see e.g.~\cite{AdaPor04:tree-coalgs,patt03:NASSLLI-notes}),
and $\Fun_\omega$ is called the finitary part of $\Fun$.
We now give a characterisation of the finitary part of $\TTwo$.
For a subset inclusion map $\inc_B: B \hookrightarrow X$ 
and $D \subseteq X$, 
note that $\inc_B^{-1}[D] =  D\cap B$. 
If $U \in \TTwoIF(X)$ and $B \subseteq X$ is such that
for all $D \subseteq X$: $D \in U \iff D\cap B \in U$,
then we call $B$ a {\em base set for $U$}.

\begin{lem}
\label{lem:TTwoIF}
Let $X$ be a set. We have:
\[ \begin{array}{l}
  \TTwoIF(X) = 
   \{ U \in \TTwo(X) \mid \exists B \subseteq_\omega X . 
           \, \forall D \subseteq X : (D \in U \iff D\cap B \in U)\}.
\end{array}\]
\end{lem}

\proof
The proof is obtained by spelling out the definitions.
\qed 

\begin{defi}\label{defi:imfi}
We define the class of image-finite neighbourhood frames
as the class $\Coalg{\TTwoIF}$ of $\TTwoIF$-coalgebras.
The class of image-finite neighbourhood models is the class of
neighbourhood models based on an image-finite neighbourhood frame.
\end{defi}

So, image-finite neighbourhood frames are the neighbourhood frames 
in which all neighbourhood collections are determined by a finite base set. 
It should be clear that a finite neighbourhood frame $\tup{S,\nu}$
is image-finite, since  for all $s \in S$, 
$S$ is a finite base set for $\nu(s)$.
In proving that image-finite neighbourhood models form a 
Hennessy-Milner class, we use the following lemma.

\begin{lem}\label{lem:coh-imficoh}
Let $S$ be a set and $\theta$ an equivalence relation on $S$.
Moreover, let $B \subseteq S$ and denote by $B_\theta \subseteq B$
a set of representatives of the $\theta$-classes
intersecting $B$.
For all $X,X' \subseteq S$,
if $X$ and $X'$ are both $\theta$-coherent,
then 
$X\cap B = X' \cap B$ iff $X \cap {B_\theta} =  X'\cap {B_\theta}$.
\end{lem}

\proof
Let $S, B$ and $B_\theta \subseteq B$ be as stated,
and assume that $X$ and $X'$ are $\theta$-coherent subsets of $S$.
It is clear 
that $X \cap B =  X' \cap B$ implies $X \cap B_\theta =  X' \cap B_\theta$.
For the other implication, assume $X \cap B_\theta = X'\cap B_\theta$.
We have: 
$s \in X \cap B$ implies 
there is an $s' \in B_\theta$ such that $s \theta s'$.
Since $X$ is $\theta$-coherent, 
$s' \in X \cap B_\theta = X' \cap B_\theta$.
Now since $X'$ is $\theta$-coherent, $s \in X'$, and thus $s \in X' \cap B$. 
Hence we have shown $X \cap B \subseteq X' \cap B$.
The other inclusion is shown similarly.
\qed


\begin{prop}\label{prop:HM-imfi-nbhd-models}
The class of image-finite neighbourhood models
is a Hen\-nes\-sy-Milner class.
\end{prop}

\proof
The class of image-finite neighbourhood models
is closed under disjoint unions,
since for any functor $\Fun$, 
the category $\Coalg{\Fun}$ has coproducts
(cf.~\cite{Rut00:TCS-univ-coal}).
By Lemma~\ref{lem:m-satur-du-hm} it suffices to show that 
in an image-finite neighbourhood model,
modal equivalence is a congruence.
So let $\nstr{M} = \tup{S,\nu, V}$ be image-finite, 
and let $s, t \in S$.
We then have finite base sets
$B_s, B_t \subseteq _\omega S$ for $\nu(s)$ and 
$\nu(t)$, respectively.
Let $B_{st} = B_s \cup B_t$.
By Lemma~\ref{lem:modcoh-imfi}
it suffices to find for any  modally coherent $X \subseteq S$,
a formula $\varphi \in \langML$ 
such that 
\begin{equation}\label{eq:Fst}
X \cap B_{st} \;=\;  \ext{\varphi}^{\nstr{M}} \cap B_{st} ,
\end{equation}
since then
$X \cap B_{s} =  \ext{\varphi}^{\nstr{M}} \cap B_{s}$
and
$X \cap B_{t} =  \ext{\varphi}^{\nstr{M}} \cap B_{t}$, and
hence 
$X \in \nu(s)$ iff $\ext{\varphi}^\nstr{M} \in \nu(s)$,
similarly for $t$, and consequently,
if $s \modeq t$, 
then $X \in \nu(s)$ if and only if $X \in \nu(t)$.

We now show how to obtain such a $\varphi$.
Let $X \subseteq$ be modally coherent and let
$B_{st}' \subseteq B_{st}$
be a set of representatives of the $\modeq$-classes intersecting
$B_{st}$.
Since $B_{st}$ is finite, so is $B_{st}'$.
Assume $B_{st}' = \{s_1, \ldots, s_n\}$.
Now there are modal formulas $\varphi_1, \ldots, \varphi_n \in \langML$
which characterise $s_1, \ldots, s_n$, respectively, 
within $B_{st}'$,
that is,
$\nstr{M}, s_i \models \varphi_j$ iff $i=j$, for $1 \leq i,j \leq n$.
Namely, for each $s_i \in B_{st}'$, 
we have for all $s_j \in B_{st}' \setminus \{s_i\}$,
$s_i \not\modeq s_j$.
Hence there is a formula $\varphi_{i,j}$ such that   
$\nstr{M}, s_i \models \varphi_{i,j}$ and $\nstr{M}, s_j \not\models \varphi_{i,j}$.
Take $\varphi_i = \bigwedge_{j=1, j\neq i}^{n} \varphi_{i,j}$, $i = 1, \ldots,n$.
We now define $\varphi = \bigvee \{\varphi_i \mid s_i \in X \cap B_{st}'\}$.
To see that $\varphi$ satisfies (\ref{eq:Fst})
it suffices by Lemma~\ref{lem:coh-imficoh} to show that
$X  \cap {B_{st}'} =  \ext{\varphi}^\nstr{M}\cap {B_{st}'}$.
Clearly, by definition of $\varphi$, if
$s_i \in X \cap B_{st}'$ then 
$s_i \in \ext{\varphi}^\nstr{M} \cap B_{st}'$. 
Conversely, if $s_j \in \ext{\varphi}^\nstr{M} \cap B_{st}'$
then $\nstr{M}, s_j \models \varphi_i$ 
for some $i$ such that $s_i \in X \cap B_{st}'$.
Since $\varphi_i$ characterises $s_i$ in $B_{st}'$, 
it follows that $s_j = s_i \in X \cap B_{st}'$.
\qed

\begin{rem}\label{rem:hm-cml}\label{rem:referee_happy}
As we already mentioned,
Proposition~\ref{prop:HM-imfi-nbhd-models}
is a consequence of a more general 
result in coalgebraic modal logic,
which we briefly explain here.
In coalgebraic modal logic, the semantics of modalities is given by
predicate liftings.
A predicate lifting for a functor $\Fun\colon\Set\to\Set$
is a natural transformation 
$\predlif\colon \Two \to \Two\circ\Fun$.
Given a set $\Lambda$ of predicate liftings for $\Fun$,
the finitary coalgebraic modal language 
$\langML(\Lambda)$ is the multi-modal language
which contains a modality $\mop{\predlif}$
for each $\predlif\in\Lambda$.
Given an $\Fun$-coalgebra $\str{X} = \tup{X,\xi}$,
the truth of formulas is defined in the standard inductive manner
for the basic Boolean connectives.
The truth of a modal formula $\mop{\predlif}\phi$ is defined by:
$\str{X},x \models \mop{\predlif}\phi$ iff 
$\xi(x) \in \predlif_X({\ext{\phi}^\str{X}})$.
Atomic propositions can also be interpreted using 
constant predicate liftings.
We refer to \cite{Patt03:coalg-ML} for details.

Using currying, every predicate lifting
$\predlif\colon \Two \to \Two\circ\Fun$
corresponds to a natural transformation
$\transp{\predlif} \colon \Fun \to \TTwo$,
called the transposite of $\predlif$.
A set $\Lambda$ of predicate liftings for $\Fun$ is called
{\em separating} if
the source of transposites 
$\{ \transp{\predlif} \mid \predlif\in\Lambda\}$
is jointly injective.
Schr\"{o}der shows 
in~\cite[Theorem 41,Corollary 45]{Schr08:TCS-expr}) that
if $\Fun\colon \Set \to \Set$ is a finitary functor,
and $\Lambda$ is a separating set of predicate liftings,
then the 
finitary coalgebraic modal language $\langML(\Lambda)$
is expressive for $\Fun$-coalgebras,
meaning that over the class of $\Fun$-coalgebras,
$\langML(\Lambda)$-equivalence implies behavioural equivalence.

We can instantiate the result for the finitary functor
$\TTwoIF\times\Pow(\At)$ and classical modal logic. 
The basic modal language and its interpretation over 
neighbourhood models
is the finitary coalgebraic modal logic given 
by $\Lambda = \{\lambda\} \cup \{\rho_i \mid i < \omega\}$,
where
$\lambda\colon \Two \to \Two \circ \TTwoIF$
is defined by
$\lambda_X(A) = \{ U \in \TTwoIF(X) \mid A \in U \}$,
and the $\rho_i$, $i < \omega$, are constant predicate liftings
that interpret the atomic propositions.
It is known that $\{\lambda\} \cup \{\rho_i \mid i < \omega\}$
is separating iff $\{\lambda\}$ is separating.
The transposite $\transp{\lambda}\colon \TTwoIF \to \TTwo$ 
is simply the inclusion map, i.e., 
$\transp{\lambda}_X = \inc_{\TTwoIF(X)}$ for all sets $X$,
so trivially $\{\transp{\lambda}\}$ is jointly injective,
hence $\{\lambda\}$ is separating.
It now follows from Schr\"{o}der's result 
that over the class of image-finite neighbourhood models,
modal equivalence implies behavioural equivalence. 
\end{rem}

We now show that the notion of image-finiteness for neighbourhood frames 
restricts to the subclasses of neighbourhood frames
that correspond with Kripke frames and monotonic neighbourhood frames, 
respectively.

Monotonic neighbourhood frames are coalgebras for the
subfunctor $\Mon$ of $\TTwo$ (cf. Remark~\ref{rem:mon-nbhd-coalg})
which sends a set $X$ to the collection of all subsets of $\Pow(X)$
which are closed under supersets.
Due to motonicity, given a function $f \colon X \to Y$,
we can describe $\Mon(f)$ in terms of the direct image of $f$,
namely, for all $V \in \Mon(X)$,  
$\Mon(f)(V) = \bigcup \{ \up{f[D]} \mid D \in V \}$. 
Recall that for a subset $B \subseteq X$, 
$\up{B} = \{ B' \subseteq X \mid B \subseteq B'\}$.
Image-finite monotonic neighbourhood frames,
are then nothing but $\MonIF$-coalgebras.
By simply working out the definitions, we find that
for a set $X$ and $U \in \Mon(X)$:
\[U \in \MonIF(X) \quad\text{ iff }  \quad
   \exists C_1, \dotsc, C_n \subseteq_\omega X : 
            U = \up{C_1} \cup \ldots \cup \up{C_n}.
\]
The neighbourhood collections in
an image-finite monotonic neighbourhood model
are thus generated by finite sets of finite neighbourhoods
which are minimal with respect to $\subseteq$ in $\Pow(X)$.
Such minimal neighbourhoods will be referred to as core neighbourhoods.
More precisely,
if $\nstr{M} = \tup{S,\nu,V}$ is a neighbourhood model, $s \in S$
and $C \in \nu(s)$ is such that for all $D \subsetneq C$,
$D \notin \nu(s)$, $C$ is called a {\em core neighbourhood} of $s$.
The collection of core neighbourhoods of $s$ is denoted $\nu^c(s)$.
This terminology follows \cite{Pau:phd,Han03:math-thesis}
where image-finite monotonic neighbourhood models
were called {\em locally core finite}.

Finally, recall that a Kripke model $\tup{S,R,V}$ is image-finite,
if for all $s \in S$, the set of $R$-successors $R[s]$ is finite.

\begin{prop}\label{p:imfi-krip-mon}
Let $\nstr{M} = \tup{S,\nu,V}$ be a neighbourhood model. 
\begin{enumerate}[\em(1)]
\item If $\nstr{M}$ is a monotonic neighbourhood model, then
$\nstr{M}$ is image-finite as a monotonic neighbourhood model iff
$\nstr{M}$ is image-finite as a neighbourhood model.
\item If $\nstr{M}$ is augmented, then
$\krp{\nstr{M}}$ is image-finite as a Kripke model iff 
$\nstr{M}$ is image-finite as a neighbourhood model.
\end{enumerate}
\end{prop}
\proof
To prove item 1, let $\nstr{M}$ be monotonic. 
Since $\Mon$ is a subfunctor of $\TTwo$, also $\MonIF$ is a subfunctor
of $\TTwoIF$. It follows that any image-finite monotonic model
is also image-finite as a neighbourhood model. 
Concretely, one can show that for all $s \in S$, 
the union of core neighbourhoods $B = \bigcup \nu^c(s)$
is a finite base set for $\nu(s)$.
For the other direction, assume $\nstr{M}$ is image-finite as a 
neighbourhood model. 
Let $s \in S$, and assume $B \subseteq_\omega S$ is a finite base set for
$\nu(s)$. 
We first show that every neighbourhood is in the upwards closure 
of some finite core neighbourhood: $U \in \nu(s)$ implies
$B \cap U \in \nu(s)$,  and since $B \cap U$ is finite,
there must be a finite $C \in \nu^c(s)$ 
such that $C \subseteq B \cap U \subseteq U$.
Suppose now that $C \in \nu^c(s)$ is an arbitrary core neighbourhood
of $s$.
As $B$ is a base set for $\nu(s)$,
$C \cap B \in \nu(s)$, and hence by $\subseteq$-minimality of $C$,
$C \subseteq B$.
It now follows from the finiteness of $B$,
that $s$ has only finitely many core neighbourhoods $C_1,\dotsc, C_n$ 
of finite cardinality,
and $\nu(s) = \up{C_1} \cup \dotsc \cup \up{C_n}$.

For item 2, let $\krp{\nstr{M}} = \tup{S,R,V}$, i.e., 
for all $s \in S$, $\nu(s) = \up{R[s]}$, and 
$\nu^c(s) = \{R[s]\}$.
This immediately shows that if $\krp{\nstr{M}}$ is image-finite
then $\nstr{M}$ is image-finite as a monotonic model,
and hence by item 1, also as a neighbourhood model.
Conversely,
if $\nstr{M}$ is image-finite, then by item 1 
$\nstr{M}$ is image-finite as a monotonic model,
hence for all $s \in S$, $\bigcup\nu^c(s)  = R[s]$ is finite.
\qed

The following example demonstrates that image-finite neighbourhood models
are not necessarily modally saturated, and it also shows that
a Kripke modally saturated model, is not necessarily modally saturated as 
a (monotonic) neighbourhood model.

\begin{exa}\label{ex:imfi}
Consider the Kripke model $\nstr{K} = \tup{S,R,V}$
where $S = \bb{N}$, the set of natural numbers, 
and $R$ is the usual relation $>$ on $\bb{N}$, 
that is, for $m,n \in \bb{N}$, $\tup{m,n} \in R$ iff $m > n$,
and $R[m] = \{ n \in \bb{N} \mid n < m \}$.
Finally, the valuation $V$ is defined as
$V(p_i) = \emptyset$, for all atomic propositions $p_i \in \At$.
$\nstr{K}$ is an image-finite Kripke model, 
hence by Proposition~\ref{p:imfi-krip-mon}
the augmented neighbourhood model 
$\aug{\nstr{K}}$ corresponding to $\nstr{K}$ is also image-finite
as a (monotonic) neighbourhood model.
Since $\nstr{K}$ is image-finite, $\nstr{K}$ is
Kripke modally saturated.
However, $\aug{\nstr{K}}$ is not modally saturated
as a neighbourhood model nor as a monotonic model. 
To see this, first note that
the set $\bb{N}$ is trivially modally coherent 
and by monotonicity $\bb{N}$ is also a
neighbourhood of every $n \in \bb{N}$.
Now,
consider the set of modal $\langML$-formulas,
$\Psi = \{ \Diamond^n\Box\bot \mid n \in \bb{N} \}$.
Note that by transitivity,
$\nstr{K}, m \models  \Diamond^n\Box\bot$ iff $m \geq n$.
Since $\nstr{K}$ and $\aug{\nstr{K}}$ are pointwise equivalent, and
every finite subset $\Psi_0 \subseteq_\omega \Psi$ is satisfiable 
in $\nstr{K}$ at the maximal $n \in \bb{N}$ such that 
$\Diamond^n\Box\bot \in \Psi_0$, it follows that
$\Psi$ is finitely satisfiable in the neighbourhood $\bb{N}$
in $\aug{\nstr{K}}$.
However, $\Psi$ is clearly not satisfiable in $\bb{N}$.
We have thus shown that $\bb{N}$ is not modally compact,
hence $\aug{\nstr{K}}$ is not (monotonic) modally saturated.
\end{exa}

\subsection{Ultrafilter extensions}
\label{ssec:ultrafilter-ext}

In this section, we prove a 
behavioural-equivalence-some\-where-else result 
by showing that any two modally equivalent states of 
neighbourhood models have behaviourally equivalent 
representatives in the ultrafilter extensions of 
these neighbourhood models. 
To this end, we define ultrafiler extensions of neighbourhood models,
and we prove analogues of results
known for ultrafilter extensions of Kripke models.
In particular, we show that ultrafilter extensions are
modally saturated. This result will be used
in our proof of Craig interpolation in
subsection~\ref{ssec:interpolation}.

Just as ultrafilter extensions of Kripke models
are obtained from algebraic duality (see e.g.~\cite{BdRV:ML-book}),
ultrafilter extensions of neighbourhood models 
are a by-product of a more general duality between 
coalgebras and certain algebras on the category of 
Boolean algebras,
as described in 
e.g.~\cite{KupKurPat:uf-ext,KurRos07:GoldblattThomason-thm}.
Our definition of ultrafilter extensions of 
neighbourhood frames is obtained 
by instantiating the more general definition of 
ultrafilter extensions of $\Fun$-coalgebras
presented in~\cite{KurRos07:GoldblattThomason-thm}
to $\Fun=\TTwo$.
The basic properties follow from
the category theoretical framework. 
With quite some effort, the
behavioural-equivalence-somewhere-else result
can be obtained as a special case of a more general theorem
in \cite{KupKurPat:uf-ext}.
However, 
instead of requiring knowledge of the (rather abstract) theory in
\cite{KupKurPat:uf-ext,KurRos07:GoldblattThomason-thm},
we have chosen to give a direct, concrete description of
ultrafilter extensions of neighbourhood models,
and to use standard model-theoretic techniques to prove
basic properties. 
We believe that such a presentation will make the results of
this section and the proof of the Craig interpolation theorem
better accessible to readers whose background is mainly
in modal logic.
For the interested reader, we give a brief summary
of the construction from~\cite{KurRos07:GoldblattThomason-thm}
in Remark~\ref{rem:ue-KurRos}.

\newcommand{\BA}{\cat{BA}}
\newcommand{\MA}{\cat{MA}}
\newcommand{\Stone}{\cat{Stone}}
\newcommand{\Kripke}{\cat{Krip}}
\newcommand{\PowBA}{\mathbb{P}}
\newcommand{\UfBA}{\mathbb{U}}
\newcommand{\Spa}{\mathbb{S}pa}
\newcommand{\KrpExt}[1]{\tilde{#1}}
\newcommand{\NbhExt}[1]{\bar{#1}}
\newcommand{\LogBA}{\mathbb{L}}

Let us begin by introducing some terminology and notation, 
and recalling some facts concerning ultrafilters.

\begin{defi}
	Let $S$ be a non-empty set. 
    A set 
	$\U \sse \Pow (S)$ is called an {\em ultrafilter over $S$} if
	$S \in \U$,
	$U_1,U_2 \in \U$ implies $U_1 \cap U_2 \in \U$,
	$U_1 \in \U$ and $U_1 \sse U_2 \sse S$ 
	implies $U_2 \in \U$, and
	for all $U\sse S$ we have: 
        $U \in \U$ iff $S \setminus U \not\in \U$.
    The collection of ultrafilters
    over $S$ will be denoted by $\Uf(S)$.    
For a set $S$ and a subset $U \sse S$, we define
\[ 
\ufhat{U} \coloneqq \{ \U \in \Uf(S) \mid U \in \U \}.
\]
For a set $S$ and $s \in S$, we define
\[ \princ_s \coloneqq \{U \sse S \mid s \in U \}.\]
It can easily be confirmed that $\princ_s \in \Uf(S)$.
The induced map $\princ\colon S  \to  \Uf(S)$
is called the {\em principal ultrafilter map}
and $\princ_s$ is the
{\em principal ultrafilter} generated by $s$.
\end{defi}

The duality betwen Stone spaces and Boolean algebras
gives rise to the following two contravariant functors.
$\PowBA\colon \SetOp\to\BA$ maps a set $X$ to its Boolean
algebra of subsets. 
The functor $\UfBA\colon \BA \to \SetOp$ 
maps a Boolean algebra 
to the set of its ultrafilters. 
Both functors can be regarded as subfunctors of the contravariant
powerset functor $\Two$, 
as they both map a morphism $f$ in their respective categories
to the inverse image function $f^{-1}$.
Composing these functors, we find that for a set $X$,
$\UfBA\PowBA(X) = \Uf(X)$, and 
for a function $f\colon X \to Y$,
$\UfBA\PowBA(f) = (f^{-1})^{-1}$.
Hence $\Uf$ can be regarded as a subfunctor of $\TTwo$.

The following definition of 
ultrafilter extensions of neighbourhood models is obtained 
by instantiating the corresponding  
coalgebraic notion for $\Fun$-coalgebras in 
\cite{KurRos07:GoldblattThomason-thm} to the case that
$\Fct{F} =\TTwo$. 
We sketch the main ideas of the construction
in Remark~\ref{rem:ue-KurRos} below.
In fact, the definition of the neighbourhood
relation of the ultrafilter extension goes back to 
the definition of the canonical neighbourhood model 
in~\cite{Segerberg71:classic-ML}. 

\begin{defi}\label{defi:uf-ext}
	Let $\nstr{M}=\tup{S,\nu,V}$ 
	be a neighbourhood model. The {\em ultrafilter
	extension} of $\nstr{M}$ is defined as the triple
	$\nstr{M}^u \coloneqq \tup{\Uf(S),\mu,V^{u}}$, where
	\begin{enumerate}[$\bullet$]  
		\item $\Uf(S)$ is the set of ultrafilters over the
		   set $S$,
		\item $\mu: \Uf(S) \to \TTwo(\Uf(S))$ is defined by
			\[ \mu(\U) \coloneqq \{ \ufhat{U} \sse \Uf(S) \mid
			   U \sse S, \; \black U \in \U \}, \]
			where for any $U \sse S$ we put
			$\black U \coloneqq \{ s \in S \mid U \in \nu(s)\}$,
		\item $V^u(p) \coloneqq \{ \U \in \Uf(S) \mid 
		   V(p) \in \U \}$.
	\end{enumerate}
\end{defi}

\begin{rem}\label{rem:ue-KurRos}
   In~\cite{KurRos07:GoldblattThomason-thm} 
   the neighbourhood functor $\TTwo$
   is denoted by $\mathcal{H}$. Given the coalgebraic
   modal logic for neighbourhood frames with one 
   predicate lifting for the interpretation 
   of the $\Box$-operator (see Remark~\ref{rem:referee_happy})
   one can define a functor $\LogBA\colon\BA \to \BA$ 
   such that the category of $\LogBA$-algebras 
   provides the algebraic semantics of the logic.
For a Boolean algebra $\nstr{A} = \tup{A,+,-,0}$,
$\LogBA(\nstr{A})$ is the free Boolean algebra generated
by $\{ \Box a \mid a \in A\}$.
Let $\cat{Alg}(L)$ be the category of $\LogBA$-algebras over $\cat{BA}$.
The functors 
$\PowBA\colon \SetOp \to \cat{BA}$
and
$\UfBA\colon \cat{BA} \to \SetOp$
are extended to functors
$\NbhExt{\PowBA}\colon \Coalg{\TTwo}^\text{op} \to \cat{Alg}(\LogBA)$
and
$\NbhExt{\UfBA} \colon \cat{Alg}(\LogBA) \to \Coalg{\TTwo}^\text{op}$.
The ultrafilter extension of a $\TTwo$-coalgebra
$\tup{S,\nu}$ is then obtained as
$\NbhExt{\UfBA}\NbhExt{\PowBA}(\tup{S,\nu})$.
The lifting of $\PowBA$ and $\UfBA$ relies on the existence
of two natural transformations:
$\delta \colon  \LogBA\PowBA \to \PowBA\TTwo$ and
$h \colon \UfBA\LogBA\to\TTwo\UfBA$
whose components at a set $X$ are defined as follows
(cf.~Def.~2.6.5 and Ex.~3.6 of
\cite{KurRos07:GoldblattThomason-thm}):
\[\begin{array}{lcl}
\delta_X (\Box U) & = & \{ N \in  \TTwo(X) \mid U \in N\}\\[1ex]
h_X(\U) & = & \{ \ufhat{U} \sse \UfBA\PowBA(X) \mid \Box U \in U \}
\end{array}\]
The liftings $\NbhExt{\PowBA}$ and $\NbhExt{\UfBA}$
are now given as follows on objects:
$\NbhExt{\PowBA}$ maps a $\TTwo$-coalgebra $\tup{X,\nu}$ 
to $\NbhExt{\PowBA}(\tup{X,\nu}) = 
\tup{\LogBA\PowBA(X),\PowBA(\nu)\circ\delta_X}$ as illustrated here: 
\[ \xymatrix{
\LogBA\PowBA(X) \ar[r]^-{\delta_X} &
\PowBA\TTwo(X) \ar[r]^-{\PowBA(\nu)} & \PowBA(X)
}\]
$\NbhExt{\UfBA}$ maps a
$\tup{\nstr{A},\alpha}$ in $\cat{Alg}(\LogBA)$
to $\NbhExt{\UfBA}(\tup{\nstr{A},\alpha}) = 
 \tup{\UfBA(\nstr{A}),h_\nstr{A}\circ\UfBA(\alpha)}$:
\[\xymatrix{
\UfBA(\nstr{A}) \ar[r]^-{\UfBA(\alpha)} &
\UfBA\LogBA(\nstr{A}) \ar[r]^-{h_\nstr{A}} &
\TTwo(\UfBA(\nstr{A}))
}\]
By working out the details,
the reader can now confirm that the composition
$\NbhExt{\UfBA}\NbhExt{\PowBA}$ yields
the ultrafilter extension of neighbourhood frames
provided in Definition~\ref{defi:uf-ext}.
\end{rem}

The construction of the ultrafilter extension
in Definition~\ref{defi:uf-ext} can be seen 
as an extension of the $\Set$-functor
$\Uf\colon \Set\to\Set$ to a functor 
$(\_)^u \colon \cat{Nbhd} \to \cat{Nbhd}$
such that for any neighbourhood model $\nstr{M}$, 
the principal ultrafilter map $\princ$ 
is truth-preserving injective map from 
$\nstr{M}$ into $\nstr{M}^u$.
In order to see that the construction $(\_)^u$ 
of the ultrafilter extension is functorial we show that
bounded morphisms between neighbourhood models induce 
bounded morphisms between the corresponding ultrafilter
extensions. 

\begin{lem}\label{lem:uf_bounded}
   Let $\nstr{M}_1=\tup{S_1,\nu_1,V_1}$ and $\nstr{M}_2=\tup{S_2,\nu_2,V_2}$ be
   neighbourhood models an let $f:S_1 \to S_2$ be a bounded
   morphism from $\nstr{M}_1$ to $\nstr{M}_2$. The function
   $f^u \coloneqq \Uf(f)$ 
   is a bounded morphism from $\nstr{M}_1^u=\tup{\Uf(S_1),
   \mu_1,V_1^u}$ to
   $\nstr{M}_2^u=\tup{\Uf(S_2),\mu_2,V_2^u}$.
\end{lem}

\proof
It can easily be confirmed that for any subset $U \subseteq S_2$:
$(f^u)^{-1}[\ufhat{U}] = \widehat{f^{-1}[U]}$ 
and
$f^{-1}[\black{U}] = \black{(f^{-1}[U])}$.
To prove that $f^u$ is a bounded morphism, 
let $\U \in \Uf(S_1)$ and $U \subseteq S_2$.
We now have:
\[ \begin{array}{rcll}
  \ufhat{U} \in \mu_2(f^u(\U)) 
 & \text{iff} & \black U \in f^u(\U) = \TTwo(f)(\U) & \\[1ex]
 & \text{iff} & f^{-1}[\black U] = 
          \black(f^{-1}[U]) \in \U 
        & \\[1ex] 
 & \text{iff} & \widehat{f^{-1}[U]} = 
          (f^u)^{-1}[\ufhat{U}] \in \mu_1(\U). 
        & 
\end{array}\]
Moreover, $f^u$ respects valuations:
$V_1(p) \in \U$ iff $f^{-1}[V_2(p)] \in \U$ iff
$V_2(p) \in f^u(\U)$.
\qed
%
The next proposition connects truth of a modal formula in the ultrafilter extension to the truth set of the formula in the original model. 

\begin{prop}\label{prop:ufsem}
   Let $\nstr{M}=\tup{S,\nu,V}$ be a neighbourhood model with 
   ultrafilter extension $\nstr{M}^u$. For all $\U \in \Uf(S)$
   and for all formulas $\varphi \in \langML$ we have
   \[ \nstr{M}^u,\U \models \varphi \quad \mbox{iff} \quad \ext{\varphi}^\nstr{M} \in \U .\]
\end{prop}
\proof
The standard proof is obtained by induction on the formula $\varphi$. 
Details are left to the reader.
\qed

Using Proposition~\ref{prop:ufsem},
we now easily show that the principal ultrafilter map
$\princ$ preserves the truth of modal
formulas.
However, it is important to note that, in general, 
$\princ$ is not
a bounded morphism from a model $\nstr{M}=\tup{S,\nu,V}$
to its ultrafilter extension $\nstr{M}^u$. 

\begin{lem}\label{lem:princip-uf-truth}
	Let $\nstr{M}=\tup{S,\nu,V}$  be a neighbourhood model
	with ultrafilter extension $\nstr{M}^u=\tup{\Uf(S),\mu,V^u}$
	and let $\princ:S \to \Uf(S)$ be the injective map
	from $S$ to $\Uf(S)$. For every modal formula
	 $\varphi$ we have $\nstr{M},s\models \varphi$ iff $\nstr{M}^u,\princ_s
	 \models \varphi$.
\end{lem}
\proof
	Let $s \in S$ and let $\varphi$ be modal formula. 
	Then $\nstr{M},s \models \varphi$ iff $s \in \ext{\varphi}^\nstr{M}$
	iff $ \ext{\varphi}^\nstr{M} \in \princ_s $ iff $\nstr{M}^u,\princ_s \models \varphi$
	where the last equivalence is a consequence of 
	Prop.~\ref{prop:ufsem}.
\qed
Another consequence of Proposition~\ref{prop:ufsem} is the fact
that ultrafilter extensions are modally saturated.
\begin{prop}\label{prop:uf-sat}   
   For any neighbourhood model $\nstr{M}$,
   the ultrafilter extension $\nstr{M}^u$ is modally saturated.
\end{prop}
\proof
  Let $\nstr{M} = \tup{S,\nu,V}$ and $\nstr{M}^u=\tup{\Uf(S),\mu,V^u}$.
   We show that any $\ufhat{U} \sse\Uf(S)$ is compact.
   This suffices since
   all neighbourhoods in $\nstr{M}^u$ are of the form 
   $\ufhat{U} \sse \Uf(S)$ and
   for any $\ufhat{U}$,  $\Uf(S)\setminus\ufhat{U}=\widehat{\cmp{U}}$. 
   Let
   $\Psi$ be a set of formulas with the property that $\Psi$
   is finitely satisfiable in $\ufhat{U}$. For any finite
   set of formulas $\{ \psi_1,\ldots,\psi_n\} \sse \Psi$ there
   exists therefore an ultrafilter $\U \in \ufhat{U}$ such that
   $\nstr{M}^u,\U \models \psi_1 \wedge \ldots \wedge \psi_n$. 
   This implies by Prop.~\ref{prop:ufsem} that 
   \[ \{ \ext{\psi_1}^{\nstr{M}},\ldots,\ext{\psi_n}^{\nstr{M}}\} 
   \cup \{U\} \sse \U \]
   Since $\U$ is closed under finite intersections
   this implies $\ext{\psi_1}^{\nstr{M}}\cap
   \ldots\cap \ext{\psi_n}^{\nstr{M}} \cap U \in \U$ and hence 
   $\ext{\psi_1}^{\nstr{M}}\cap \ldots\cap \ext{\psi_n}^{\nstr{M}} \cap U 
   \not= \emptyset$.
   As the set $\{\psi_1,\ldots ,\psi_n\}$ was arbitrary
   we conclude that the set $X \coloneqq
   \{U\} \cup \{ \ext{\psi}^{\nstr{M}} \mid \psi \in \Psi \}$ has the    
   finite intersection property. Hence by the ultrafilter theorem,
   there exists some
   ultrafilter $\U'\in \Uf(S)$ 
   such that $X \sse \U'$. By construction we get
   $\U' \in \ufhat{U}$ and again by
   Prop.~\ref{prop:ufsem},
   that $\Psi$ is satisfiable at $\U' \in \ufhat{U}$.
\qed
We are now able to prove that the class of ultrafilter
extensions of neighbourhood models is a Hennessy-Milner class.
\begin{prop}\label{prop:hm-uf}
   The class 
   $\mathbf{U}\coloneqq
   \{ \nstr{M}^u \mid \nstr{M} \in \cat{Nbhd}\}$ 
   of ultrafilter extensions of neighbourhood models
   is a Hennessy-Milner class.
\end{prop}
\proof
Let $\nstr{M}_1$ and $\nstr{M}_2$ be arbitrary neighbourhood models.
   By Lemma~\ref{lem:m-satur-du-hm} it suffices to
   show that modal equivalence is a congruence
   on the disjoint union
   $\nstr{M}_1^u + \nstr{M}_2^u$ of their ultrafilter extensions. 
   By Proposition~\ref{prop:uf-sat}, $(\nstr{M}_1 + \nstr{M}_2)^u$
   is modally saturated, hence the quotient map 
   $\varepsilon\colon (\nstr{M}_1 + \nstr{M}_2)^u \to 
              (\nstr{M}_1 + \nstr{M}_2)^u/\!\modeq$ 
   is a bounded morphism. 
   Furthermore,   
   denote by $\inc_i \colon \nstr{M}_i \to \nstr{M}_1 + \nstr{M}_2$,
   $i \in \{1,2\}$, the canonical inclusion morphisms.
   By Lemma~\ref{lem:uf_bounded},
   $\inc_i^u \colon \nstr{M}_i^u \to (\nstr{M}_1 + \nstr{M}_2)^u$,
   $i \in \{1,2\}$, are bounded morphisms,
   hence there exists, by the universal
   property of the disjoint union $\nstr{M}_1^u + \nstr{M}_2^u$, 
   a bounded morphism $g$
   such that the following diagram commutes:
\[ \xymatrix{\nstr{M}_1 \ar@{-->}[r] & 
         \nstr{M}_1^u \quad \ar@{^(->}[r] 
	    \ar[rd]_{\inc_1^u} & 
        \;\nstr{M}_1^u + \nstr{M}_2^u \; \ar[d]^g 
        & \quad \nstr{M}_2^u  
        \ar[ld]^{\inc_2^u}  \ar@{_(->}[l] & \nstr{M}_2 \ar@{-->}[l] \\
        & & (\nstr{M}_1+\nstr{M}_2)^u \ar[d]^\varepsilon & \\
	&  & (\nstr{M}_1+\nstr{M}_2)^u/\!\!\modeq & }
\]
Hence $\vareps\circ g \colon 
\nstr{M}_1^u + \nstr{M}_2^u \to (\nstr{M}_1+\nstr{M}_2)^u/\!\!\modeq$
is a bounded morphism, and
two ultrafilters in $\nstr{M}_1^u + \nstr{M}_2^u$
are modally equivalent if and only if they are
identified by $\vareps\circ g$.
It follows that on $\nstr{M}_1^u + \nstr{M}_2^u$, 
the modal equivalence relation is the kernel of
$\vareps\circ g$, and hence a congruence. 
\qed

As a corollary we obtain the 
behavioural-equivalence-somewhere-else result.
\begin{thm}\label{thm:somewhere}
	Let $\nstr{M}_1=\tup{S_1,\nu_1,V_1}$ and 
	$\nstr{M}_2=\tup{S_2,\nu_2,V_2}$ be neighbourhood
	models with the respective ultrafilter extensions 
	$\nstr{M}_1^u$ and 
	$\nstr{M}_2^u$.
	For all states $s_1 \in S_1$ and $s_2 \in S_2$
    we have
	\[ \nstr{M}_1,s_1 \equiv \nstr{M}_2,s_2 \quad \Rightarrow \quad \nstr{M}_1^u,
	\princ_{s_1} \beh \nstr{M}_2^u,\princ_{s_2}.\]
%
\end{thm}
\proof
    Let $s_1$ and $s_2$ be modally equivalent states 
    in $\nstr{M}_1$ and $\nstr{M}_2$, respectively. By 
    Lemma~\ref{lem:princip-uf-truth} the states $\princ_{s_1}$
    and $\princ_{s_2}$ of the ultrafilter extensions
    $\nstr{M}_1^u$ and $\nstr{M}_2^u$ are modally equivalent as well.
    The claim is now a direct consequence of Prop~\ref{prop:hm-uf}.
\qed


\section{Model-theoretic results}
\label{sec:model-theory}

\subsection{The classical modal fragment of first-order logic}
\label{ssec:frag-FOL}

We will now prove that the three equivalence notions described in 
section~\ref{sec:equiv-notions} all characterise the modal fragment 
of first-order logic over the class of neighbourhood models
(Theorem~\ref{thm:char}).
This result is an analogue of Van Benthem's characterisation theorem
for normal modal logic (cf. \cite{Benthem:Correspondence}):
\emph{On the class of Kripke models, 
modal logic is the Kripke bisimulation-invariant fragment of 
first-order logic}.
It is well known that, when interpreted over Kripke models,
the basic modal language $\langML$
can be seen as a fragment of a first-order language which has 
a binary predicate $\ax{R}_\Box$, and a unary predicate $\ax{P}$ for each 
atomic proposition $p$ in the modal language.
Formulas of this first-order language can be interpreted in 
Kripke models in the obvious way. 
Van Benthem's theorem tells us that a first-order formula
$\alpha(x)$ is invariant under Kripke bisimulation if and only if 
$\alpha(x)$ is equivalent to a modal formula.

The first step towards a Van Benthem-style characterisation theorem 
for classical modal logic is to show how $\langML$ can be viewed as a 
fragment of first-order logic.
We will translate modal formulas into a two-sorted first-order language
$\langFO$, which has previously been employed in proving a 
Van Benthem style characterisation theorems for
topological modal logic \cite{CateGabSus:topoML} and
monotonic modal logic \cite{Pau99:mon-bis},
and for reasoning
about topological models more generally
\cite{FlumZiegler:topo-model-theory}.
In Remark~\ref{rem:mon-char-thm}) we will give a more detailed
comparison between our characterisation theorem and the
characterisation theorem for monotonic modal logic 
given in \cite{Pau99:mon-bis}.
The two sorts of the language $\langFO$
are denoted $\sort{s}$ and $\sort{n}$. 
Terms of sort $\sort{s}$ are intended to represent states,
whereas terms of sort $\sort{n}$ are intended to represent 
neighbourhoods. 
We assume there are countable sets of variables of each sort.
To simplify notation, 
we will not state the type of variables explicitly.
Instead we use the following conventions: 
$x,y,x',y',x_1,y_2,\ldots$ denote variables of sort $\sort{s}$
({\em state variables})
and $u,v,u',v',u_1,v_1,\ldots$ denote variables of sort $\sort{n}$ 
({\em neighbourhood variables}).
Furthermore, the language $\langFO$ contains a unary predicate 
$\fof{P}_i$ (of sort $\sort{s}$) for each $i\in\omega$, 
a binary relation symbol $\fof{N}$ relating elements of 
sort $\sort{s}$ to elements of sort $\sort{n}$, and 
a binary relation symbol $\fof{E}$ relating elements of 
sort $\sort{n}$ to elements of sort $\sort{s}$.  
The intended interpretation of $x\fof{N}u$ is 
``$u$ is a neighbourhood of $x$'',
and the intended interpretation of $u\fof{E}x$ is 
``$x$ is an element of $u$''. 
The language $\langFO$ is generated by the following grammar: 
\[ \varphi,\psi \quad ::=\quad x=y\ |\ u=v\ |\ \fof{P}_ix\ |\ x\fof{N}u\ |\ u\fof{E} x\ |\ \neg\varphi\ |\ \varphi\wedge\psi\ |\ \exists x\varphi\ |\ \exists u\varphi \]
where $i\in\omega$; $x$ and $y$ are state variables of sort $\sort{s}$; 
and $u$ and $v$ are neighbourhood variables of sort $\sort{n}$.  
The usual abbreviations (eg. $\forall$ for $\neg\exists\neg$) apply.


Formulas of $\langFO$ are interpreted in two-sorted first-order structures 
 of the type
$\fstr{M}=\langle D^\sort{s},D^\sort{n}, \{P_i\mid i\in\omega\}, N, E\rangle$ where 
$D^\sort{s}$ and $D^\sort{n}$ are the carrier sets of sort $\sort{s}$ and
sort $\sort{n}$, respectively, and
each $P_i\subseteq D^\sort{s}$, $N\subseteq D^\sort{s}\times D^\sort{n}$ and 
$E\subseteq D^\sort{n}\times D^\sort{s}$.  
The usual definitions of free and bound variables apply.
Truth of sentences (formulas with no free variables)  $\varphi\in\langFO$ in a structure $\fstr{M}$ (denoted $\fstr{M}\models\varphi$) is defined as expected.  
If $x$ is a free state variable in $\varphi$ (denoted $\varphi(x)$), 
then we write $\fstr{M}\models\varphi[s]$ to mean that $\varphi$ is 
true in $\fstr{M}$ when $s\in D^\sort{s}$ is assigned to $x$.   
Note that $\fstr{M}\models\exists x\varphi$ iff 
there is an element $s\in D^\sort{s}$ such that $\fstr{M}\models\varphi[s]$. 
If $\Psi$ is a set of $\langFO$-formulas, and $\fstr{M}$ is an $\langFO$-model,
then $\fstr{M} \models \Psi$ means that 
for all $\psi \in \Psi$, $\fstr{M} \models \psi$.
Given a class $\cla{K}$ of $\langFO$-models, 
we denote the {\em semantic consequence relation over $\cla{K}$}
by $\models_\cla{K}$. 
In particular, for $\Psi(x) \cup \{\varphi(x)\} \subseteq \langFO$,
$\Psi(x) \models_\cla{K} \varphi(x)$ if for all $\fstr{M} \in \cla{K}$
and all $s$ of sort $\sort{s}$ in $\fstr{M}$,
$\fstr{M} \models \Phi[s]$ implies $\fstr{M} \models \varphi[s]$.
Moreover, a set of formulas $\Phi(x)$ is $\cla{K}$-consistent 
($\Phi(x) \not\models_\cla{K} \bot$) if
there exists an $\fstr{M} \in \cla{K}$
and an $s$ of sort $\sort{s}$ in $\fstr{M}$ such that
$\fstr{M} \models \Phi[s]$.

We can now translate modal  $\langML$-formulas
and neighbourhood models to the first-order setting in a natural way: 
\begin{defi}\label{fotrans} 
Let $\nstr{M}=\langle S, \nu, V\rangle$ be a neighbourhood model. 
The {\em first-order translation} of $\nstr{M}$ is the structure $\fom{\nstr{M}}=\langle D^\sort{s}, D^\sort{n},\{P_i\mid i\in\omega\},R_\nu,R_\ni\rangle$ where 
\begin{enumerate}[$\bullet$]
\item $D^\sort{s}=S$, $D^\sort{n}=\nu[S]= \bigcup_{s\in S}\nu(s)$
\item $P_i=V(p_i)\;$ for each $i\in\omega$,
\item $R_\nu=\{\tup{s,U}\ | s\in D^\sort{s}, U\in \nu(s)\}$,
\item $R_\ni=\{\tup{U,s}\ | s\in D^\sort{s}, s\in U\}$.
\end{enumerate}
\end{defi}

\begin{defi}\label{fotransforms} 
The {\em standard translation} of the basic modal language is a 
family of functions $st_x:\langML\rightarrow\langFO$ 
defined as follows: 
$\st_x(\bot) = \lnot(x=x)$,
$\st_x(p_i)=\fof{P}_ix$, 
$\st_x(\lnot\varphi) = \lnot\st_x(\varphi)$, $\st_x(\varphi\land\psi) =
\st_x(\varphi) \land \st_x(\psi)$, and
\[\st_x(\Box\varphi)=\exists u(x\fof{N} u\wedge (\forall
y(u\fof{E} y\leftrightarrow \st_y(\varphi))).
\]
\end{defi}
This translation preserves truth; the easy proof is left to the reader.

\begin{lem}\label{lem:st-truth} 
Let $\nstr{M}$ be a neighbourhood model and $\varphi\in\langML$.   
For each $s\in S$, 
$\nstr{M},s\models\varphi\text{ iff }\fom{\nstr{M}}\models st_x(\varphi)[s]$. 
\end{lem}

In the Kripke case,
every first-order model for the language with $\fof{R}_\Box$
can be seen as Kripke model.
However, it is not the case that every $\langFO$-structure is 
the translation of a neighbourhood model. 
Luckily, we can axiomatize the subclass of neighbourhood models
up to isomorphism.
Let $\ax{NAX}$ be the following axioms
\begin{description}
\item[($\ax{A1}$)] $\forall u\exists x(x\fof{N} u)$
\item[($\ax{A2}$)] $\forall u,v ((\forall x ( u\fof{E} x \leftrightarrow v\fof{E}x)) \ra u=v)$
\end{description}
It is not hard to see that if $\nstr{M}$ is a neighbourhood model, 
then $\fom{\nstr{M}}\models\ax{NAX}$.  
The next result states that, in fact, 
$\ax{NAX}$ completely characterises the class
$\cla{N} \coloneqq \{\fstr{M}\ |\ \fstr{M}\cong\fom{\nstr{M}}\text{
  for some neighbourhood model $\nstr{M}$}\}$,
where $\cong$ denotes isomorphism of $\langFO$-models.

\begin{prop}\label{prop:nbhd-axioms} 
Suppose $\fstr{M}$ is an $\langFO$-model and $\fstr{M} \models \ax{NAX}$.
Then there is a neighbourhood model $\nbm{\fof{\fstr{M}}}$ 
such that $\fstr{M}\cong\fom{(\nbm{\fof{\fstr{M}}})}$.  
\end{prop}
\proof
Let $\fstr{M} = \tup{D^\sort{s},D^\sort{n},\{P_i \mid i \in \omega\},N,E}$ 
be an $\langFO$-model such that $\fstr{M} \models \ax{NAX}$.
We will construct from $\fstr{M}$ a neighbourhood model 
$\nbm{\fstr{M}} = \tup{S,\nu,V}$
such that $\fstr{M} \cong \fom{(\nbm{\fstr{M}})}$.
In case $D^\sort{s} = \emptyset$ we also have $D^\sort{n} = \emptyset$ by axiom $\ax{A1}$ 
and hence we define $\nbm{\fstr{M}}$
to be the empty neighbourhood model.  
In the case $D^\sort{s} \not= \emptyset$ we first 
define a map $\eta: D^\sort{n} \to \Pow(D^\sort{s})$ by
$\eta(u) = \{ s \in D^\sort{s} \mid u E s \}$.
We take $S = D^\sort{s}$. 
Now define for each $s \in S$ and each $X \subseteq S$:
$X \in \nu(s)$ iff there is a $u \in D^\sort{n}$ such that $s N u$
and $X = \eta(u)$, and
define for all $i \in \omega$, 
$V(p_i)= \{s \in S \mid \fstr{M} \models \fof{P}_i[s] \}$.
Then $\nbm{\fstr{M}}$ is clearly a well-defined neighbourhood model, and it is not hard
 to see that  the maps $\id:D^\sort{s} \to D^\sort{s}$ and $\eta:  D^\sort{n} \to   \bigcup_{s \in D^\sort{s}} \nu(s)$
 yield an 
isomorphism from $\fstr{M}$ to 
$\fom{(\nbm{\fstr{M}})} = \tup{S,\nu[S],\{P'_i \mid i \in \omega\},R_\nu,R_\ni}$ (cf.~Definition~\ref{fotrans}).  The details 
are left to the reader. 
\qed

Thus, in a precise way, we can think of models in $\cla{N}$ 
as neighbourhood models. 
In particular, if $\fstr{M}$ and $\fstr{N}$ are in $\cla{N}$
we will write $\fstr{M}+\fstr{N}$ by which we (strictly speaking) 
mean the $\langFO$-model $\fom{(\nbm{\fstr{M}} + \nbm{\fstr{N}})}$
(which is also in $\cla{N}$).
Furthermore, Proposition \ref{prop:nbhd-axioms} implies that we can 
work relative to $\cla{N}$ while still preserving nice first-order properties 
such as compactness 
and the existence of countably saturated models.
These properties are essential in the proof of Theorem~\ref{thm:char}.


\subsection{Characterisation theorem}
\label{ssec:char-thm}
We are now able to formulate our characterisation theorem.
Let $\sim$ be a relation on model-state pairs.
Over the class $\cla{N}$, 
an $\langFO$-formula $\alpha(x)$ is {\em invariant under $\sim$},
if for all models 
$\fstr{M}_1$ and $\fstr{M}_2$ in $\cla{N}$
and all sort $\sort{s}$-domain elements $s_1$ and $s_2$
of $\fstr{M}_1$ and $\fstr{M}_2$, respectively,
we have 
$\fstr{M}_1,s_1 \,\sim\,\fstr{M}_2,s_2$ implies
$\fstr{M}_1 \models \alpha[s_1]\;$ iff $\;\fstr{M}_2 \models \alpha[s_2]$.
Over the class $\cla{N}$, 
an $\langFO$-formula 
{\em $\alpha(x)$ is equivalent to the translation of a modal formula} 
if there is a modal formula $\varphi \in \langML$ such that
for all models $\fstr{M}$ in $\cla{N}$, 
and all $\sort{s}$-domain elements $s$ in $\fstr{M}$,
$\fstr{M} \models \alpha[s]\;$ iff $\;\fstr{M} \models \st_x(\varphi)[s]$.

\begin{thm}\label{thm:char}
Let $\alpha(x)$ be an $\langFO$-formula.
Over the class $\cla{N}$ 
the following are equivalent:
\begin{enumerate}[\em(1)]
\item $\alpha(x)$ is equivalent to the translation of
  a modal formula,
\item $\alpha(x)$ is invariant under behavioural equivalence,
\item $\alpha(x)$ is invariant under \precocongs,
\item $\alpha(x)$ is invariant under $\TTwo$-bisimilarity.
\end{enumerate}
\end{thm}

Our proof of Theorem~\ref{thm:char} uses essentially the same
ingredients as the proof of Van Benthem's theorem 
(see e.g.~\cite{BdRV:ML-book}) where the main steps are:
\begin{enumerate}[(1)]
\item Given a Kripke model $\nstr{M}$ we can obtain 
 a modally saturated, elementary extension $\nstr{M}^*$ of $\nstr{M}$.
\item Between modally saturated Kripke models, modal
equivalence is a Kripke bisimulation.
\end{enumerate}
Together, 1 and 2 imply that modally equivalent states 
$\nstr{M},s$ and $\nstr{N},t$ are Kripke bisimilar in their
modally saturated, elementary extensions 
$\nstr{M}^*, s^*$ and $\nstr{N}^*,t^*$.
Our analogue of 2 is that
in a modally saturated neighbourhood model,
modal equivalence is a congruence, which we have shown in
Proposition~\ref{prop:m-satur-mod-congr}.
If we can show an analogue of 1, it follows that
if $\nstr{M},s$ and $\nstr{N},t$ are modally equivalent, then
they have behaviourally equivalent representatives in a 
modally saturated, elementary extension of $\nstr{M}+\nstr{N}$.

As in the Kripke case, we can obtain
an $\omega$-saturated, elementary extension of any $\langFO$-model
in the form of an ultrapower 
using standard first-order logic techniques
(see e.g.~\cite{ChangKeisler73}).
It then only remains to show that an $\omega$-saturated neighbourhood
model (viewed as a $\langFO$-model) is modally saturated.
Before we state and prove this lemma,
we recall (cf.~\cite{ChangKeisler73})
the definition of $\omega$-saturation.
Let $\fstr{M}$ be a first-order $\langFO$-model with domain $M$.
For a subset $C \subseteq M$, the {\em $C$-expansion $\langFO[C]$ of $\langFO$}
is the two-sorted first-order language obtained from $\langFO$
by adding a constant $\underline{c}$ for each $c \in C$.
Now $\langFO[C]$-formulas are interpreted in $\fstr{M}$
by requiring that a new constant $\underline{c}$ is interpreted 
as the element $c$.
The $\langFO$-model $\fstr{M}$ is {\em $\omega$-saturated},
if for every finite $C \subseteq_\omega M$, and every collection $\Gamma(x)$
of $\langFO[C]$-formulas with one free variable $x$ the following holds:
If $\Gamma(x)$ is finitely satisfiable in $\fstr{M}$  (equivalently, if $\Gamma(x)$ is consistent with the $\langFO[C]$ theory of $\fstr{M}$),
then $\Gamma(x)$ is satisfiable in $\fstr{M}$.  It is a classic result 
of model theory that every model has an $\omega$-saturated elementary 
extension (cf.~\cite{ChangKeisler73})

\begin{lem}\label{lem:countable-satur-m-satur}
Let $\fstr{M}$ be a model in $\cla{N}$,
and let $\nbm{\fstr{M}}$ be its corresponding neighbourhood model. 
If $\fstr{M}$ is $\omega$-saturated, 
then $\nbm{\fstr{M}}$ is modally saturated.
\end{lem}
\proof
Let $\fstr{M}$ be an $\langFO$-model in $\cla{N}$,
$\nbm{\fstr{M}} = \tup{S,\nu,V}$ its corresponding neighbourhood model
(cf. Proposition~\ref{prop:nbhd-axioms}),
and assume that $\fstr{M}$ is $\omega$-saturated.
Let $\Psi$ be a set of modal $\langML$-formulas,
and let $U \subseteq S$ be a neighbourhood of some state $s$.
Then $U$ corresponds to a domain element $u \in D^\sort{n}$ of 
$\fstr{M}$ via the isomorphism  $\fstr{M} \cong \fom{(\nbm{\fstr{M}})}$. 
If $\Psi$ is finitely satisfiable in $U$ in $\nbm{\fstr{M}}$,
then the set of $\langFO[\{{u}\}]$-formulas 
$\{\underline{u} \fof{N} x \} \cup \{ \st_x(\psi) \mid \psi \in \Psi\}$
is finitely satisfiable in $\fstr{M}$,
and hence satisfiable, which implies that $\Psi$ 
is satisfiable in $U$.
Similarly, if $\Psi$ is finitely satisfiable in $\cmp{U}$,
then the set of $\langFO[\{{u}\}]$-formulas 
$\{
\lnot \underline{u}\fof{N} x \} \cup \{ \st_x(\psi) \mid \psi \in \Psi\}$
is finitely satisfiable in $\fstr{M}$,
and hence satisfiable, which implies that $\Psi$ 
is satisfiable in $\cmp{U}$.
\qed

We are now ready to prove Theorem~\ref{thm:char}.

\begin{proofof}{Theorem~\ref{thm:char}}
It is clear that {\em 2 $\Ra$ 3 $\Ra$ 4} 
(cf. Proposition~\ref{p:equiv-compare}). 
To see that {\em 4 $\Ra$ 2}, we only need to recall 
(cf.~\cite{Rut00:TCS-univ-coal}) that
graphs of bounded morphisms are $\TTwo$-bisimulations.
Furthermore, as truth of modal formulas is preserved by 
behavioural equivalence, {\em 1 $\Ra$ 2} is clear.
We complete the proof by showing that {\em 2 $\Ra$ 1}.

Let $\MOC_\cla{N}(\alpha)=\{\st_x(\varphi) \mid \varphi \in \langML,\, \alpha(x)\models_{\cla{N}} \st_x(\varphi)\}$
be the set of modal consequences of $\alpha(x)$ over the class $\cla{N}$.
It suffices to show that $\MOC_\cla{N}(\alpha)\models_\cla{N} \alpha(x)$,
since then by compactness 
there is a finite subset 
$\Gamma(x) \subseteq \MOC_\cla{N}(\alpha)$ 
such that $\Gamma(x) \models_\cla{N} \alpha(x)$ and
 $\alpha(x) \models_\cla{N} \bigwedge\Gamma(x)$.
It follows that over $\cla{N}$, 
$\alpha(x)$  is equivalent
to  $\bigwedge\Gamma(x)$, which is the translation of a modal formula.
So suppose $\fstr{M}$ is a model in $\cla{N}$
and $\MOC_\cla{N}(\alpha)$ is satisfied at some element $s$ in $\fstr{M}$.
We must show that $\fstr{M} \models \alpha[s]$.
Consider the set
$T(x)=\{\st_x(\varphi)\ \mid \nbm{\fstr{M}}, s \models\varphi\}\cup\{\alpha(x)\}$.
$T(x)$ is $\cla{N}$-consistent, since
suppose to the contrary that $T(x)$ is $\cla{N}$-inconsistent, 
then by compactness, there is a finite collection of modal formulas 
$\varphi_1, \ldots ,\varphi_n$ such that 
$\nbm{\fstr{M}}, s \models \varphi_i$ for all $i=1, \ldots,n$ and
$ \alpha(x) \models_\cla{N} \lnot \bigwedge_{i=1}^{n} \st_x(\varphi_i)$,
which implies that 
$\lnot \bigwedge_{i=1}^{n} \st_x(\varphi_i) \in \MOC_\cla{N}(\alpha)$.
But this contradicts the assumption that 
$\fstr{M} \models \MOC_\cla{N}(\alpha)[s]$
and
$\fstr{M} \models \st_x(\varphi_i)[s]$ for all $i=1, \ldots,n$.
Hence $T(x)$ is satisfied at an element $t$ in some $\fstr{N} \in \cla{N}$,
and 
by construction, $s$ and $t$ are modally equivalent:
For all modal formulas $\varphi \in \langML$, 
$\fstr{M} \models \st_x(\varphi)[s]$ implies
$\st_x(\varphi) \in T(x)$, and hence $\fstr{N} \models \st_x(\varphi)[t]$.
Conversely, $\fstr{M} \not\models \st_x(\varphi)[s]$ iff 
$\fstr{M} \models \lnot\st_x(\varphi)[s]$ which implies
$\st_x(\lnot\varphi) = \lnot\st_x(\varphi) \in T(x)$, 
and hence $\fstr{N} \not\models \st_x(\varphi)[t]$.

Take now an $\omega$-saturated, elementary extension $\fstr{U}$
of $\fstr{M}+\fstr{N}$. Note that $\fstr{U} \in \cla{N}$,
since validity of $\ax{NAX}$ is preserved under elementary
extensions.
Moreover, the images $s_U$ and $t_U$ in $\fstr{U}$ 
of $s$ and $t$, respectively, 
are also modally equivalent, since modal truth is transferred by 
elementary maps. Now since $\fstr{U}$ is $\omega$-saturated and 
thus by Lemma~\ref{lem:countable-satur-m-satur}, 
$\nbm{\fstr{U}}$ is modally saturated,
it follows from Proposition~\ref{prop:m-satur-mod-congr} that 
$s_U$ and $t_U$ are behaviourally equivalent.
The construction is illustrated in the following diagram;
$\preceq$ indicates that the map is elementary. 
\[ \xymatrix{
{\MOC_\cla{N}(\alpha)[s] \modeledby \fstr{M}} \ar@{->}[r]^-{i} 
 & \fstr{M}+\fstr{N} \ar[d]^-{\preceq} 
 & {\fstr{N} \models \alpha[t]} \ar@{->}[l]_-{j}\\
 & {\fstr{U}}
}
\] 
Finally, we can transfer the truth of $\alpha(x)$ from $\fstr{N},t$
to $\fstr{M},s$ by using the invariance of modal formulas under
bounded morphisms and standard translations {\em (bm+$\st$)};
elementary maps {\em (elem)};
and the assumption that $\alpha(x)$ is invariant under 
behavioural equivalence {\em ($\alpha(x)$-beh-inv)}.
\[\begin{array}{rclcl}
\fstr{N} \models \alpha[t] 
 & \iff & \fom{(\nbm{\fstr{M}} + \nbm{\fstr{N}})} \models \alpha[j(t)]
   && \text{\small\em (bm+$\st$)}\\
& \iff &\fstr{U} \models \alpha[t_U] && \text{\small\em (elem)} \\ 
& \iff &\fstr{U} \models \alpha[s_U] && \text{\small\em ($s_U \beh t_U$ and $\alpha(x)$-beh-inv)} \\ 
 & \iff & \fom{(\nbm{\fstr{M}} + \nbm{\fstr{N}})} \models \alpha[i(s)] 
   && \text{\small\em (elem)} \\
& \iff &\fstr{M} \models \alpha[s] &\phantom{xx}& \text{\small\em (bm+$\st$)} 
\end{array}\]
\end{proofof}

\begin{rem}
Note that in the proof of Theorem~\ref{thm:char}, 
we could have assumed $\alpha(x)$ to be invariant for any of the 
three equivalence notions, since Proposition~\ref{p:allonone}
tells us that also $s_U \bis t_U$ and  $s_U \rbeh t_U$.
\end{rem}

\begin{rem}\label{rem:mon-char-thm}
An analogue of Van Benthem's theorem for monotonic modal logic was
proved
by Pauly (see~\cite{Pau99:mon-bis,Han03:math-thesis}).
Although the translation of monotonic modal logic and monotonic
neighbourhood models is very similar to ours, 
Pauly's approach is slightly different to the present one, since
his result is not formulated relative to the class of first-order models
which are the translation monotonic models.
Rather, he defines a notion of monotonic bisimulation which applies
to all first-order $\langFO$-models, and shows that
translations of monotonic modal formulas are invariant under this
bisimulation notion, even if the first-order models involved are not
necessarily translations of monotonic models.
This means his result concerns a stronger notion of invariance. 
The converse is shown using $\omega$-saturation and monotonic modal
saturation, and is similar to the proof of the Van Benthem theorem.
We do not get a characterisation theorem for monotonic modal logic 
(relative to translations of monotonic models)
as a direct corollary of Theorem~\ref{thm:char},
but we believe it is possible to prove one
using the same line of argumentation and constructions.
\end{rem}

\begin{rem}\label{rem:char-thm-generalisation}
It seems straightforward to generalise Theorem~\ref{thm:char}
to multi-modal classical modal logic 
with polyadic modalities of finite arity.
Multi-modal neighbourhood models are of interest
in coalgebraic modal logic due to the following:

It is not always possible to find a collection
of separating unary, predicate liftings for a functor 
$\Fun\colon\Set\to\Set$. However,
Schr\"{o}der showed in \cite{Schr08:TCS-expr} that any 
finitary functor $\Fun$ has a separating 
set of finitary, {\em polyadic} predicate liftings, i.e.,
there exists a finitary coalgebraic modal logic
with polyadic modalities which is
expressive for $\Fun$-coalgebras. 
A $k$-ary predicate lifting 
$\predlif\colon {(2^{(-)})}^\k \to 2^{\Fun(-)}$
has transposite
$\transp{\predlif}_X \colon  \Fun(X) \to \Nbk(X)$,
where $\Nbk$ denotes the functor
$\Nbk = 2^{(-)} \circ (2^{(-)})^\k$. 
Note that a map $X \to \Nbk(X)$ is a $\k$-ary
neighbourhood function.
If $\Lambda$ is a separating set
of $\k$-ary predicate liftings for $\Fun$, 
then for all sets $X$, the source of transposites
$\{ \transp{\predlif}_X \colon 
  \Fun(X) \to \Nbk(X) \mid \predlif \in \Lambda\}$
yields a natural embedding.
\begin{equation}\label{eq:emb-Fun-NbK}
\tup{\transp{\predlif}}_{\predlif\in\Lambda}
 \colon \Fun \to \Pi_{{\Lambda}} \Nbk,
\end{equation}
where $\Pi_{{\Lambda}} \Nbk$ is the $\card{\Lambda}$-fold 
product of $\Nbk$.
Hence for every finitary functor $\Fun$,
an $\Fun$-coalgebra can transformed into a pointwise equivalent 
multi-modal, polyadic neighbourhood frame.

\end{rem}


\subsection{Interpolation}
\label{ssec:interpolation}

In this section we show that the results on ultrafilter extensions
from the previous section can be used to prove 
Craig interpolation for classical modal logic.
For several normal and monotonic modal logics, 
Craig interpolation can be proved using superamalgamation in the 
corresponding variety of modal algebras, 
see e.g.~\cite{GabMak:interp,HanKup04:CMCS-UpP,Madarasz-1,Madarasz-2,Marx:phd}.
We believe similar proofs can be carried out for 
classical modal logic.
Our proof, however, is based on the ideas used in
the  proof of Craig interpolation
for normal modal logic presented in \cite{AndrekaBenNem98:bounded-frags}.
The proof in \cite{AndrekaBenNem98:bounded-frags} 
uses first-order model-theoretic arguments similar to those 
employed in the proof of the Van Benthem characterisation theorem,
but Theorem~\ref{thm:somewhere} allows us to prove
Craig interpolation in a purely modal setting, 
without the use of $\omega$-saturated models or the explicit use of 
algebraic duality.
All that is needed is that modal truth is invariant under
ultrafilter extensions (Lemma~\ref{lem:princip-uf-truth}), 
and that ultrafilter extensions are modally saturated
(Proposition~\ref{prop:uf-sat}).

So far we have worked with a fixed a set $\At$ of atomic propositions,
giving rise to the language $\langML = \langML(\At)$.
In the current section we need to generalise our notions of 
bounded morphism and modal saturation
to sublanguages $\langML(\At')$ of $\langML(\At)$
generated by a specific subset $\At'$ of atomic propositions.
We point out that all models are always models for the full
language $\langML(\At)$.
This generalisation is straightforward, but in the interest of clarity
we provide the details and the exact results we need.
Let $\At' \subseteq \At$, and
let $\nstr{M}_1=\tup{ S_1, \nu_1, V_1}$ and
 $\nstr{M}_2=\tup{ S_2, \nu_2, V_2}$
be neighbourhood $\langML(\At)$-models.
 A function $f\colon S_1 \to S_2$ is a 
{\em bounded $\langML(\At')$-morphism from $\nstr{M}_1$ to $\nstr{M}_2$}
(notation: $f\colon \nstr{M}_1 \to_{\langML(\At')} \nstr{M}_2$)
if
$f$ is a bounded (frame) morphism from
$\tup{S_1,\nu_1}$ to $\tup{S_2,\nu_2}$, and
for all $p \in \At'$, and all $s \in S_1$:
$s\in V_1(p)$ iff $f(s) \in V_2(p)$. 
An $\langML(\At')$-congruence is the kernel of
a bounded $\langML(\At')$-morphism.
Two states $s_1 \in S_1$ and $s_2 \in S_2$ are 
{\em modally $\langML(\At')$-equivalent} 
(notation: $s_1 \modeq_{\langML(\At')} s_2$), 
if they satisfy the same $\langML(\At')$-formulas.
Given a neighbourhood $\langML(\At)$-model $\nstr{M} = \tup{S,\nu,V}$, 
a subset $X \subseteq S$ is {\em modally $\langML(\At')$-compact}
if for all sets $\Psi$ of modal $\langML(\At')$-formulas,
$\Psi$ is satisfiable in $X$, whenever $\Psi$ is finitely satisfiable
in $X$, and $\nstr{M}$ is 
{\em modally $\langML(\At')$-saturated}
if for every $\modeq_{\langML(\At')}$-coherent neighbourhood 
$X$, both $X$ and $\cmp{X}$ are modally $\langML(\At')$-compact.

\begin{lem}\label{lem:restricted-notions}
Let $\At' \subseteq \At$.
\begin{enumerate}[\em(1)]
\item If $\nstr{M}_1$ and
 $\nstr{M}_2$
are $\langML(\At)$-neighbourhood models, and
 $f\colon \nstr{M}_1 \to_{\langML(\At')} \nstr{M}_2$,
then for all $s$ in $\nstr{M}_1$, and all $\varphi \in \langML(\At')$:
$\nstr{M}_1, s \models \varphi$ iff $\nstr{M}_2, f(s) \models \varphi$.
\item  If $\nstr{M} = \tup{S,\nu,V}$ is a neighbourhood
  $\langML(\At)$-model, and
$R \subseteq S \times S$ is an equivalence relation,
then $R$ is an $\langML(\At')$-congruence on $\nstr{M}$
iff $R$ is a congruence on the underlying frame $\tup{S,\nu}$,
and for all $\tup{s,t} \in R$, and all $p\in \At'$: 
$s \in V(p)$ iff $t \in V(p)$.
\item If a neighbourhood $\langML(\At)$-model $\nstr{M}$ is modally
  $\langML(\At')$-saturated, then all $\modeq_{\langML(\At')}$-coherent
subsets are definable by an  $\langML(\At')$-formula.
\item If a neighbourhood $\langML(\At)$-model $\nstr{M}$ is modally
  $\langML(\At')$-saturated, then $\modeq_{\langML(\At')}$ is
an $\langML(\At')$-congruence.
\item  If $\nstr{M}$ is neighbourhood $\langML(\At)$-model,
then its ultrafilter extension $\nstr{M}^u$ 
is modally $\langML(\At')$-saturated.
\end{enumerate}
\end{lem}
\proof
As usual, 1 can be proved by straightforward formula induction.
Item 2 is immediate.
Item 3 can be proved by retracing the argument used in 
Lemma~\ref{lem:m-satur-definability}.
Item 4 follows from item 3 and essentially the same
argument used in Lemma~\ref{lem:modcoh-definable}.
Item 5 can be proved in the same way as Proposition~\ref{prop:uf-sat}.   
\qed

For a formula $\varphi \in \langML$, we denote by $\Prop(\varphi)$
the set of atomic propositions occurring in $\varphi$.
Recall that for $\Phi \cup \{\varphi\} \subseteq \langML$,
we write $\Phi \models \varphi$ if 
$\varphi$ is a local semantic consequence of $\Phi$ over the class of 
all neighbourhood models.
Note that compactness of $\models$ follows from the compactness of 
$\models_\cla{N}$, the first-order consequence relation over the class of 
neighbourhood models.

\begin{thm}[Craig interpolation]\label{thm:interpolation}
Let $\varphi_1, \varphi_2 \in \langML$.
If $\;\models \varphi_1 \ra \varphi_2$, then there exists
a formula $\chi \in \langML$ with 
$\Prop(\chi) \subseteq \Prop(\varphi_1) \cap \Prop(\varphi_2)$
such that $\models \varphi_1 \ra \chi$ and $\models \chi \ra \varphi_2$.
\end{thm}

\newcommand{\incM}{\iota}
\newcommand{\incN}{\kappa}

\proof
Assume that $\models \varphi_1 \ra \varphi_2$.
Let $\At_i = \Prop(\varphi_i)$, $i=1,2$, and $\At_0 = \At_1 \cap \At_2 $.
Denote by 
$\Cons_{\langML(\At_0)}(\varphi_1) = \{ \chi \in \langML(\At_0) \mid \varphi_1 \models \chi \}$
the set of modal $\langML(\At_0)$-consequences of $\varphi_1$.
It suffices to show that $\Cons_{\langML(\At_0)}(\varphi_1) \models \varphi_2$,
since then by compactness, there are 
$\chi_1, \ldots, \chi_n \in \Cons_{\langML(\At_0)}(\varphi_1)$
such that $\chi_1 \land \ldots \land\chi_n \models \varphi_2$,
and  $\varphi_1 \models \chi_1 \land \ldots \land\chi_n$, i.e,
$\chi = \chi_1 \land \ldots \land\chi_n$ is a Craig interpolant for 
$\varphi_1 \ra \varphi_2$.

So, assume $\nstr{M}$ is an $\langML(\At)$-model and $s$ is a state 
in $\nstr{M}$ such that
$\nstr{M}, s \models \Cons_{\langML(\At_0)}(\varphi_1)$, and let
$\Psi = \{ \psi \in \langML(\At_0) \mid \nstr{M}, s \models \psi \}$. 
Now $\Psi \cup \{\varphi_1\}$ is consistent, since otherwise there would exist
$\{ \psi_1, \ldots, \psi_n\} \subseteq \Psi$ such that 
$\models \psi_1 \land \ldots \land \psi_n  \ra \lnot \varphi_1$,
hence
$\models \varphi_1 \ra \lnot\psi_1 \lor \ldots \lor\lnot\psi_n$,
which would imply that 
$\lnot\psi_1 \lor \ldots \lor\lnot\psi_n \in \Cons_{\langML(\At_0)}(\varphi_1)$
contradicting the assumption that 
$\nstr{M}, s \models \Cons_{\langML(\At_0)}(\varphi_1)$.

By definition of $\models$, $\Psi \cup \{\varphi_1\}$ is satisfiable in some
neighbourhood $\langML(\At)$-model $\nstr{N}$ at a 
state $t$ in $\nstr{N}$, i.e., $\nstr{N}, t\models \Psi \cup \{\varphi_1\}$.
Then by construction $s \modeq_{\langML(\At_0)} t$, and
as truth is preserved by the injections
$\incM\colon \nstr{M} \to \nstr{N} +\nstr{M}$ and
$\incN\colon \nstr{N} \to \nstr{N} +\nstr{M}$, and when passing to
ultrafilter extensions, the principal ultrafilters 
generated by $\incM(s)$ and $\incN(t)$ are 
also modally $\langML(\At_0)$-equivalent
in $\nstr{U} = \tup{U,\mu,V} = {(\nstr{N} + \nstr{M})}^u$,
i.e., $\princ_{\incM(s)} \modeq_{\langML(\At_0)} \princ_{\incN(t)}$.
Now since ultrafilter extensions are modally $\langML(\At_0)$-saturated 
(Lemma~\ref{lem:restricted-notions}(5))
it follows from 
Lemma~\ref{lem:restricted-notions}(4) that
$\modeq_{\langML(\At_0)}$ is an
$\langML(\At_0)$-congruence on $\nstr{U}$.
For ease of notation, we denote the relation $\modeq_{\langML(\At_0)}$
on $\nstr{U}$ by $Z$ in the rest of this proof.
We have, in particular, 
$Z$ is a congruence on the underlying frame $\tup{U,\mu}$ of $\nstr{U}$,
and by Proposition~\ref{p:allonone}
$Z$ is also a $\TTwo$-bisimulation on $\tup{U,\mu}$. 
This means there exists a
coalgebra map $\zeta\colon Z \to \TTwo(Z)$ such that the projections 
$\pi_i\colon \tup{Z,\zeta} \to \tup{U,\mu}$, $i=1,2$, 
are bounded frame morphisms.
We now define a valuation $V'$ on $\tup{Z,\zeta}$ 
to obtain a neighbourhood $\langML(\At)$-model 
$\nstr{Z} = \tup{Z,\zeta,V'}$ such that 
$\pi_1\colon \nstr{Z} \to \nstr{U}$ 
is a bounded $\langML(\At_1)$-morphism
and 
$\pi_2\colon \nstr{Z} \to \nstr{U}$ 
is a bounded $\langML(\At_2)$-morphism.
Let $p \in \At$ and $\tup{u_1,u_2} \in Z$, 
then we define
\[ \tup{u_1,u_2} \in V'(p) \iff \left\{\begin{array}{ll}
u_1 \in V(p) & \text{if } p \in \At_1,\\
u_2 \in V(p) & \text{if } p \in \At_2,\\
\text{never} & \text{if } p \in \At\setminus (\At_1 \cup \At_2).
\end{array}\right.
\]
Note that $V'$ is well-defined due to
Lemma~\ref{lem:restricted-notions}(2).
The construction is illustrated below. The dashed arrow going to $\nstr{U}$
indicates that the principal ultrafilter map $\princ$ is not a 
bounded morphism, still $\princ$ does preserve modal truth
(Lemma~\ref{lem:princip-uf-truth}).
\[ \xymatrix{
  {  \varphi_1 \modeledby \nstr{N}, t} \ar@{->}[r]^-{\incN}
 & \nstr{N}+\nstr{M} \ar@{-->}[d]^-{\modeq}_-{\princ} 
 & {\nstr{M}, s \models \Cons_{\langML(\At_0)}(\varphi_1)} \ar@{->}[l]_-{\incM} \\
 & {\nstr{U}}\\
 & \nstr{Z} \ar@/^/[u]^-{\pi_1} \ar@/_/[u]_-{\pi_2}
}
\]
Now we have:
$\nstr{N}, t \models \varphi_1$ implies
$\nstr{U}, \princ_{\incN(t)} \models \varphi_1$.
Since $\tup{\princ_{\incN(t)},\princ_{\incM(s)}} \in Z$ and 
$\pi_1$ is a bounded  $\langML(\At_1)$-morphism
from $\nstr{Z}$ to $\nstr{U}$,
we have
\mbox{$\nstr{Z}, \tup{\princ_{\incN(t)},\princ_{\incM(s)}} \models \varphi_1$}. 
By the main assumption that 
\mbox{$\models \varphi_1 \ra \varphi_2$}, we get that
$\nstr{Z}, \tup{\princ_{\incN(t)},\princ_{\incM(s)}} \models \varphi_2$, and now since
$\pi_2$ is a bounded $\langML(\At_2)$-morphism from 
$\nstr{Z}$ to $\nstr{U}$, we get
$\nstr{U}, \princ_{\incM(s)} \models \varphi_2$ and hence
$\nstr{M}, s \models \varphi_2$.
\qed


\section{Conclusion and related work}

In the first part of this paper we 
discussed and compared different notions of equivalence between
neighbourhood structures.
We gave back-and-forth style characterisations
of $\TTwo$-bisimulations and \precocongs, and
showed that, as expected, 
behavioural equivalence is the only one of the three notions 
that allows us to prove a Hennessy-Milner theorem for
image-finite neighborhood models
(cf.~Section~\ref{sec:HM-classes}). 
Furthermore, we showed that 
for an {\em arbitrary} $\Set$-functor $\Fun$,
{\precocongs} capture behavioural equivalence on a single
$\Fun$-coalgebra (Theorem~\ref{t:on}). 
For functors $\Fun$ that weakly preserve kernel pairs, such as $\TTwo$,
this is already achieved with $\Fun$-bisimulations
\cite{GummSchr05:types-coal},
but we believe that {\precocongs}
could be an interesting alternative to $\Fun$-bisimulations
for functors which lack this property.
A first indication of this is \cite{Kup:term-seq-games}
where {\precocongs} are used
to obtain a game-theoretic characterisation of
behavioural equivalence.

After having reached a good understanding of 
state equivalence over neighbourhood structures,
we focused on generalising two well-known 
model-theoretic results to the setting
 of neighbourhood models: the Van Benthem Characterisation Theorem 
(Theorem~\ref{thm:char}) and Craig Interpolation 
(Theorem~\ref{thm:interpolation}). 
Our proof of Theorem~\ref{thm:char} builds on ideas from the 
original proof of the Van Benthem characterisation theorem 
(\cite{Benthem:Correspondence}).  Closely related to our work 
are also the invariance results by 
Pauly~(\cite{Pau99:mon-bis}) on monotonic modal logic,   
and Ten Cate et al.~(\cite{CateGabSus:topoML})
on topological modal logic.

A number of other model-theoretic results are worth exploring.
Perhaps the most interesting one is a generalisation of the 
Goldblatt-Thomason Theorem (see e.g.~\cite{BdRV:ML-book}).
The classic result for Kripke models can be proved
using model-theoretic constructs or by using algebraic duality.
The algebraic duality proof has already been
generalised to the coalgebraic setting by
Kurz \& Rosick\'{y}'s~\cite{KurRos07:GoldblattThomason-thm}.
Indeed, a special case of their main result is the result we are after: 
a Goldblatt-Thomason Theorem for neighbourhood models 
(cf. \cite{KurRos07:GoldblattThomason-thm}, Corollary 3.17(2) and Remark 3.18).
Given the formal machinery we have developed in this paper 
(e.g., the ultrafilter extensions from Section~\ref{ssec:ultrafilter-ext}), 
one may hope for a model-theoretic proof of this result 
(see e.g., Section 3.8 in \cite{BdRV:ML-book}).
Such a model-theoretic proof has been given for 
topological models (which are special cases of neighbourhood models)
by Ten Cate et al.~(\cite{CateGabSus:topoML}).
However, 
an important ingredient in the model-theoretic proof for the Kripke case
is the fact that any Kripke model is bisimilar to the 
disjoint union of its generated submodels.  
This is not true for an arbitrary neighbourhood model 
(cf.~\cite{Gumm01:func-coalg}), and at the moment, it is not clear which
alternative construction could be used in its place.

A second model-theoretic issue raised by the results in this paper concerns 
our translation of the modal language into a two-sorted first-order language 
(cf.~Definition~\ref{fotransforms}). As is well-known, with respect to 
Kripke structures, the basic modal language can be translated into the 
{\em guarded fragment} of first-order logic 
(cf.~\cite{AndrekaBenNem98:bounded-frags}).  
This fact has been used to explain a number of the important  properties of 
modal logic (see, for example, \cite{AndrekaBenNem95:modal-classical} 
for an extensive discussion).
The question is whether classical modal logic is also
contained in some kind of guarded fragment.
Our translation of $\Box\varphi$ does not fall into the 
guarded fragment of two-sorted first-order logic. However, it is not 
difficult to see that over the class $\cla{N}$ of neighbourhood models
viewed as first-order structures, 
$\st_x(\Box\varphi)$ is equivalent to the following 
single-sorted first-order formula: 
 $$\exists u (Nbhd(u) \wedge x\fof{N}u \wedge \forall y(u\fof{E}y \rightarrow \st_y(\varphi)) \wedge\forall  y(State(y) \rightarrow (\neg (\st_y(\varphi)) \vee u\fof{E}y)))$$
where $Nbhd$ and $State$ are designated predicates 
intended to mean ``...is a neighbourhood'' and ``...is a state'', 
respectively.   This formula is in the (loosely) guarded fragment.

Our characterisation theorem for classical modal logic
leads to a number of interesting research questions.
For example,
we would like to explore the possibility of proving our result 
using game-theoretic techniques 
similar to the ones exploited by Otto (\cite{Otto06:bis-inv-finite}).
Furthermore, neighbourhood structures can also be seen as a type of
Chu spaces. 
We would like to relate our characterisation theorem
to Van Benthem's characterisation of the Chu transform invariant 
fragment of a two-sorted first-order logic
in~\cite{Benthem:Chu}.

Finally,
it would be interesting to find out if
our characterisation theorem
can be generalised to coalgebraic modal logic
for an arbitrary finitary functor $\Fun\colon\Set\to\Set$,
using the embedding of $\Fun$-coalgebras into
multi-modal, $\k$-ary neighbourhood frames
as described in Remark~\ref{rem:char-thm-generalisation}.
It might be possible to prove that, under certain assumptions, 
the coalgebraic modal logic over $\Fun$-coalgebras can be viewed
as the bisimulation invariant fragment of some many-sorted
first-order logic.
Initial investigations suggest that this is possible for
functors of the form $A^{(2^\k)^{(-)}}$ where $A$ is a finite set
and $\k$ is a natural number.
An $A^{(2^\k)^{(-)}}$-coalgebra can be seen as a 
multi-modal, polyadic neighbourhood frame
$\tup{X,\{\nu_a \mid a \in A\}}$
given by an $A$-indexed collection of $\k$-ary neighbourhood functions
$\nu_a \colon X \to 2^{(2^X)^\k}$
such that for each $\k$-tuple of subsets $\tup{U_1,\dotsc,U_\k}$
and each state $x \in X$,
$\tup{U_1,\dotsc,U_\k} \in \nu_a(x)$ for exactly one $a \in A$.
We must leave the details of this result as future work.


\section*{Acknowledgements}
We would like to thank H. Peter Gumm for many fruitful discussions,
and
Yde Venema for initiating our cooperation on this subject. Special 
thanks also goes to the anonymous referees for useful comments and 
corrections.

\bibliography{nbhd-lmcs} 

\begin{thebibliography}{10}

\bibitem{AczMen:final}
P.~Aczel and N.P. Mendler.
\newblock A final coalgebra theorem.
\newblock In D.H. Pitt, D.E. Rydeheard, P.~Dybjer, A.M. Pitts, and
  A.Poign{\'e}, editors, {\em Category Theory and Computer Science}, volume 389
  of {\em Lecture Notes in Computer Science}, pages 357--365, 1989.

\bibitem{AdaHerStr90:ACC}
J.~Ad\'{a}mek, H.~Herrlich, and G.E. Strecker.
\newblock {\em Abstract and Concrete Categories: The Joy of Cats}.
\newblock J. Wiley and Sons, 1990.
\newblock Online version: \nolinkurl{http://katmat.math.uni-bremen.de/acc}.

\bibitem{AdaPor04:tree-coalgs}
J.~Ad{\'a}mek and H.-E. Porst.
\newblock On tree coalgebras and coalgebra presentations.
\newblock {\em Theoretical Computer Science}, 311:257--283, 2004.

\bibitem{AlurHenzKupf02:ATL}
R.~Alur, T.A. Henzinger, and O.~Kupferman.
\newblock Alternating-time temporal logic.
\newblock {\em Journal of the ACM}, 49(5):672--713, 2002.

\bibitem{AndrekaBenNem95:modal-classical}
H.~Andr\'{e}ka, J.~van Benthem, and I.~N\'{e}meti.
\newblock Back and forth between modal logic and classical logic.
\newblock {\em Logic Journal of the {IGPL}}, 3:685 -- 720, 1995.

\bibitem{AndrekaBenNem98:bounded-frags}
H.~Andr\'{e}ka, J.~van Benthem, and I.~N\'{e}meti.
\newblock Modal languages and bounded fragments of predicate logic.
\newblock {\em Journal of Philosophical Logic}, 27(3):217--274, 1998.

\bibitem{Benthem:phd}
J.~{van} Benthem.
\newblock {\em Modal Correspondence Theory}.
\newblock PhD thesis, Mathematical Institute, University of Amsterdam, 1976.

\bibitem{Benthem:Correspondence}
J.~{van} Benthem.
\newblock Correspondence theory.
\newblock In D.~Gabbay and F.~Guenthner, editors, {\em Extensions of Classical
  Logic}, volume~II of {\em Handbook of Philosophical Logic}, pages 167--247.
  Reidel, Dordrecht, 1984.

\bibitem{Benthem:Chu}
J.~{van} Benthem.
\newblock Information transfer across {C}hu spaces.
\newblock {\em Logic Journal of the IGPL}, 8(6):719--731, 2000.

\bibitem{BdRV:ML-book}
P.~Blackburn, M.~{de} Rijke, and Y.~Venema.
\newblock {\em Modal Logic}.
\newblock Cambridge University Press, 2001.

\bibitem{CateGabSus:topoML}
B.~{ten} Cate, D.~Gabelaia, and D.~Sustretov.
\newblock Modal languages for topology: Expressivity and definability.
\newblock {\em To appear in Annals of Pure and Applied Logic}.
\newblock Preprint available at: \url{http://arxiv.org/abs/math/0610357}.

\bibitem{ChangKeisler73}
C.~Chang and H.~Keisler.
\newblock {\em Model Theory}.
\newblock North-Holland, 1973.

\bibitem{Chellas}
B.F. Chellas.
\newblock {\em Modal Logic - An Introduction}.
\newblock Cambridge University Press, 1980.

\bibitem{Dosen:nbhd-dua}
K.~Do{\v{s}}en.
\newblock Duality between modal algebras and neighbourhood frames.
\newblock {\em Studia Logica}, 48:219--234, 1989.

\bibitem{FlumZiegler:topo-model-theory}
J.~Flum and M.~Ziegler.
\newblock {\em Topological {M}odel {T}heory}, volume 769 of {\em Lecture Notes
  in Mathematics}.
\newblock Springer Verlag, 1980.

\bibitem{GabMak:interp}
D.M. Gabbay and L.~Maksimova.
\newblock {\em Interpolation and Definability: Modal and Intuitionistic Logic}.
\newblock Number~46 in Oxford Logic Guides. Oxford University Press, 2005.

\bibitem{Goble:Murder}
L.~Goble.
\newblock Murder most gentle: The paradox deepens.
\newblock {\em Philosophical Studies}, 64(2):217--227, 1991.

\bibitem{Gumm01:func-coalg}
H.P. Gumm.
\newblock Functors for coalgebras.
\newblock {\em Algebra Universalis}, 45:135--147, 2001.

\bibitem{GummSchr05:types-coal}
H.P. Gumm and T.~Schr{\"o}der.
\newblock Types and coalgebraic structure.
\newblock {\em Algebra universalis}, 53:229--252, 2005.

\bibitem{Han03:math-thesis}
H.H. Hansen.
\newblock Monotonic modal logic ({M}aster's thesis).
\newblock Research Report PP-2003-24, Institute for Logic, Language and
  Computation. University of Amsterdam, 2003.

\bibitem{HanKup04:CMCS-UpP}
H.H. Hansen and C.~Kupke.
\newblock A coalgebraic perspective on monotone modal logic.
\newblock In {\em Proceedings of the 7th Workshop on Coalgebraic Methods in
  Computer Science (CMCS)}, volume 106 of {\em Electronic Notes in Theoretical
  Computer Science}, pages 121--143. Elsevier Science Publishers, 2004.

\bibitem{HanKupPac07:CALCO-nbis}
H.H. Hansen, C.~Kupke, and E.~Pacuit.
\newblock Bisimulation for neighbourhood structures.
\newblock In T.~Mossakowski et~al., editor, {\em Proceedings of the 2nd
  Conference on Algebra and Coalgebra in Computer Science (CALCO 2007)}, volume
  4624 of {\em Lecture Notes in Computer Science}, pages 279--293. Springer,
  2007.

\bibitem{Kup:term-seq-games}
C.~Kupke.
\newblock Terminal sequence induction via games.
\newblock In {\em Proceedings of the 7th International Tbilisi Symposium on
  Language, Logic and Computation}, 2008.

\bibitem{KupKurPat:uf-ext}
C.~Kupke, A.~Kurz, and D.~Pattinson.
\newblock Ultrafilter extensions for coalgebras.
\newblock In J.L. Fiadero, N.~Harman, M.~Roggenbach, and J.J.M.M. Rutten,
  editors, {\em Proceedings of the 2nd Workshop on Algebraic and Coalgebraic
  Methods in Computer Science (CALCO 2005)}, volume 3629 of {\em Lecture Notes
  in Computer Science}, pages 263--277. Springer, 2005.

\bibitem{Kurz:personal}
A.~Kurz.
\newblock Personal communication.

\bibitem{Kurz:diss}
A.~Kurz.
\newblock {\em {Logics for Coalgebras and Applications to Computer Science}}.
\newblock PhD thesis, {Ludwig-Maximilians-Universit{\"a}t}, 2000.

\bibitem{KurRos07:GoldblattThomason-thm}
A.~Kurz and J.~Rosick\'{y}.
\newblock The {G}oldblatt-{T}homason-theorem for coalgebras.
\newblock In T.~Mossakowski et~al., editor, {\em Proceedings of the 2nd
  Conference on Algebra and Coalgebra in Computer Science (CALCO)}, volume 4624
  of {\em Lecture Notes in Computer Science}, pages 342--355. Springer, 2007.

\bibitem{Madarasz-1}
J.~Madar\'{a}sz.
\newblock Interpolation and amalgamation; pushing the limits. {P}art {I}.
\newblock {\em Studia Logica}, 61:311--345, 1998.

\bibitem{Madarasz-2}
J.~Madar\'{a}sz.
\newblock Interpolation and amalgamation; pushing the limits. {P}art {II}.
\newblock {\em Studia Logica}, 62:1--19, 1999.

\bibitem{Marx:phd}
M.~Marx.
\newblock {\em Algebraic Relativization and Arrow Logic}.
\newblock PhD thesis, University of Amsterdam, 1995.

\bibitem{Montague:univ-grammar}
R.~Montague.
\newblock Universal grammar.
\newblock {\em Theoria}, 36:373--398, 1970.

\bibitem{Otto06:bis-inv-finite}
M.~Otto.
\newblock Bisimulation invariance and finite models.
\newblock In W.~Pohlers Z.~Chatzidakis, P.~Koepke, editor, {\em Logic
  Colloquium '02}, volume~27 of {\em Lecture Notes in Logic}, pages 276--298.
  Association for Symbolic Logic, 2006.

\bibitem{PadGovSu07:KRAQ}
V.~Padmanabhan, G.~Governatori, and K.~Su.
\newblock Knowledge assesment: A modal logic approach.
\newblock In {\em Proceedings of the 3rd Int. Workshop on Knowledge and
  Reasoning for Answering Questions (KRAQ)}, 2007.

\bibitem{patt03:NASSLLI-notes}
D.~Pattinson.
\newblock An introduction to the theory of coalgebras.
\newblock Lecture notes accompanying the course at NASSLLI 2003. Available at
  \nolinkurl{http://www.pst.ifi.lmu.de/~pattinso/Publications/nasslli.all.ps.g%
z}.

\bibitem{Patt03:coalg-ML}
D.~Pattinson.
\newblock Coalgebraic modal logic: Soundness, completeness and decidability of
  local consequence.
\newblock {\em Theoretical Computer Science}, 309(1--3):177--193, 2003.

\bibitem{Pau99:mon-bis}
M.~Pauly.
\newblock Bisimulation for general non-normal modal logic.
\newblock Manuscript (unpublished), 1999.

\bibitem{Pau:phd}
M.~Pauly.
\newblock {\em Logic for Social Software}.
\newblock PhD thesis, University of Amsterdam, 2001.

\bibitem{Pau02:ML-coal-pow}
M.~Pauly.
\newblock A modal logic for coalitional power in games.
\newblock {\em Journal of Logic and Computation}, 12(1):149--166, 2002.

\bibitem{Rut00:TCS-univ-coal}
J.J.M.M. Rutten.
\newblock Universal coalgebra: a theory of systems.
\newblock {\em Theoretical Computer Science}, 249:3--80, 2000.

\bibitem{Schr08:TCS-expr}
L.~Schr\"{o}der.
\newblock Expressivity of coalgebraic modal logic: The limits and beyond.
\newblock {\em Theoretical Computer Science}, 390:230--247, 2008.

\bibitem{Scott:advice}
D.~Scott.
\newblock Advice on modal logic.
\newblock In K.~Lambert, editor, {\em Philosophical Problems in Logic}, pages
  143--173. Reidel, Dordrecht, 1970.

\bibitem{Segerberg71:classic-ML}
K.~Segerberg.
\newblock {\em An Esssay in Classical Modal Logic}.
\newblock Number~13 in Filosofiska Studier. Uppsala Universitet, 1971.

\bibitem{Vardi86:epistemic}
M.Y. Vardi.
\newblock On epistemic logic and logical omniscience.
\newblock In J.~Halpern, editor, {\em Proceedings of the 1986 Conference on
  Theoretical Aspects of Reasoning about Knowledge (TARK)}, pages 293--305.
  Morgan Kaufmann, 1986.

\bibitem{Ven06:Handbook-ML-AC}
Y.~Venema.
\newblock Algebras and coalgebras.
\newblock In P.~Blackburn, J.~van Benthem, and F.~Wolter, editors, {\em
  Handbook of Modal Logic}, volume~3 of {\em Studies in Logic and Practical
  Reasoning}, pages 331--426. Elsevier, 2006.

\end{thebibliography}
\bibliographystyle{plain}

\end{document}